\documentclass{mn2e}

\usepackage{mathtools}
\usepackage{graphicx}
\usepackage{subfig}
\usepackage{epstopdf}
\usepackage{natbib} 
\newcommand{\rvir}{R$_{\rm{vir}}$}
\newcommand{\logN}{log($N_{\rm{HI}}$)}
\newcommand{\NHI}{$N_{\rm{HI}}$}
\newcommand{\lya}{Ly$\alpha$}

\newcommand{\Enzo}{\textsc{Enzo}}
\newcommand{\fcover}{f$_{\rm{cover}}$}

\begin{document}

%

\title[Warm Gas in Galaxy Clusters]{Warm Gas in and Around Simulated Galaxy Clusters as Probed by Absorption Lines}
\author[A. Emerick et. al.]{A.~Emerick,$^1$\thanks{e-mail: emerick@astro.columbia.edu} G.~Bryan,$^1$ M.E.~Putman$^1$\\
 $^1$ Department of Astronomy, Columbia University, New York, NY 10025}

\maketitle
\begin{abstract}
Understanding gas flows into and out of the most massive dark matter structures in our Universe, galaxy clusters, is fundamental to understanding their evolution. Gas in clusters is well studied in the hot ($>$ 10$^{6}$ K) and cold ($<$ 10$^{4}$ K) regimes, but the warm gas component (10$^{4}$ - 10$^{6}$ K) is poorly constrained.  It is challenging to observe directly, but can be probed through \lya\  absorption studies. We produce the first systematic study of the warm gas content of galaxy clusters through synthetic \lya\  absorption studies using cosmological simulations of two galaxy clusters produced with \Enzo.  We explore the spatial and kinematic properties of our cluster absorbers, and show that the majority of the identified absorbers are due to fast moving gas associated with cluster infall from IGM filaments. Towards the center of the clusters, however, the warm IGM filaments are no longer dominant and the absorbers tend to have higher column densities and metallicities, representing stripped galaxy material. We predict that the absorber velocity distribution should generally be bi-modal and discuss the effects of cluster size, mass, and morphology on the properties of the identified absorbers, and the overall cluster warm gas content. We find tentative evidence for a change in the well known increasing \NHI\  with decreasing impact parameter for the most massive dark matter halos. Our results are compared directly to observations of \lya\  absorbers in the Virgo cluster, and provide predictions for future \lya\ absorption studies.
\end{abstract}

\begin{keywords}
Methods: numerical -- Galaxies: clusters -- intergalactic medium 
\end{keywords}

\section{Introduction}
Galaxy clusters are the largest virialized structures in our universe. Clusters host dark matter potential wells around 10$^{14}$-10$^{15}$M$_{\odot}$ that are able to heat the infalling gas to temperatures of 10$^{7}$-10$^{8}$K. This hot  intracluster medium (ICM), constitutes the most significant baryonic component of galaxy clusters and is well observed in thermal X-ray emission \citep[][and references therein]{Voit2005}.  The flow of gas into and out of galaxies within clusters, and into the cluster itself, is strongly affected by the cluster environment.  Our understanding of the competing processes involved is still forming, from galaxy level processes such as galactic outflows and inflows to cluster specific processes such as ram-pressure stripping, galaxy mergers, galaxy harassment, and starvation \citep[e.g.,][]{Cayette1990, Vollmer2001, Lewis2002, Roediger2005, Chung2007, TonnesenBryan2007, Y12}. In addition, our understanding of gas flows onto clusters from IGM filaments is incomplete.  Much of this gas is expected to exist in a warm, diffuse phase at temperatures near T $\sim$ 10$^{4}$ - 10$^{6}$ K \citep[e.g.,][]{Burns2010, Werner2008, Kravstov2012}. Although much of the baryonic mass in the Universe exists in this phase, it remains relatively unexplored in a cluster environment. This is partly because this regime is too diffuse to observe directly via 21cm emission, and too cool to be seen in X-ray. However, \lya\ absorption along lines of sight towards background QSO's presents a valuable means to probe the warm gas \citep[e.g.,][]{Tripp2005}. 

\lya\ absorption along lines of sight towards background QSO's has long been a useful probe of the intergalactic medium (IGM) at high redshifts (z $\sim$ 2 - 6) (see \cite{Rauchlya} and references therein). A majority of the baryons in the Universe reside in a diffuse state that comprises the IGM. \lya\ absorption can account for nearly all of the high redshift baryons in the Universe \citep{Weinberg1997}. Our knowledge of the \lya\ forest in the nearby Universe is comparatively lacking, although attempts have been made recently towards understanding this redshift regime both observationally \citep[e.g.,][]{Danforth2014} and with simulations \citep[e.g.,][]{Dave2010,Juna}. \lya\ absorption at z $\la$ 2 is possible with UV spectrographs aboard HST --- most recently with the Cosmic Origins Spectrograph (COS). Although \lya\ has often been used to study the IGM and warm gas associated with galaxies \citep[e.g.,][]{Tumlinson2013}, there has been little examination of warm, \lya\ absorbers in the cluster environment. To date \cite{Y12} (hereafter Y12) has been the only systematic study of warm gas via \lya\ absorption in galaxy clusters. Y12 examined 43 \lya\ absorbers along 23 sightlines towards the Virgo cluster. They found that the warm gas in Virgo is preferentially located in the cluster outskirts and associated with substructures. They conclude that the observed gas primarily represents gas flow into the Virgo cluster. To extend the observations of Y12 to morphologically different galaxy clusters, there are ongoing observations of \lya\ absorption along 9 sightlines towards the Coma cluster (Yoon et al., in preparation) and planned observations of single sightlines through 11 clusters (PI: Tejos, HST PID:13833).

The Virgo and Coma clusters provide a unique laboratory to probe the warm gas content in two distinct clusters. The Virgo cluster is the closest cluster to our Milky Way, at 16.5 Mpc \citep{virgo-vel-disp}. It is an irregular, cool-core cluster comprised of two distinct primary substructures, with a few associated minor substructures \citep{virgoVI}. The Virgo cluster has been well observed at all wavelengths \citep[e.g.,][]{Sandage1985, Boselli2003, Urban2011}. This cluster has a virial mass and radius of 1.5 - 6.0 $\times$ 10$^{14}$M$_{\odot}$ and 1.6 Mpc respectively (see Section~\ref{sec:enzo} for our definition of these quantities and Table~\ref{table:cluster properties} for references).  The Coma cluster, located at z = 0.023, is larger and more massive than the Virgo cluster, with a virial mass and radius of 1.3 - 1.4 $\times$ 10$^{15}$M$_{\odot}$ and 2.9 Mpc.  The Coma cluster exhibits clear infalling substructure, as indicated by the dynamics of the cluster galaxies \citep{Coma_substructure}, and the presence of a radio relic \citep{BrownRudnick2011}. The comparison between these two clusters allows for a better determination of how cluster mass and morphology affects its warm gas content. 

In this paper, we use the adaptive-mesh refinement code \Enzo~\citep{Enzo2014} to simulate a Virgo like and a Coma like galaxy cluster, with corresponding masses and virial radii (\rvir) of 3.3 $\times$ 10$^{14}$ M$_{\odot}$ and 14.5 $\times$ 10$^{14}$ M$_{\odot}$, and 1.85 Mpc and 3.0 Mpc. The physics incorporated in these cosmological simulations include the metagalactic ionizing background, primordial and metal radiative cooling, a primordial chemistry network, and star formation and feedback. Using the \textit{yt} analysis toolkit ~\citep{ytmethod}, we observe \lya\ absorption via 2500 sight lines in each cluster. Generating synthetic spectra accounts for the observability of the HI gas by including thermal Doppler broadening effects, gas line of sight velocities, and Gaussian random noise. We have developed a method to automatically identify absorption features and fit Voigt-profiles adapted from previous works \citep{Egan2014,AUTOVP,lanzetta}; this method performs well for weak to saturated lines, and even for the occasional somewhat blended feature found at z = 0. We analyze our results with the goal of comparing to observables in the Virgo and Coma clusters. In addition, we develop an understanding of the properties and environment of the gas seen in absorption that can only be attained through simulation. 

This paper is structured as follows. In Section~\ref{sec:computational methods}, we discuss \Enzo\ and our simulation setup, the cluster properties, and our synthetic spectra. We present our results in terms of understanding the warm gas distribution in galaxy clusters in Section~\ref{sec:gas properties}. In Section~\ref{sec:comparison to observation} we compare our simulated clusters directly to the Virgo cluster observations in Y12 and provide predictions for future observations. Finally, we discuss the results in Section~\ref{sec:discussion}.

\section{Computational Methods}
\label{sec:computational methods}
We outline the computational methods used for our study here, beginning with an overview of the simulation code we used to produce our clusters, \Enzo, in Sec.~\ref{sec:enzo}, along with the included physics. We discuss the properties of our simulated galaxy clusters in Sec.~\ref{sec:cluster properties}, and compare each to the Virgo and Coma clusters. Our synthetic observations are discussed in Sec.~\ref{sec:spectra}, from the generation of the spectra in Sec.~\ref{sec:generating spectra}, to the automatic fitting procedure used to identify absorbers in Sec.~\ref{sec:fitting spectra}. 

\subsection{Enzo Simulation}
\label{sec:enzo}

\begin{table*} 
 \centering
 \begin{tabular}{ c c c c  c  c  c}
    \hline
    \hline
      & Virgo & ref. & Virgo-like & Coma & ref. & Coma-like \\
    \hline 
    \rvir (Mpc) & 1.6 & \cite{virgo-prop}& 1.85 & 2.9 & \cite{coma-prop} & 3.0 \\
    Virial Mass (10$^{14}$ M$_{\odot}$) & 2.2 & \cite{virgo-prop} & 3.3 & 14 & \cite{coma-prop} & 14.5 \\
    T (keV) & $\sim$ 2.5 & \cite{Shibata2001} & 2.0 & $\sim$ 8.5 & \cite{Arnaud2001} & 6.4 \\
    $\sigma_{\rm{v}}$ (km s$^{-1}$) & 544 & \cite{virgo-vel-disp} & 517 & 1082 & \cite{coma-vel-disp} & 972\\
    \hline
 \end{tabular}
 \caption{The virial radius, virial mass, ICM gas temperature, and velocity dispersion of the Virgo and Coma galaxy clusters in comparison with the simulated Virgo-like and Coma-like clusters.}
 \label{table:cluster properties}
\end{table*}

We used the adaptive mesh refinement hydrodynamics + N-body code \Enzo~\citep{Enzo2014} to produce cosmological simulations of two separate galaxy clusters in order to compare the effect of cluster mass and morphology on the observed \lya\ absorption. The simulated clusters were first found in a low resolution, fixed grid simulation to approximately match the mass and size of the Virgo and Coma clusters. Once selected, we performed a zoom-in simulation on each galaxy cluster in order to achieve high spatial and mass resolution in the region of interest (the galaxy clusters).  We refer to our simulated clusters as the Virgo-like and Coma-like clusters throughout this paper. The properties of these two simulations are outlined below.

Our Virgo-like and Coma-like galaxy clusters are simulated in periodic boxes 80 $h^{-1}$ Mpc and 160 $h^{-1}$ Mpc on a side, respectively. We use cosmological parameters adopted from the WMAP5 $\Lambda$CDM results \citep{WMAP5}. Specifically, $\Omega_{\rm{m}}$ = 0.314, $\Omega_{\Lambda}$ = 0.686, h $\equiv$ $H_{0}$/100 km s$^{-1}$ Mpc$^{-1}$ = 0.671, $\Omega_{\rm{b}}$ = 0.049, $n$ = 0.962, and $\sigma_{8}$ = 0.9. The initial conditions are generated at z = 50 with the \cite{powerspectrum} power spectrum. The zoom-in volumes for each cluster were constructed to have similar spatial and mass resolutions, with a maximum spatial resolution for each cluster at z = 0 of 7.3 kpc and a dark matter particle mass resolution of 4.2$\times$10$^{8}$ M$_{\odot}$.  

The galaxy cluster halos were identified using a publicly available version of the HOP halo finding algorithm \citep{HOP} implemented in $\textit{yt}$ \citep{ytmethod}. In this paper, we use $R_{\rm vir}$ to define the virial radius of a galaxy cluster, such that the mean density inside this radius is 100 times the critical density of the universe at that redshift, consistent with the definitions in \citet{BryanNorman1998}.  

The HI fractions used to generate the \lya\ absorption are obtained directly from the simulation, as we self-consistently solve a chemical reaction network of nine species: e$^-$, H, H$^+$, He, He$^+$, He$^{++}$, H$^{-}$, H$_{2}$, and H$_{2}^+$ \citep{chemical-network}.    We use primordial radiative cooling based upon the non-equilibrium chemistry, and include metal line cooling as implemented in \cite{Smith2008,Smith2011}.  In solving the chemical rate equations, we adopt the UV background from \cite{HM12}. We discuss the influence of our choice of ionizing background on our results in Sec.~\ref{sec:ionizing background}.

Our simulations include a method of star formation and thermal feedback which contains a simple prescription for purely thermal stellar feedback.  The method and tests are described in detail elsewhere \citep{Enzo2014,Smith2011}, but we briefly summarize it here.  A star formation rate is computed in each cell as $\dot{\rho}_{\rm SFR} = \epsilon_{\rm SF} \rho / t_{\rm dyn}$ where $t_{\rm dyn}$ is the local dynamical time and $\epsilon_{\rm SF} = 0.02$ is the star-formation efficiency per free-fall time.  Once created, star particles inject gas, thermal energy and metals into their local cell over a time-scale of a few dynamical times.  We assume $10^{51}$ erg per 55 solar masses of stars created, and a yield of 2\%.

Although we expect to be able to sufficiently capture the relevant physics for studying the warm and hot gas content of our clusters, we note that our simulations do not contain detailed physical processes that may be involved in the production or destruction of cold, dense gas in galaxies.  In particular, we do not expect to be able to accurately reproduce observed stellar and gas properties of galaxies in our simulations. This is because: (i) our maximum resolution is limited to 7.3 kpc; (ii) the thermal feedback model we adopt is recognized to rapidly radiate the injected supernovae feedback and so fails to regulate star formation or generate galactic winds, and (iii) we do not include AGN feedback.  The possible implications of some of these processes to our work will be discussed in greater detail in Sec.~\ref{sec:uncertainties}, but to look ahead, this will primarily affect our results for higher column density absorbers.   

We also note that we are unable use the simulations used in Y12 for comparisons as the simulation data used in that paper was corrupted in storage and is now lost. However, a recent reappraisal of the existing data from Y12 showed that the original analysis was incorrectly carried out on a $z=0.56$ data set, rather than $z=0$. This, combined with the use of a post-processing step that assumed a $z=0$ ionizing background, would have resulted in an over prediction of the column densities by a factor of approximately $(1+0.56)^{5.1} \approx 10$. This adjustment would bring the simulation results of that paper in approximate agreement with those found here. In addition to correcting this error, our current simulations are an improvement over that previous work since we track non-equilibrium effects on the HI fraction directly, while Y12 assumed ionization equilibrium and calculated the HI fraction via post processing with \textit{CLOUDY} \citep{CLOUDY, Cloudy2013}. In addition, we implement the most up to date ionizing background from \cite{HM12}, which was not available at the time the Y12 simulations were made.


\subsubsection{Simulated Cluster Properties: Comparison to Virgo and Coma}
\label{sec:cluster properties}

\begin{figure*}
    \captionsetup[subfigure]{labelformat=empty,position=top}
    \centering
    \subfloat[Virgo-like]{\includegraphics[width=0.42\linewidth]{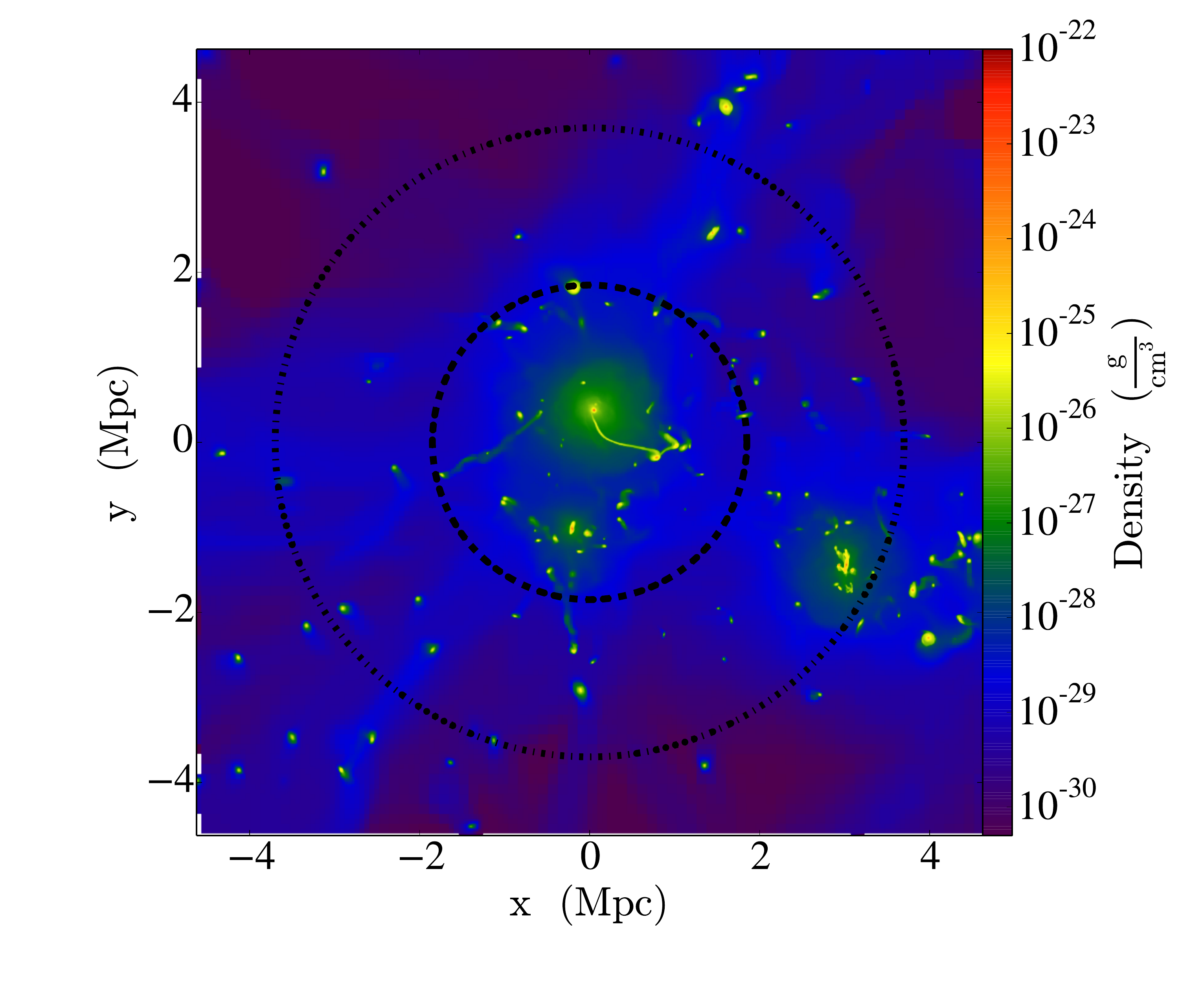}}
    \subfloat[Coma-like]{\includegraphics[width=0.42\linewidth]{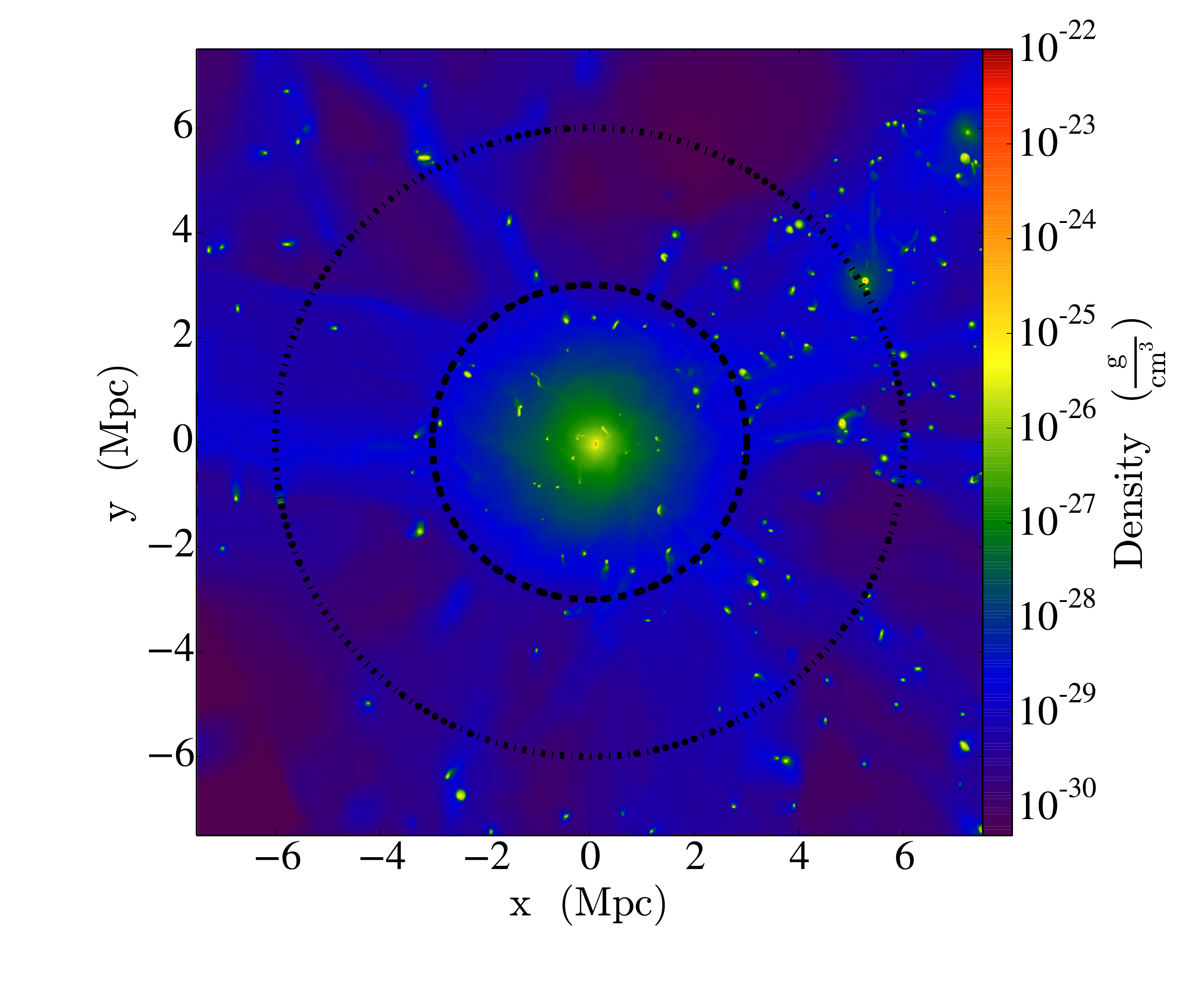}}
    
    \subfloat{\includegraphics[width=0.42\linewidth]{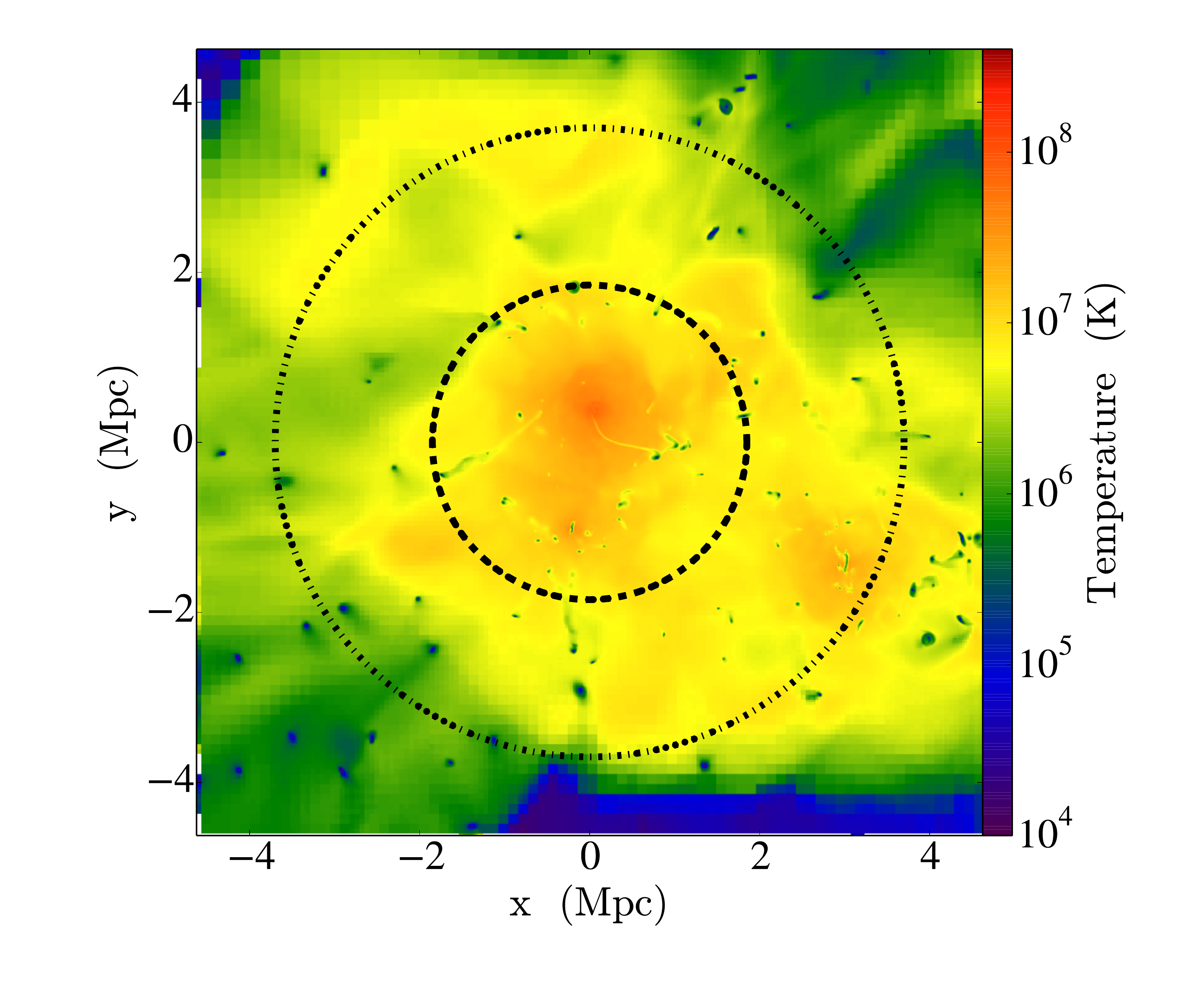}}
    \subfloat{\includegraphics[width=0.42\linewidth]{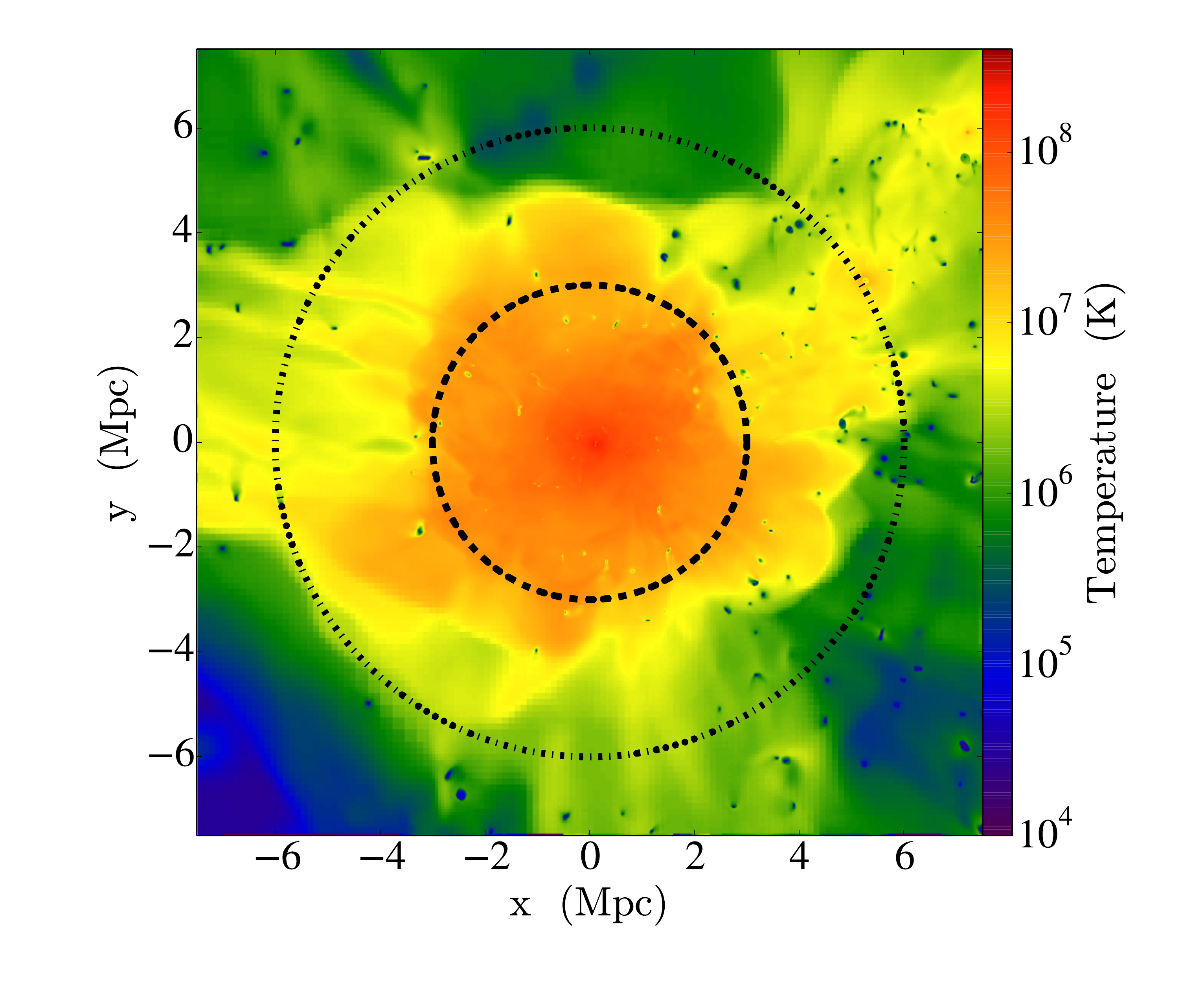}}
    
    \subfloat{\includegraphics[width=0.42\linewidth]{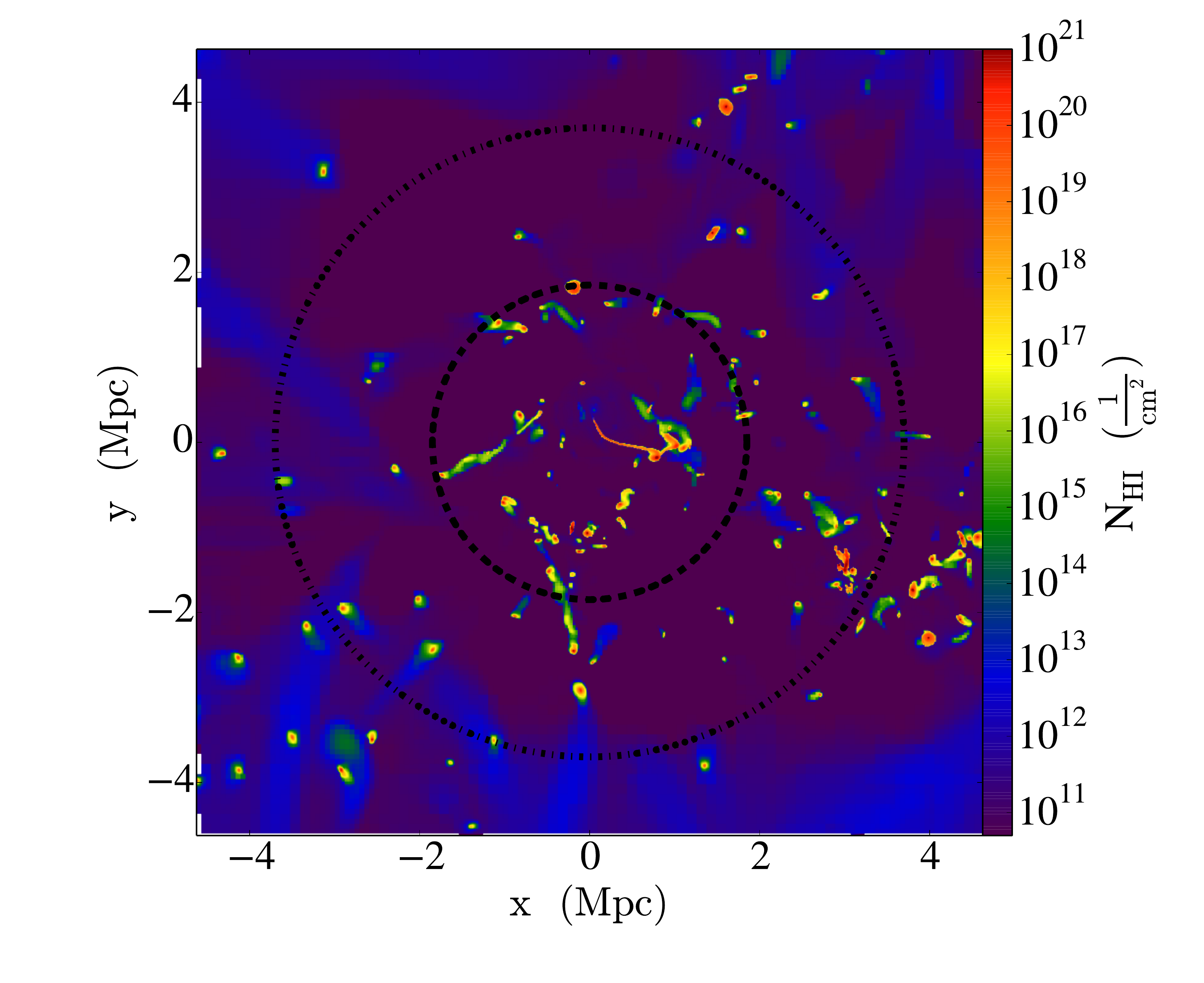}}
    \subfloat{\includegraphics[width=0.42\linewidth]{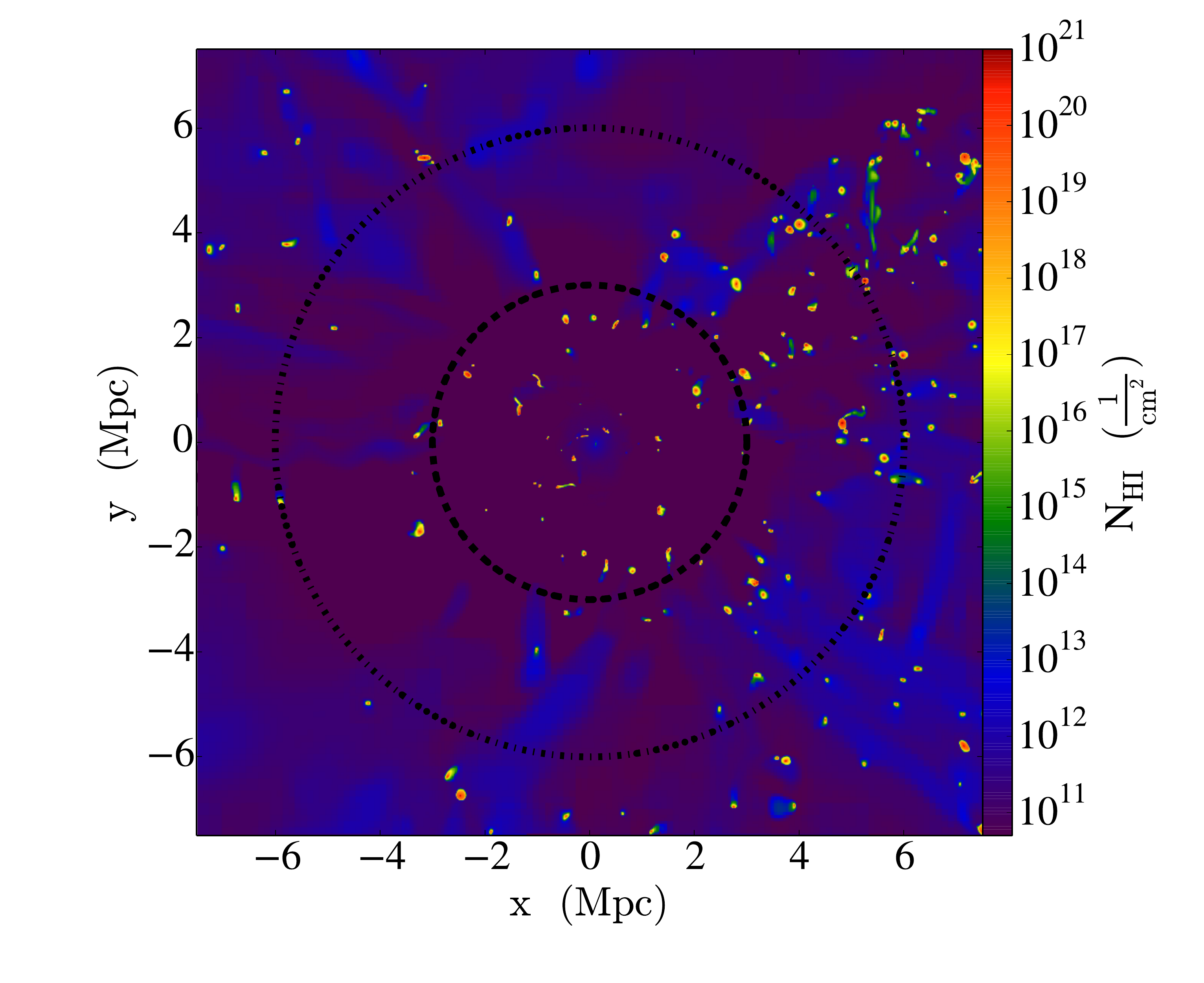}}
\caption{Density weighted projections of the gas density (top), temperature (middle) and unweighted \NHI\ (bottom) for the Virgo-like and Coma-like galaxy clusters. Note the difference in axes scales between clusters. All projections represent a cubical volume centered on the cluster center, 5.0 \rvir\ on a side. The dashed and dot-dashed circles represent 1 and 2 \rvir\ respectively.}
\label{fig:projections}
\end{figure*}

Table~\ref{table:cluster properties} provides a comparison of cluster properties between our simulated clusters and the Virgo and Coma clusters. Our Virgo-like cluster is slightly more massive than the Virgo cluster mass adopted in \cite{virgo-prop}, but lies in the middle of the range 1.5 - 6.0$\times$10$^{14}$ M$_{\odot}$ given in \cite{Bohringer1994}. The virial radius of the simulated cluster is slightly larger than that of the Virgo cluster. We therefore consider our Virgo-like cluster to be a somewhat larger and more massive version of the Virgo cluster. However, the most important similarities are morphological. The Virgo cluster is highly irregular, with several dynamically significant and currently merging substructures that have been well studied \citep{Binggeli1987}. As shown in the density projections (top left) in Figure~\ref{fig:projections}, our Virgo-like galaxy cluster has multiple obvious and infalling substructures. This cluster is then dynamically young.

Our Coma-like galaxy cluster also has similar mass to (but slightly larger than) the Coma cluster.  As shown in Figure~\ref{fig:projections} (top right), its nearly spherical shape offers a contrast to the irregular Virgo-like cluster. Although our Coma-like cluster is comparable in mass to Coma, it does have some  morphological differences. Our Coma-like cluster has no obvious infalling substructure, yet the Coma cluster has a well observed infalling galaxy group, and possibly other substructures \citep[e.g.,][]{Briel1992,Colless1996,Coma_substructure,Neumann2003}.

The middle two panels for Figure~\ref{fig:projections} give the density weighted temperature projections for the two clusters. It is apparent that the ICM in the more massive Coma-like cluster is hotter, with temperatures well above 10$^{7}$~K out to \rvir. The temperature distribution in each cluster tends to follow the density projection, as the Coma-like cluster is more spherical than the Virgo-like cluster, and the substructures seen in density in the Virgo-like cluster are apparent as temperature enhancements. The density weighted projections allow for easy visual identification of galaxies in each cluster. The filaments feeding the Coma-like cluster are obvious also in temperature projection, but are obscured in the still merging Virgo-like cluster. The lower two panels show the integrated HI column density projection for each galaxy cluster. As expected, the largest concentrations of HI in each cluster correspond directly to a galaxy in the given cluster, or tails of gas associated with a galaxy. The majority of the low column density (towards \NHI\ $= 10^{13.0}$ cm$^{-2}$) regions reside outside the virial radius of each cluster, and primarily towards 2 \rvir. It also appears that the low column density HI generally avoids any cluster regions above a few million degrees. Although these projections give a nice qualitative illustration of the cluster morphological properties, they are, of course, not directly observed.  We resort, then, to synthetic observations of our galaxy clusters in order to account for the observability of the HI gas seen in Figure~\ref{fig:projections}, and thus see how it would reveal itself in a real observation. We discuss the methods of our synthetic observations in the following section.

\subsection{Synthetic Spectra}
\label{sec:spectra}
As discussed, Y12 used Ly$\alpha$ absorption spectra from background QSO's to study the warm gas content in the Virgo cluster. They compared the inferred neutral hydrogen column densities and covering fractions to a simulation by calculating the total integrated HI column density along lines of sight in a simulated galaxy cluster. In order to make a more robust comparison between simulation and observation, we expand upon this analysis by producing synthetic absorption spectra along lines of sight through the simulated galaxy clusters.  The following section discusses our method for generating synthetic spectra which takes into account the thermal Doppler broadening of lines as well as the projected line of sight velocity of the absorbing gas. We then automatically fit Voigt profiles with a procedure described in Sec.~\ref{sec:fitting spectra}. 

\subsubsection{Generating Absorption Spectra}
\label{sec:generating spectra}
We use the \textit{yt} toolkit~\citep{ytmethod} to generate absorption spectra from the simulated galaxy clusters. A complete discussion of these methods can be found in \cite{Egan2014}. We outline the procedure below.

Each absorption spectrum is generated along a line of sight through a galaxy cluster at a specified projected distance from the cluster center (or impact parameter, $\rho$). The lines of sight generated are all perpendicular to the cluster x-y plane (in the same viewing direction as Figure~\ref{fig:projections}); there is no overlap between lines of sight. We investigated the effects of changing viewing angle on the results, and the differences are insignificant for the global absorber properties. However, projection effects do become important in the observed velocity distribution of absorbers, as discussed in Sec.~\ref{sec:observed velocity}. For each cell along the line of sight, the HI column density is calculated as N$_{\rm{HI}}$ = n$_{\rm{HI}}\times l$, where n$_{\rm{HI}}$ is the HI number density and $l$ the length of each cell. The thermal Doppler broadening parameter is calculated with the gas temperature in the cell, $T$,  as $b=\sqrt{2~\rm{k}_{\rm{B}}\rm{T}/\rm{m}_{\rm{H}}}$, where k$_{\rm{B}}$ is Boltzman's constant, and m$_{\rm{H}}$ is the mass of the Hydrogen atom. Observed absorption of Ly$\alpha$ will manifest itself as a Voigt profile, having a normalized flux given as $F(\tau) = e^{-\tau}$, where $\tau$ is the line optical depth. 


\begin{figure*}
\centering
\includegraphics[width=0.42\linewidth]{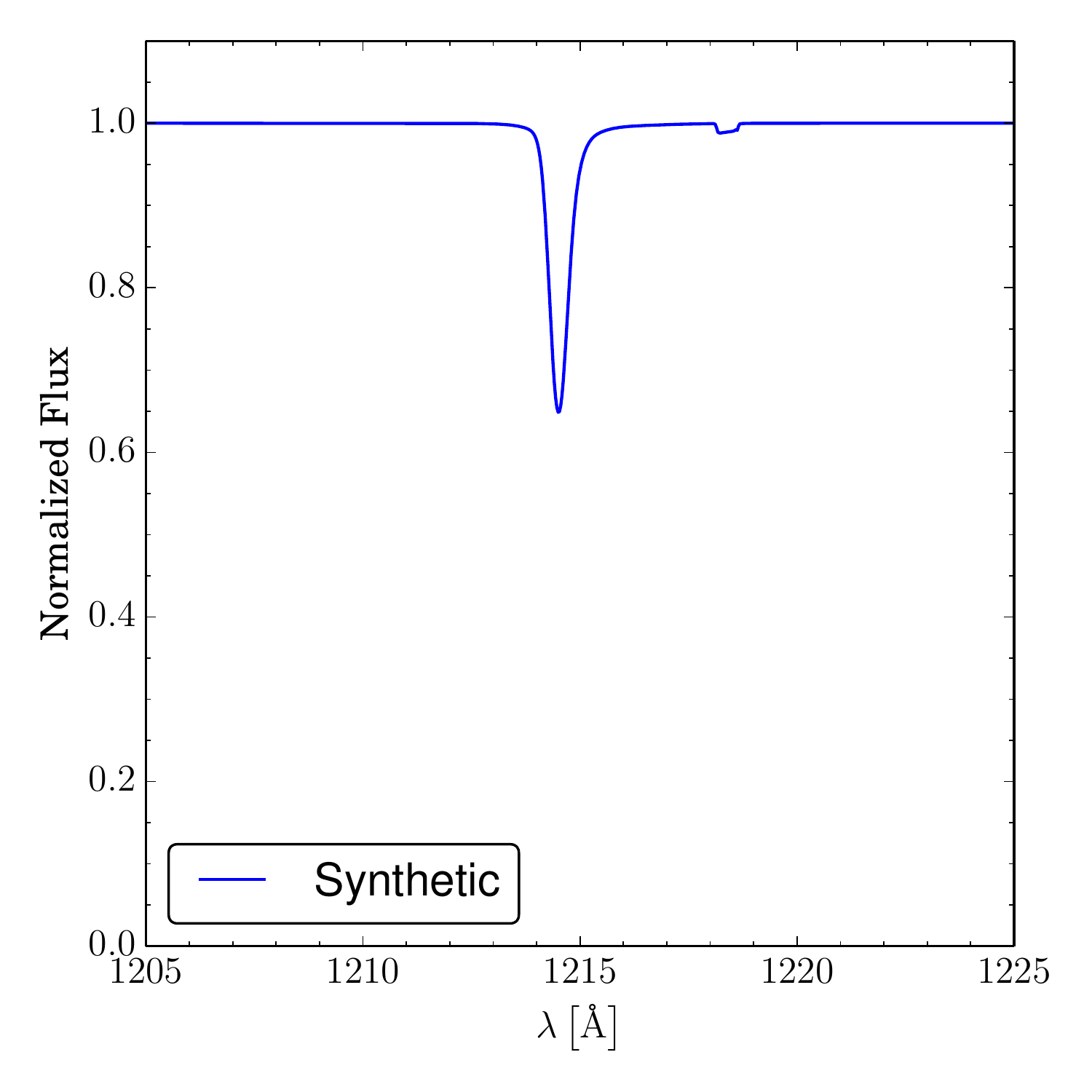} \hspace{0.4cm}
\includegraphics[width=0.42\linewidth]{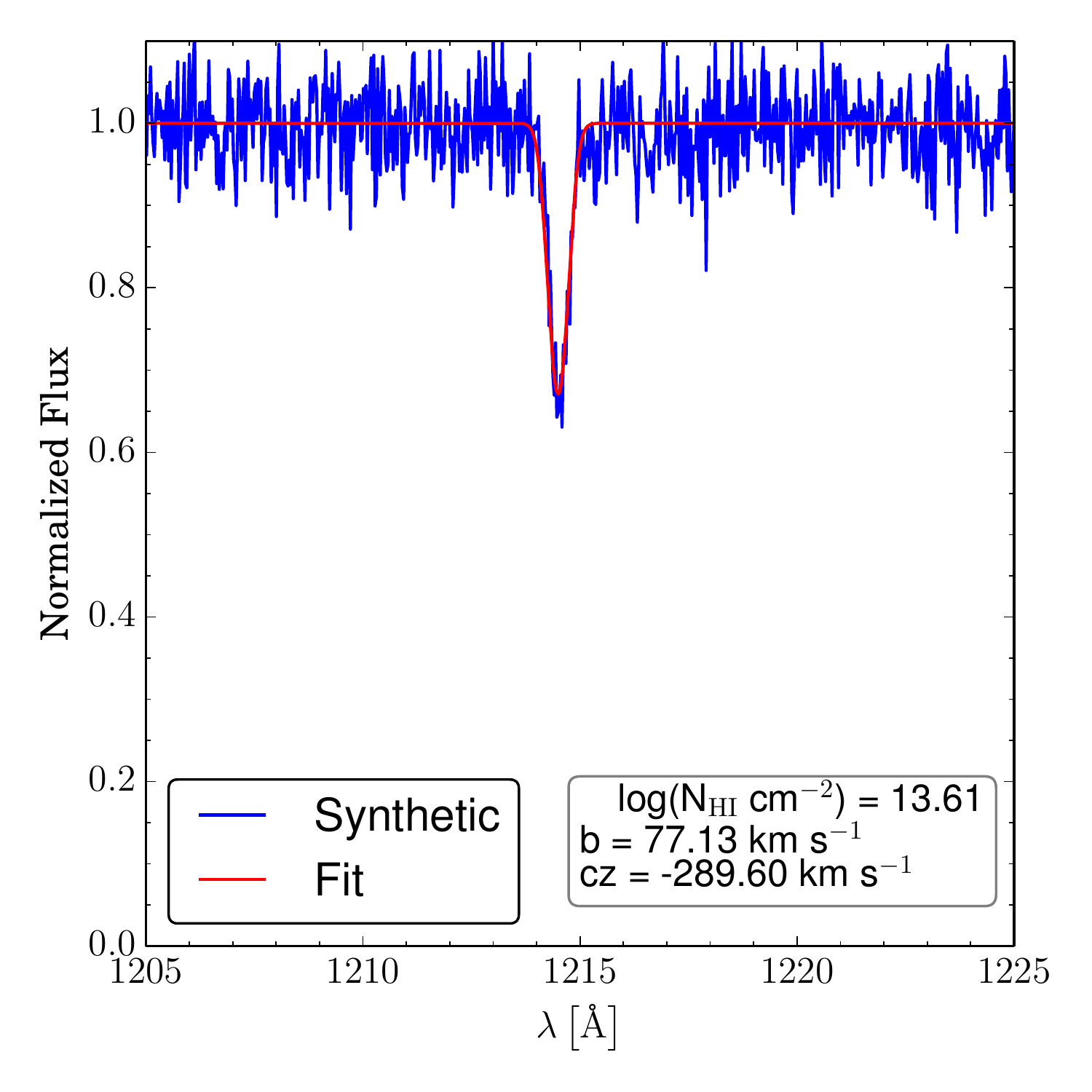} \hspace{0.6cm} \\
\vspace{0.1cm} 

\hspace{0.4cm}\includegraphics[width=0.45\linewidth]{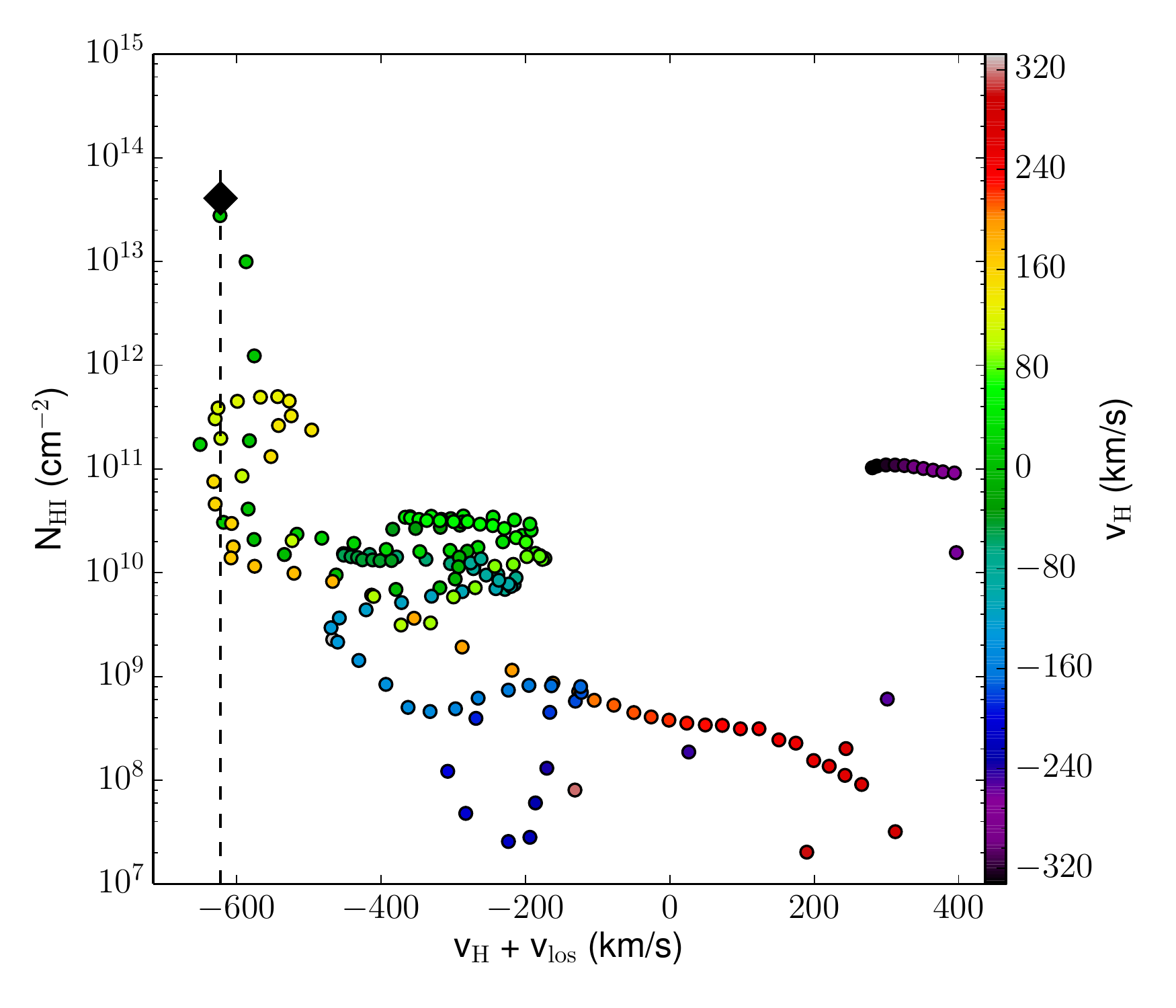}
\includegraphics[width=0.45\linewidth]{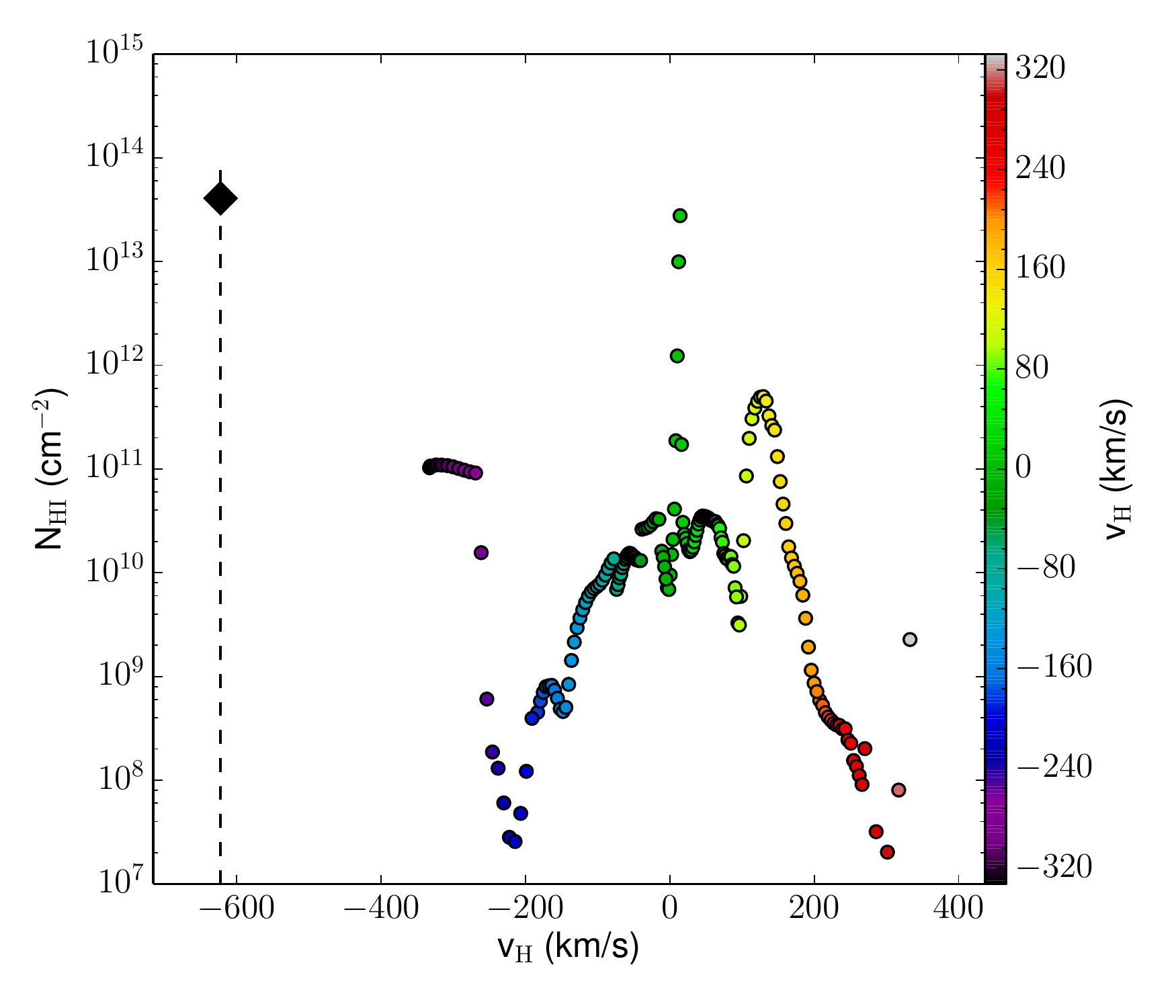}
\caption{Top: Example spectra made along a line of sight through the Virgo-like cluster, showing spectra with (right) and without (left) noise included with a SNR of 20.0.  As shown, the noise obscures the small feature visible at 1218.2 \AA. Bottom: The \NHI\ in each simulation cell along the line of sight (points), given in velocity space (left), which is the combination of the Hubble flow velocity (v$_{\rm{H}}$) and projected kinematic velocity, and in the de-projected physical space (right), in terms of just the Hubble flow velocity. The points are colored by v$_{\rm{H}}$ to identify where a cell shifts from the left to right plots. These plots illustrate how the feature fitted in the top spectra is matched with a cell in the simulation, as explained in Sec.~\ref{sec:identifying absorbers}. The dashed line and black diamond give the cz and \NHI\ of the fitted absorber in the top right plot. In the bottom plots, v$_{\rm{H}}$ for each cell is shifted by the cluster Hubble flow velocity such that cells at v$_{\rm{H}}$ = 0 are the same distance from the observer as the distance from observer to cluster center.}
\label{fig:sample los}
\end{figure*}

A Voigt profile is computed for each cell along the line of sight. Absorption occurs at the natural line wavelength of Ly$\alpha$ in the frame of reference of the absorbing gas. Thus redshift effects and line shifts due to the 3D velocity of the gas projected along the line of sight are taken into account.

For each cluster we observe a cubic volume 5 \rvir\ on a side, centered on the cluster center.  This corresponds to volumes of 791.5 Mpc$^{3}$ and 3375 Mpc$^{3}$ for the Virgo-like and Coma-like clusters respectively, closely matching the observed volume in Y12 and the ongoing Coma observations, which extend out to $\rho$ = 2.19~\rvir\ from the cluster center. A 50$\times$50 grid of sight lines is placed on each cluster. We generate spectra with a constant wavelength resolution, comparable to the COS G130M grating resolution at $\lambda$=1300 \AA (used in Y12).  The G130M grating operates over the wavelength range 1150-1450 \AA\ and has a resolving power ($\lambda/\Delta \lambda$) R $\sim$ 16,000-21,000. This corresponds to a bin width ($\Delta \lambda$) of 70~m\AA. To better compare our synthetic observations to COS, we add Gaussian random noise to the pure spectra. Although this noise component is simple compared to the complicated systematics involved in actual COS spectra, it is sufficient to allow a comparison of well-resolved column density distributions.  We specify a desired signal-to-noise ratio (SNR) of 20, slightly better than the quality of observations made in Y12. Modifying the SNR primarily affects the ability to identify low column density features. Changing the SNR does not significantly affect the results for absorbers above \NHI\ $\sim 10^{13.0}$ cm$^{-2}$. 

Figure~\ref{fig:sample los} gives an example of the pure spectrum (blue) generated using this procedure on the top left, with the noisy spectrum (blue) given in the top right. As shown, there are two absorption features in the pure spectrum, the smaller of which is masked completely by the included noise. Obtaining the shown fit (red) is discussed in Section~\ref{sec:fitting spectra}. 

\subsubsection{Voigt Profile Fitting}
\label{sec:fitting spectra}
For each spectrum, regions of absorption are identified using an algorithm adopted from \cite{lanzetta} outlined below. An equivalent width for the absorption in a given wavelength bin, or e$_{i}$, is calculated using the bin size, $\Delta \lambda$, flux, F$_{i}$, and continuum flux C$_{i}$ as
\begin{equation}
e_{i} = \Delta \lambda \left( 1 - F_{i}/C_{i} \right).
\end{equation}
Since the generated spectra are normalized to the continuum, the continuum flux, C$_{i}$, is unity for each bin. Using the uncertainty in each bin, denoted as $\sigma_{F_{i}}$, a standard deviation is computed for each equivalent width as
$
\sigma_{e_{i}} = \Delta \lambda  \sigma_{F_{i}} / C_{i}.
$
Using a passband of 5 bins, a total equivalent width (E$_{i}$) is calculated for a given bin by summing the equivalent width of that bin and the 4 adjacent bins. The associated uncertainty is calculated from summing the individual variances. Thus,
\begin{equation}
E_{i} = \sum^{i+2}_{n=i-2} e_{n} \qquad {\rm and} \qquad
\sigma_{E_{i}}^{2} = \sum^{i+2}_{n=i-2} \sigma_{e_{i}}^{2}.
\end{equation}

A region is flagged as an absorption feature if each bin in that region has a total equivalent width larger than the associated uncertainty by a specified significance. A spectral region is considered a feature if E$_{i} > 5 \sigma_{E_{i}}$ \citep{lanzetta}.  Each identified absorption feature is fit to a Voigt profile with three free parameters: \NHI, $b$, and $z$. 

Before fitting a given feature, we make initial guesses for the absorption region utilizing the procedure discussed in \cite{AUTOVP}. Choosing reasonable initial guesses is important for obtaining a realistic fit to the absorption feature. Each absorption feature is first smoothed using a Gaussian filter.  A line is placed at the minimum flux of the smoothed profile, or the center of a saturated region for saturated lines. For each unsaturated line, \NHI\ and $b$ are iterated over from larger to smaller values until the guessed line is above (less absorption) the smoothed feature. For saturated lines, \NHI\ and $b$ are iterated over until the guessed line is above the smoothed feature at the edges of the saturated region.

A fit is found from the initial guesses by minimizing $\chi^{2}$, given as
\begin{equation}
\chi^{2} = \sum \frac{ \left(F_{i} - F_{i}^{fit} \right)^{2}}{\sigma_{F_{i}}^{2}},
\end{equation}
where F$_{i}$ is the flux in a given bin, F$_{i}^{fit}$ is the fitted flux, and $\sigma_{F_{i}}^{2}$ is the variance given by the uncertainty in each bin. This non-linear least squares problem is solved using an implementation of the Levenberg-Marquardt algorithm in the SciPy library \citep{scipy}. Once a line is fit, the resulting fitted flux is subtracted away from the spectrum. Then the region is passed through the same absorption feature identification algorithm discussed above. If there is still identifiable flux remaining, a new line is added to the feature and the fitting procedure repeats. This is done either until no identifiable flux remains, or until the addition of a new line worsens the resulting fit. We consider an ideal fit as one with a reduced $\chi^{2}$ value close to unity. Once lines are fit to all of the individually identified features, all of the lines are refit simultaneously to the full spectra in a final fitting step in order to improve fits on the transitions between adjacent lines, especially in very blended features. 

Occasionally, the automatic fitting procedure has trouble separating blended features, attempting to fit the entire feature (incorrectly) with a single line. To mitigate this problem, we have included a step to identify and separate multiple peaks in an identified absorption region. This method is able to properly account for any absorption feature found at redshift zero. At high redshift, features become extremely blended, with potentially several individual components in a given complex. Our method currently has trouble for lines at $z > 2$, requiring manual intervention. However, this redshift regime is beyond the scope of this study.

\subsection{Identifying Absorbing Gas in the Simulation}
\label{sec:identifying absorbers}
In addition to making accurate comparisons to observation, we are motivated by the ability to use the simulations to better understand the physical environment and 3D distribution of the observed warm gas component. We do this by associating each identified absorber with a physical region in our galaxy cluster simulations. Doing this, however, has two potential issues: 1) multiple cells along the line of sight contribute at various levels to the observed absorption, and 2) in reality, the HI absorber seen in spectra depends strongly on the geometry and orientation of the HI region. This creates an issue in trying to associate a structure/region in physical space as an absorber from an identified spectral feature (see \cite{Egan2014} for a discussion of this issue).  However, at the spatial resolution used here, we find that a single cell in the simulation contributes the majority of the observed HI column.  Although several cells may ultimately constitute the physical absorber, the environment of these cells does not vary dramatically from the cell with most of the HI content, therefore (for simplicity) we only associate an identified absorption feature with a single cell in our simulations, calling this the physical absorber.

The absorber(s) identified in a given spectra are located along the line of sight in velocity space, rather than physical space. The velocity space is the sum of the Hubble flow velocity of the gas (which is directly related to distance along the line of sight from the observer) and the peculiar gas velocity projected along the line of sight. Because of this, translation to the physical position of the absorbers is impossible without knowing the peculiar gas velocity. This is not possible in real observations, but is possible from the simulations since we have complete knowledge of the gas motions.

The process of identifying the gas motions is shown in the bottom of Figure~\ref{fig:sample los}. On the left, we plot the column density of each grid cell along the line of sight as a function of the line of sight velocity of the gas, given as $v_{\rm{los}} = v_{\rm{H}} + v_{\rm{pec}}$, where $v_{\rm{H}} = H_0 D$ is the Hubble velocity of the gas in a cell at distance $D$ along the line-of-sight, and $v_{\rm{pec}}$ is the gas peculiar velocity projected along the line of sight.  This plot shows where all of the gas would appear to the observer if it was detectable with spectral observations. The spectrum on the top right, then, is the sum of the Voigt profiles associated with the points on the lower left plot. The Doppler shift formula gives a one-to-one correspondence between the horizontal axis in the lower left plot, and the wavelength in the spectrum in the upper right. The black diamond in the lower left plot gives the fitted column density and Doppler shifted velocity of the feature shown in the spectrum.  As this plot illustrates, one cell dominates the total HI column density in the region, even though the feature is the sum of many cells. 

Since we have full kinematic information in the simulation, we can calculate the component velocities, $v_{\rm{H}}$ and $v_{\rm{pec}}$, separately. We can then de-project the gas by subtracting $v_{\rm{pec}}$ from $v_{\rm{los}}$. This gives the plot on the bottom right, where the gas cell column density is plotted as a function of the Hubble flow velocity, v$_{\rm{H}}$; in other words, this panel simply shows the physical location of the gas (expressed in terms of the Hubble velocity).  Both of these plots are colored by the Hubble flow velocity in each cell to readily match the cells between the two plots. Comparing these two plots shows just how much projection effects shift the gas in velocity space. As shown, although the absorber responsible for the feature in the spectrum at the top right has a total Doppler shifted velocity just below -600 km s$^{-1}$ (far to the left of most of the gas cells in the lower left plot), it is physically located near the cluster center (v$_{\rm{H}} \approx 0$, lower right plot), and is therefore in the middle of the line of sight through the cluster.

\section{Properties of Warm Gas in Galaxy Clusters}
\label{sec:gas properties}
We study the warm gas content of two simulated galaxy clusters from the point of view of observable \lya\  absorbers. 
We discuss the spatial distribution of our cluster warm gas and the identified absorbers in Sec.~\ref{sec:spatial dist}, the kinematic properties of the absorbers in Sec.~\ref{sec:kinematics}, and the thermal properties in Sec.~\ref{sec:thermal properties}.

\begin{figure*}
\centering
\includegraphics[width=0.45\linewidth]{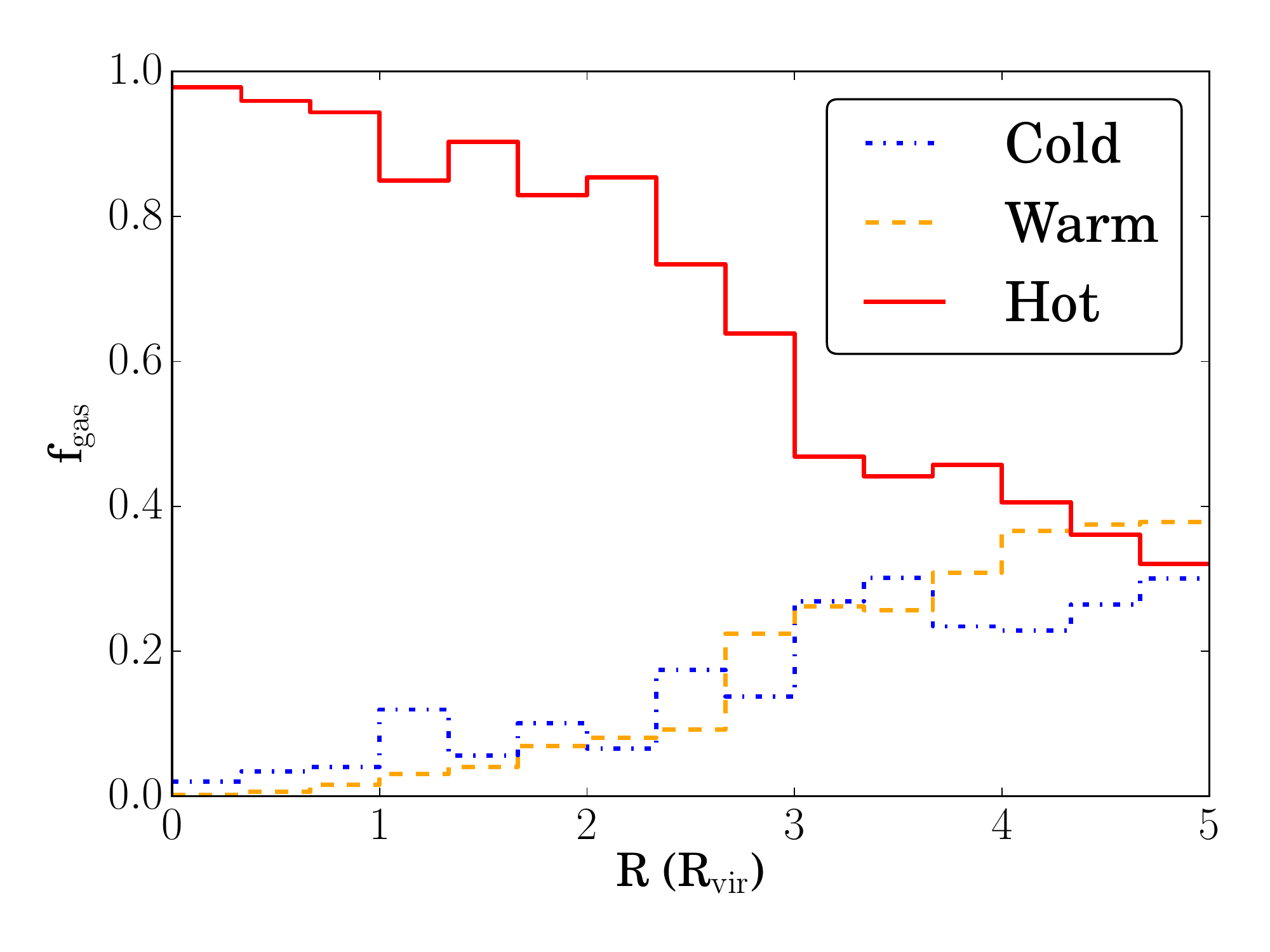}
\includegraphics[width=0.45\linewidth]{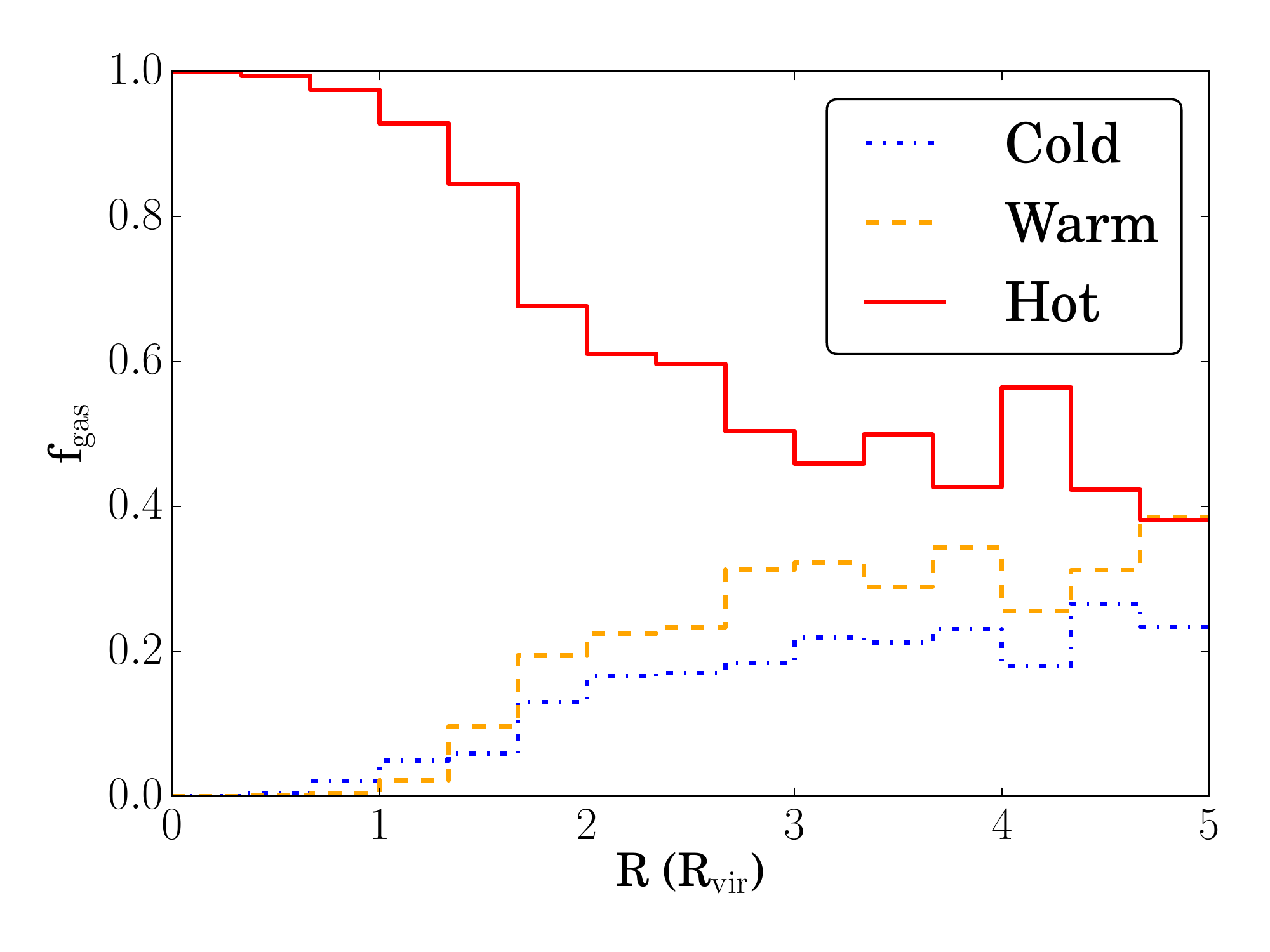}
\caption{The radial profile of the diffuse gas mass fractions for the Virgo-like (left) and Coma-like (right) clusters, as separated into three temperature regimes, cold (blue, dash-dotted, $T < 2 \times 10^{4}$ K), warm (yellow, dashed, $2 \times 10^{4}$ K $< T < 10^{6}$ K), and hot (red, solid, $T > 10^{6}$ K).}
\label{fig:gas mass fraction}
\end{figure*}

\subsection{Spatial Distribution of Warm Gas}
\label{sec:spatial dist}
\subsubsection{Cluster Gas Mass Fraction}
\label{sec:gas content}
We begin by characterizing the diffuse gas around our simulated clusters. Hot gas (T $>$ 10$^{6}$ K) is the largest baryonic component of galaxy clusters, comprising $\sim$80\% or more of the cluster baryon mass \citep{Voit2005}.  For non-cool core clusters, the average ICM temperature generally decreases radially outward from cluster center \citep{DeGrandi2002}. In the top of Figure~\ref{fig:gas mass fraction}, we plot the radial profile of the gas mass fraction for each of our simulated clusters, separated into the three temperature regimes: cold ($T < 2 \times 10^{4}$ K), warm ($2 \times 10^{4}$ K $< T < 10^{6}$ K), and hot ($T > 10^{6}$ K). Here we would like to only examine the diffuse gas associated with the cluster ICM and surroundings, and ignore gas bound in galaxies by placing a cutoff at $n < 0.1$ cm$^{-3}$ for the profiles given in Figure~\ref{fig:gas mass fraction}. This serves primarily to smooth out the cold gas mass fraction curve by reducing fluctuations caused by the presence/absence of galaxies in a given radial bin. We note that we do not mask this gas in our synthetic observations. 

For each cluster, the hot gas dominates (above 80\%) within the virial radius, and decreases gradually to about 30 - 40\% at the farthest distance from the cluster shown (5 \rvir). Over this range, the warm and cold gas fractions increase from, at most, a few percent, to about 30\% at 5 \rvir.  The overall trend of the gas fractions is similar in the two clusters, but shows a few differences.  The Coma-like cluster has more warm gas (and less hot gas) around 1-3 \rvir.  It also has a generally lower fraction of cold gas than the Virgo-like cluster at all radii.

\subsubsection{Projected map of absorbers}
\label{sec:projected map}

\begin{figure*}
\centering
\includegraphics[width=0.45\linewidth]{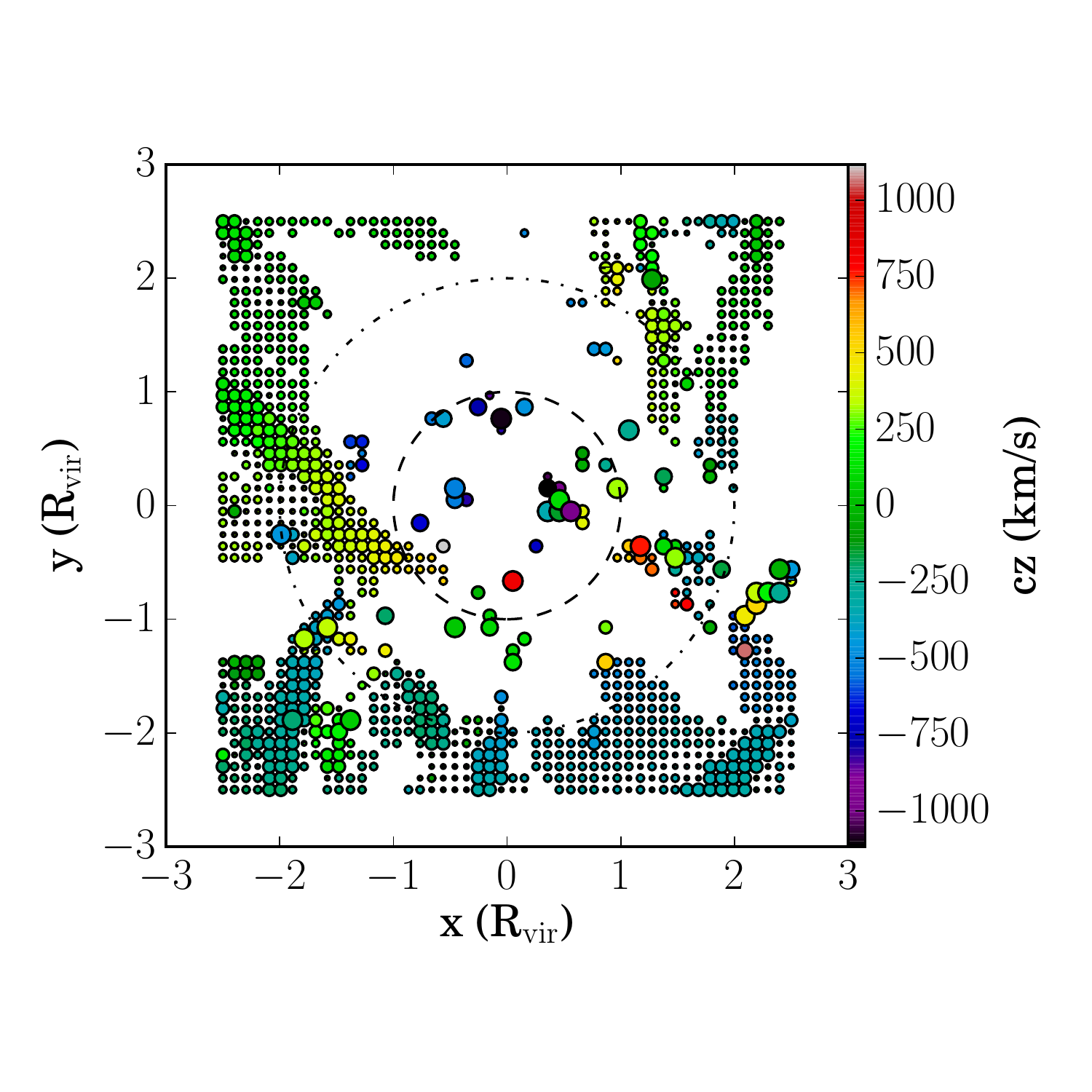}
\includegraphics[width=0.45\linewidth]{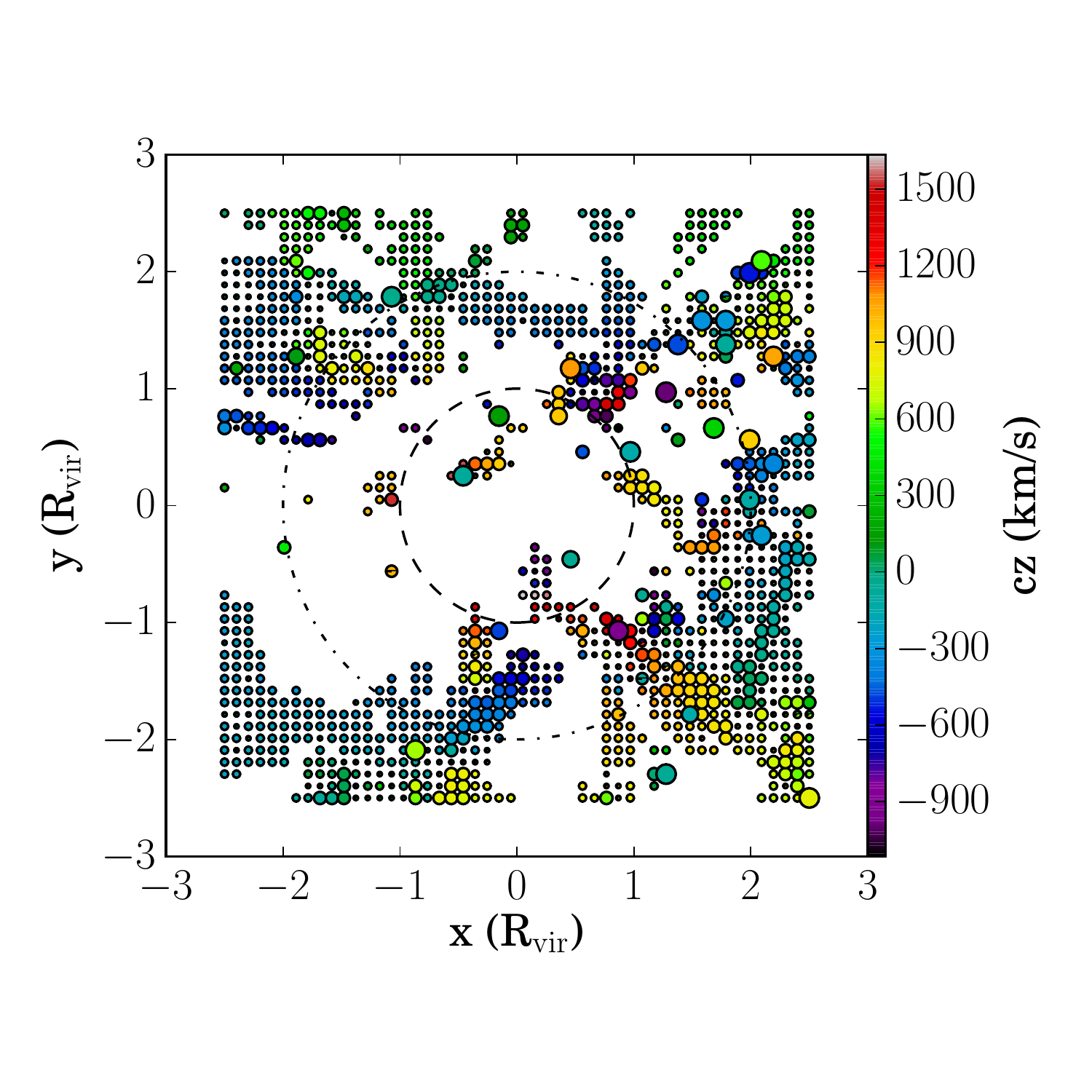}
\caption{Projected map of \lya\ absorbers identified in our simulated galaxy clusters, Virgo-like (left) and Coma-like (right). Each identified absorber is represented by a circle with size corresponding to the identified column density in one of four \NHI\ column density bins: $10^{12 - 13}$, $10^{13 - 15}$, $10^{15 - 17}$, and greater than $10^{17}$ cm$^{-3}$. We note that gas at HI column densities around 10$^{12}$-10$^{13}$ cm$^{-2}$ is at and just below the detection limit in the current observations of the Virgo and Coma clusters. The absorbers are colored by their observed velocity, normalized to the cluster redshift.  Plotted circles are subdivided if more than one absorber is present in a column density bin. In each cluster, the dash and dash-dotted lines denote 1 and 2 \rvir. These absorber maps have the same orientation as the projection in Figure~\ref{fig:projections}.}
\label{fig:absorber map}
\end{figure*}

We now focus our discussion of the cold/warm gas content of our simulated galaxy clusters in terms of what is actually observable through absorption. We show the projected distribution of identified \lya\ absorbers for the Virgo-like (left) and Coma-like (right) galaxy clusters in Figure~\ref{fig:absorber map}.  A circle is plotted to represent an identified absorber along a given line of sight, with the circle size indicating the column density and the color the observed velocity, normalized to the cluster redshift.  If a given line of sight has multiple absorbers within a column density bin, the circle is subdivided evenly among those absorbers. 

One can immediately see the correspondence between Figure~\ref{fig:absorber map} and the total HI column density projections in Figure~\ref{fig:projections}. In both clusters, there are several regions of high column density seen in the projected map that also appear as absorbers in the synthetic observations.  In addition, the projected map clearly traces filamentary structures at larger distances from the cluster center (outside and around 1 - 2 \rvir); these are visible also in Figure~\ref{fig:projections}. There is a noticeable velocity gradient along groups of absorbers in each cluster. The gradients and proximity of absorbers to each other in this case immediately suggest a coherent flow or complex of absorbers, with a velocity gradient created by a combination of a true physical velocity gradient and projection effects (i.e. coherent flow at an angle to the line of sight). An obvious example is the green-yellow structure towards the left hand side of the Virgo-like cluster, located mostly between 1 and 2 \rvir, but extending out past 2 \rvir. 

\subsubsection{Column Density Distribution}
\label{sec:cdd}
\begin{figure}
\centerline{\includegraphics[width=\columnwidth]{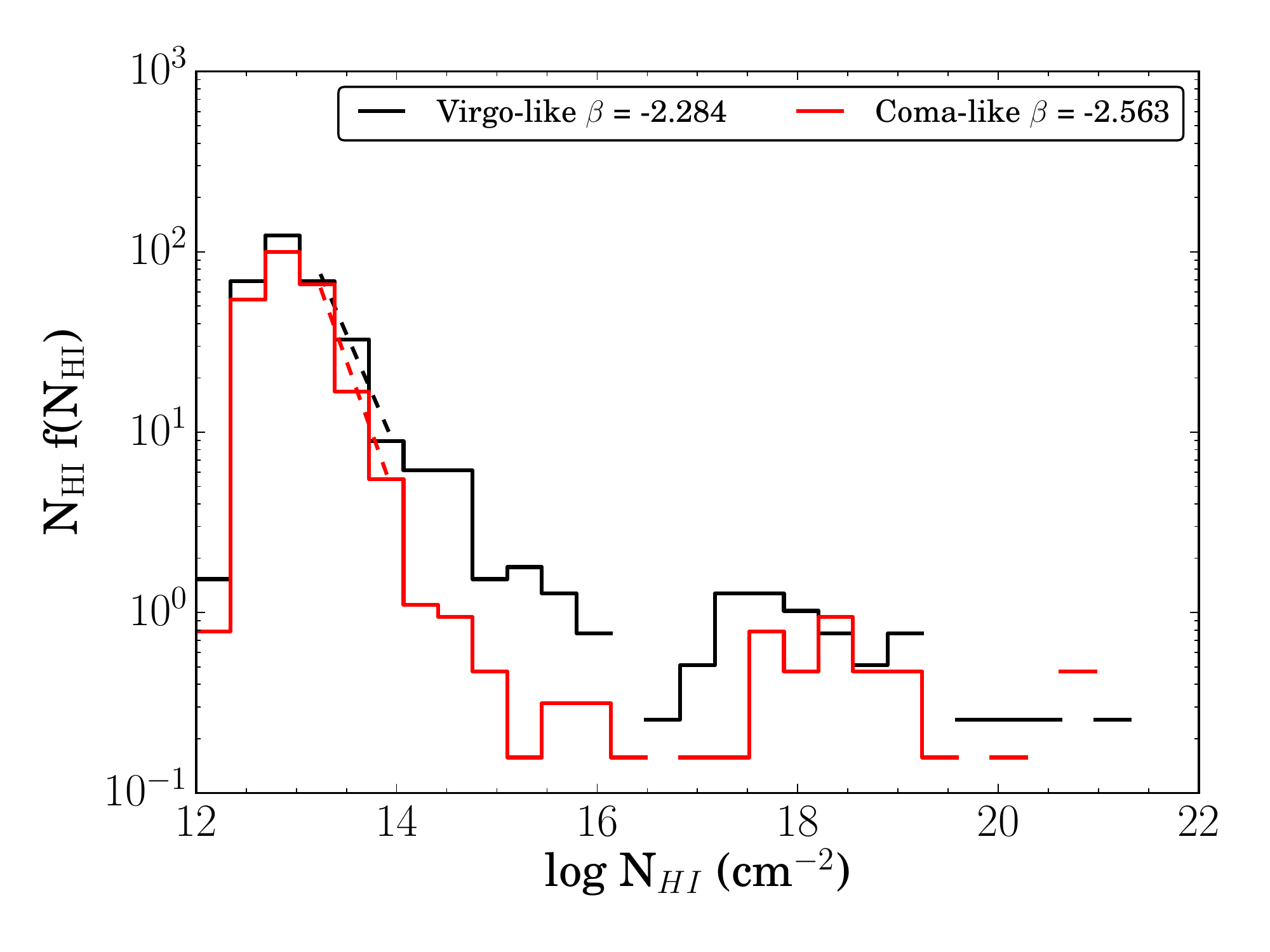}}
\caption{The column density distribution of identified absorbers in the Virgo-like (black) and Coma-like (red) clusters given as \NHI~f(\NHI), where f(\NHI) = $\partial^2 \mathcal{N} / \partial N  \partial z$. The power law fits over the range $13 <$ \logN\  $< 14$ are given for each cluster as the dashed lines, with slopes ($\beta$) shown in the legend.}
\label{fig:cluster NfN}
\end{figure}

We next turn to a more quantitative analysis of the absorbers.  The column density distribution (CDD) function, f(\NHI), is defined as the number of absorbers per column density and per redshift interval, or f(\NHI)~=~$\partial^2\mathcal{N} / \partial N  \partial z$. The column density distribution is often used to characterize \lya\ absorbers and the associated warm gas in the IGM \citep[e.g.,][]{Penton2000, Danforth2014}. Environmental effects in a galaxy cluster, including the interactions with the hot cluster ICM, should result in a noticeable departure from a typical IGM distribution of \lya\  absorbers at $z = 0$. We plot \NHI f(\NHI) in Figure~\ref{fig:cluster NfN} instead of f(\NHI) to emphasize small differences between the Virgo-like cluster (red) and Coma-like cluster (black) distributions. The turnover of the CDD at low column densities is an artifact, corresponding to where noise begins to dominate over the \lya\ absorption signal.

Overall, the Coma-like cluster CDD lies below the Virgo-like cluster at almost all column densities. The two are comparable only at low column (around \NHI\ $\sim 10^{13}$ cm$^{-2}$) and at high column (around \NHI\ $ \sim 10^{18.5}$ cm$^{-2}$). At the highest observed column densities, \NHI\ $\geq 10^{19}$ cm$^{-2}$, both distributions suffer from low number statistics. Examining Figure~\ref{fig:projections}, it is apparent that the Virgo-like cluster contains much more absorption at and above \NHI\ $> 10^{14}$ cm$^{-2}$ due to large area coverage of long, extended HI features. These features reside primarily within \rvir\ of the cluster (in projection), and correspond to long HI tails that can originate from galaxy-galaxy interactions in the cluster, or the ram-pressure stripping of gas, which can create tails observable in 21cm emission up to 100 kpc \citep{RPSexample}. The observable lifetime of these tails in emission (21 cm and H$\alpha$) can be up to at least 500 Myr \citep{TonnesenBryan2010}. Although there has been no study of their observability in \lya\ absorption, presumably the observable lifetime of these tails in absorption will be even longer as the detectable column densities are substantially lower: \NHI\ $\sim 10^{13}$ cm$^{-2}$, as opposed to \NHI\ $\sim 10^{19}$ cm$^{-2}$. These features are absent in the more relaxed Coma-like cluster, perhaps because these tails have already been heated and mixed with the hot ICM. 

We can approximately characterize the CDD as a power law with f(\NHI) $\propto$ N$_{\rm{HI}}^{\beta}$.  We fit this (dashed lines in Figure~\ref{fig:cluster NfN}) to each of our distributions over the range $13 < $ \logN\ $< 14$.  For the Virgo-like and Coma-like clusters, $\beta$ = -2.284 and $\beta$ = -2.563 respectively. This slope is significantly steeper than that observed in the $z = 0$ IGM, or $\beta$ = -1.68 $\pm$ 0.02 \citep{Danforth2014} (we note this slope was fit over the range $12.2 < $ \logN\ $< 17$).

\subsubsection{Number Density of Absorbers}
\label{sec:number density}
\begin{figure*}
\centering
\includegraphics[width=0.45\linewidth]{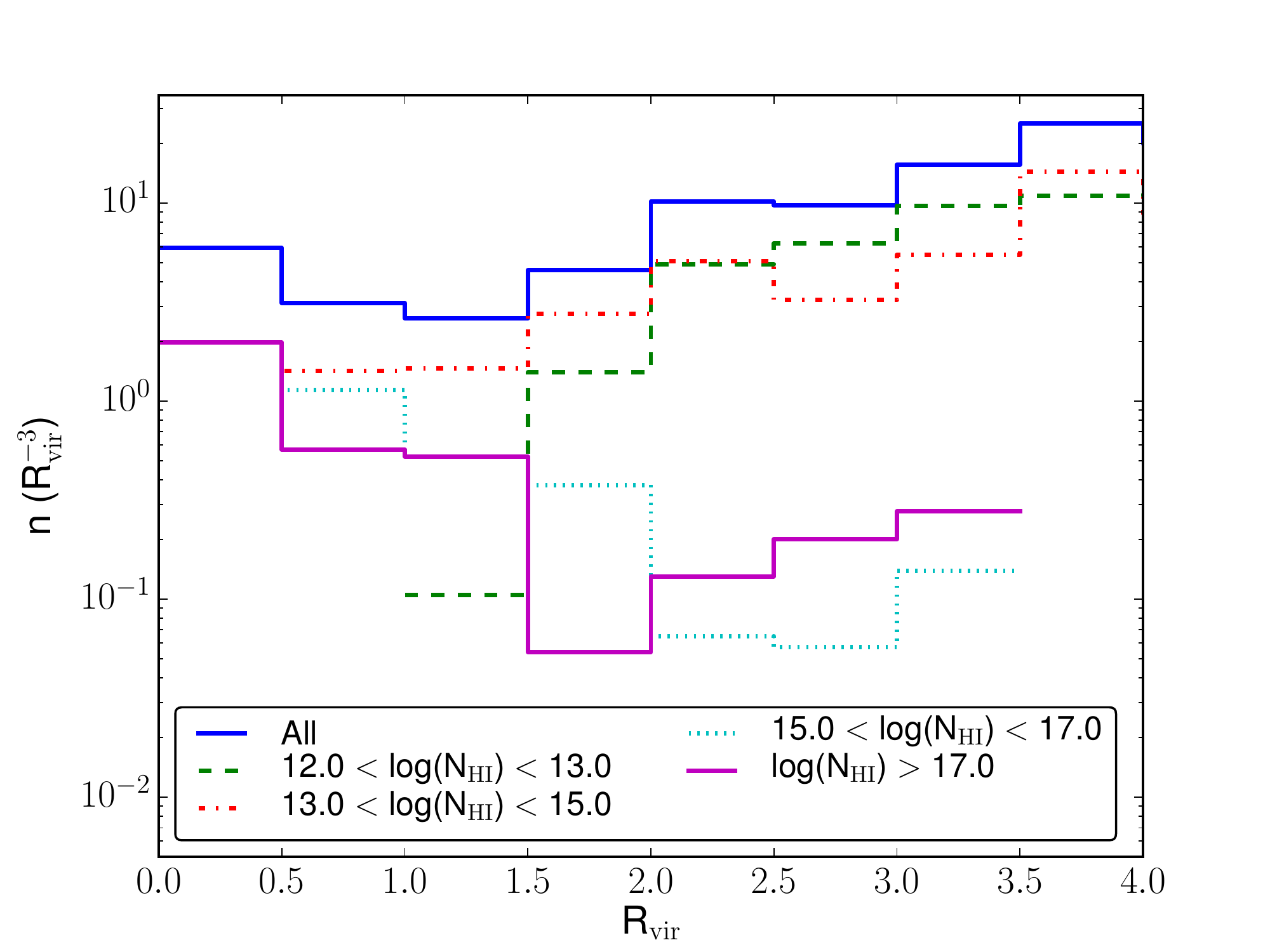}
\includegraphics[width=0.45\linewidth]{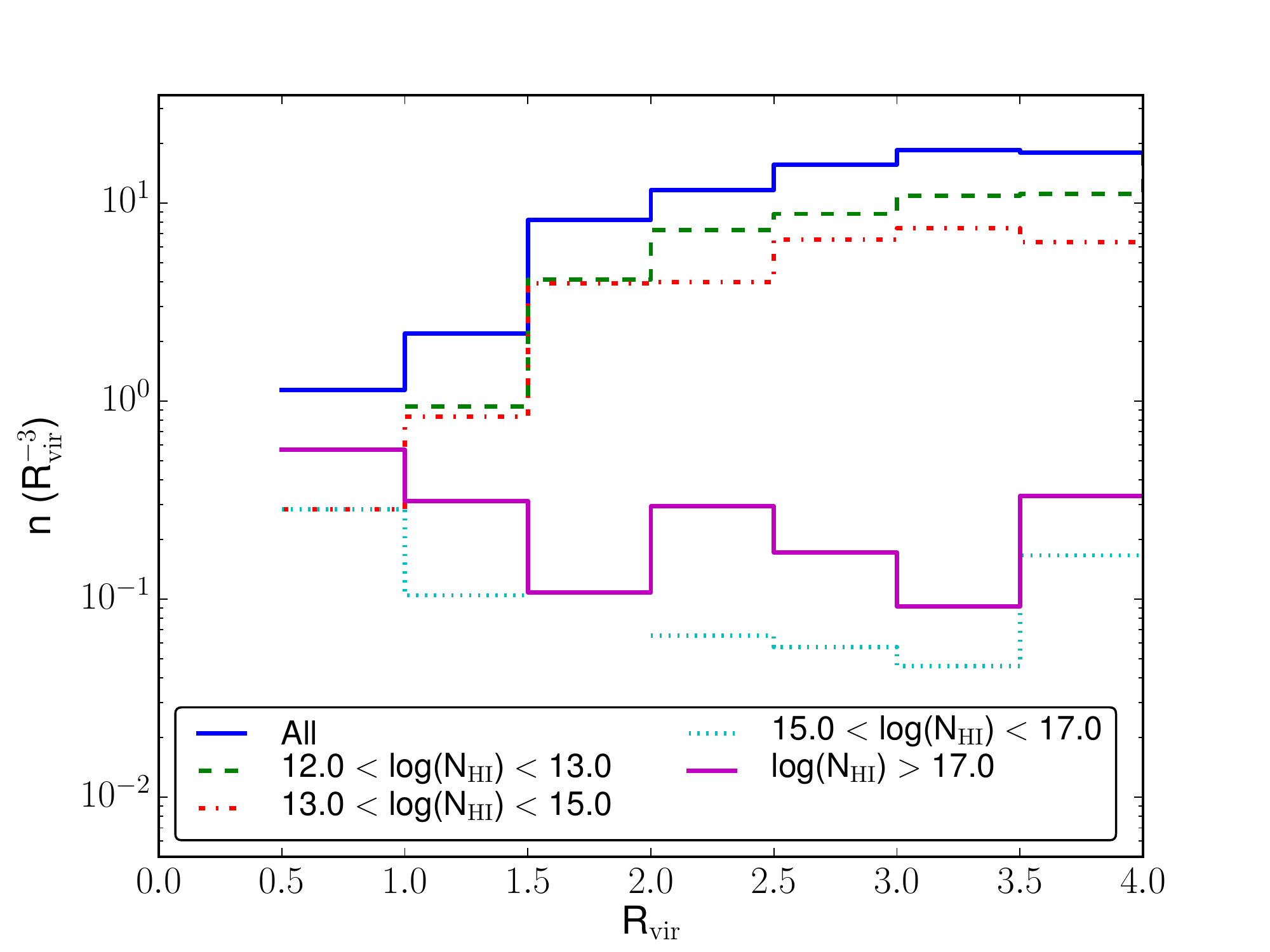}
\caption{The number density of absorbers (in units of inverse cubic virial radii) as a function of distance from the cluster center for the Virgo-like (left) and Coma-like (right) clusters. The distribution is given for all absorbers (blue, solid), and those with column densities of 12.0 $<$ \logN~$<$ 13.0 (green, dashed), 13.0 $<$ \logN~$<$ 15.0 (red, dot-dashed), 15.0 $<$ \logN~$<$ 17.0 (cyan, dotted), and \logN~$>$ 17.0 (purple,solid).}
\label{fig:number density}
\end{figure*}

The observations for the Virgo cluster and of our simulated clusters both show that low column density absorbers generally avoid the hot X-ray gas and the central \rvir\ of the galaxy cluster in projection. However, we would like to understand the distribution of absorbers as a function of the physical distance from cluster center. In order to address biases in our synthetic observations, in Fig~\ref{fig:number density} we compute the three dimensional number density of absorbers in radial bins for various absorber column densities. The number density is given in units of R$_{\rm{vir}}^{-3}$, as opposed to Mpc$^{-3}$, to account for the fact that we scale our sightline grid in terms of R$_{\rm{vir}}$, and to readily identify if properties scale with the cluster virial radius. We compute the volume probed within a given radial bin by summing $l \times$d$\delta^2$ for all sightlines, where $l$ is the length of a given sightline that falls within a given radial bin, and d$\delta$ is the spacing between sightlines (a constant). This avoids the problem of taking volume in the i$^{\rm{th}}$ bin as proportional to R$_{i+1}^3$ - R$_{i}^3$, as the number density would then scale with the number of sight lines. One could compute the volume by adding up the volume of ``observed" grid cells within the radial bins, but this introduces an under-sampling bias in the central cluster regions, where the AMR has created many more grid cells than in the cluster outskirts.

Figure~\ref{fig:number density} shows the number density of all absorbers (blue, solid), and absorbers separated into the same column density bins as used in Fig.~\ref{fig:absorber map}. In both clusters, the greatest number density of absorbers lies well outside the cluster region, into the IGM, where the absorber count is dominated by low column density absorbers. Outside about 1.5 \rvir\ for each cluster, the number density trends are very similar. By this point, the number density of high column absorbers, \NHI\ $> 10^{15}$ cm$^{-3}$, has decreased (this is more dramatic for the Virgo-like cluster), and the number density of low column absorbers begins to dominate, increasing to the edge of the sampled region. High column absorbers show no strong trend with radius beyond \rvir\ for the Coma-like cluster, although there is an uptick beginning around 2.0 \rvir\ for the Virgo-like cluster. This is likely due to the infalling substructure visible towards the lower right of the Virgo-like cluster in Figure~\ref{fig:projections}. Each cluster has surprisingly similar total absorber number densities outside \rvir\, although the number density is slightly higher in the Coma-like cluster.

The behavior of low column density absorbers within \rvir\ in each cluster is similar, dropping towards the cluster center, although the drop is more dramatic in the Coma-like cluster. In each case, there are \textit{no} absorbers with column densities at \NHI $< 10^{17}$ cm$^{-2}$ found within 0.5 \rvir. In fact, for the Coma-like cluster, there are no absorbers at all within 0.5 \rvir. 

It is the absorbers at \NHI\ $> 10^{15}$ cm$^{-3}$ that make the clusters very distinct. The Virgo-like cluster shows an enhancement of these absorbers, such that, while the number density for the Coma-like cluster decreases towards cluster center, the number density for the Virgo-like cluster actually increases within \rvir. This is consistent with the idea that higher column density absorbers are associated with galaxies, and should be preferentially located within \rvir. \rvir\ also corresponds roughly to where galaxy gas loss effects from ram-pressure stripping become significant \citep{TonnesenBryan2007}. This may explain why the Virgo-like cluster shows an enhancement in absorbers within \rvir, and the Coma-like cluster does not. As the Virgo-like cluster is more dynamically active, the ram-pressure stripped gas may not have yet had time to thermalize with the cluster ICM. This gas is obvious in Figure~\ref{fig:projections} as long tails of HI gas. However, without a greater sample of simulated clusters and without better resolution of the ram-pressure stripping, it is hard to generalize this apparent difference in morphology.

\subsection{Kinematical Properties}
\label{sec:kinematics}

\begin{figure*}
\centering
\includegraphics[width=0.45\linewidth]{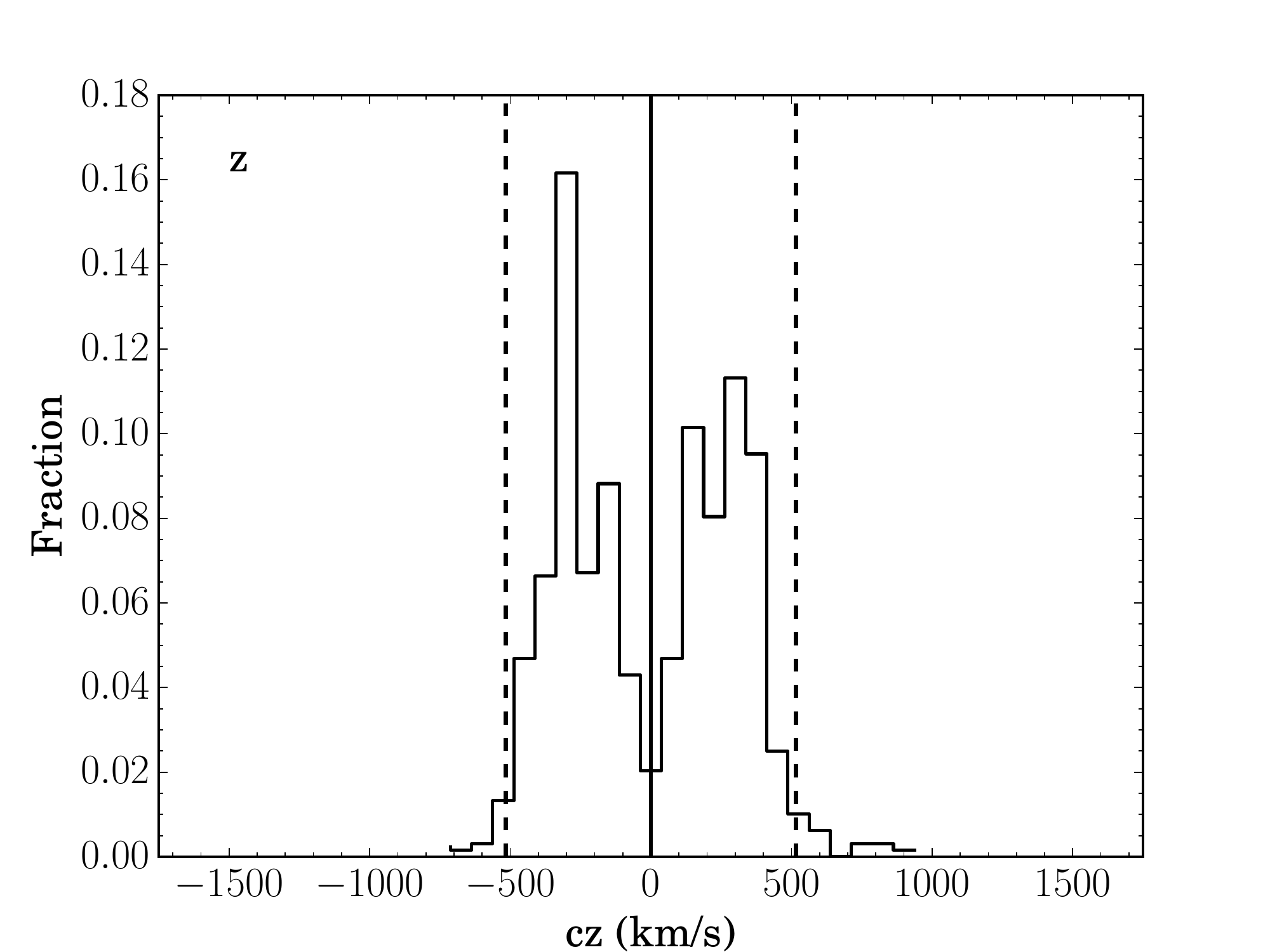}
\includegraphics[width=0.45\linewidth]{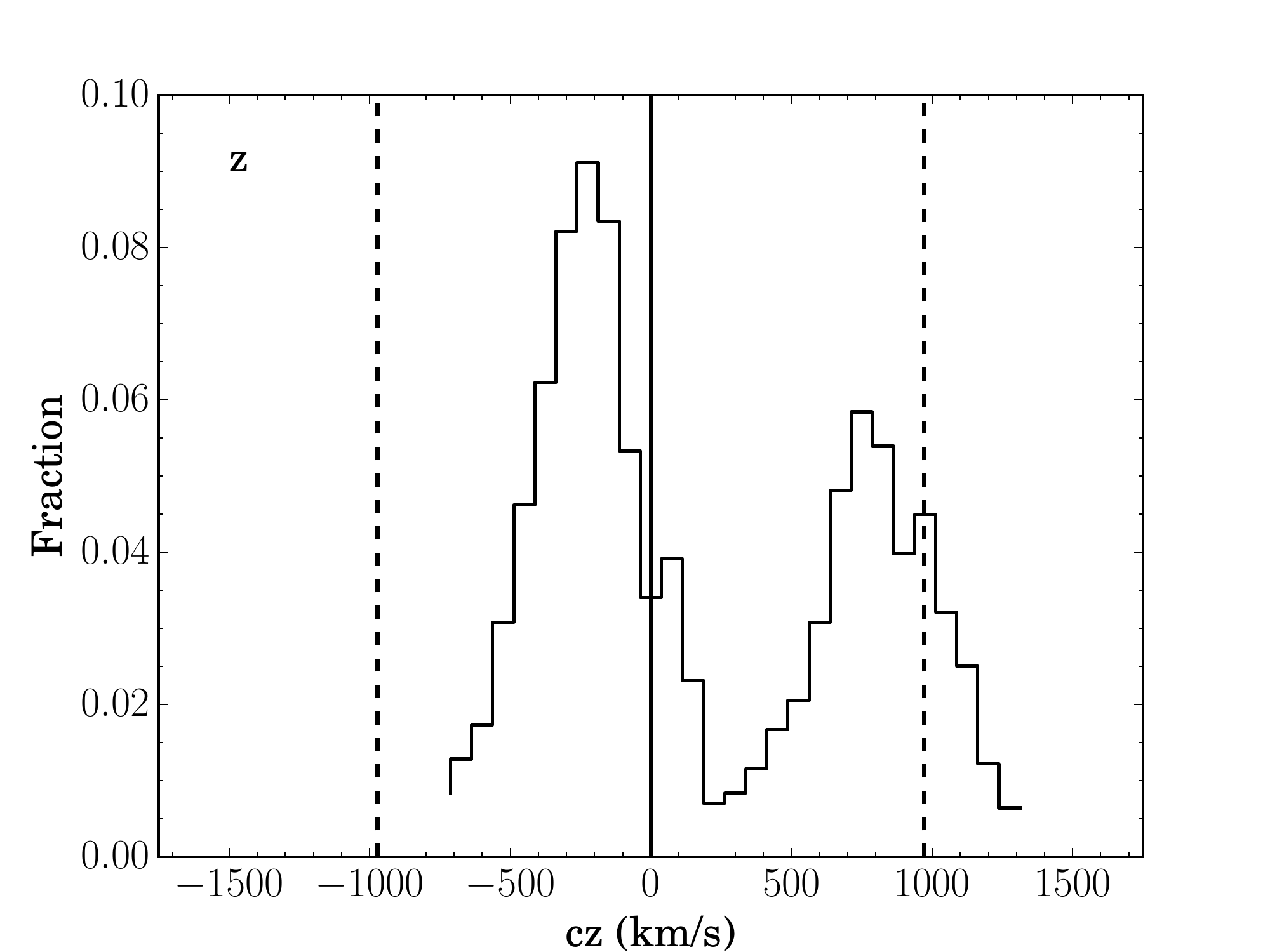}
\vspace{0.1cm}
\includegraphics[width=0.45\linewidth]{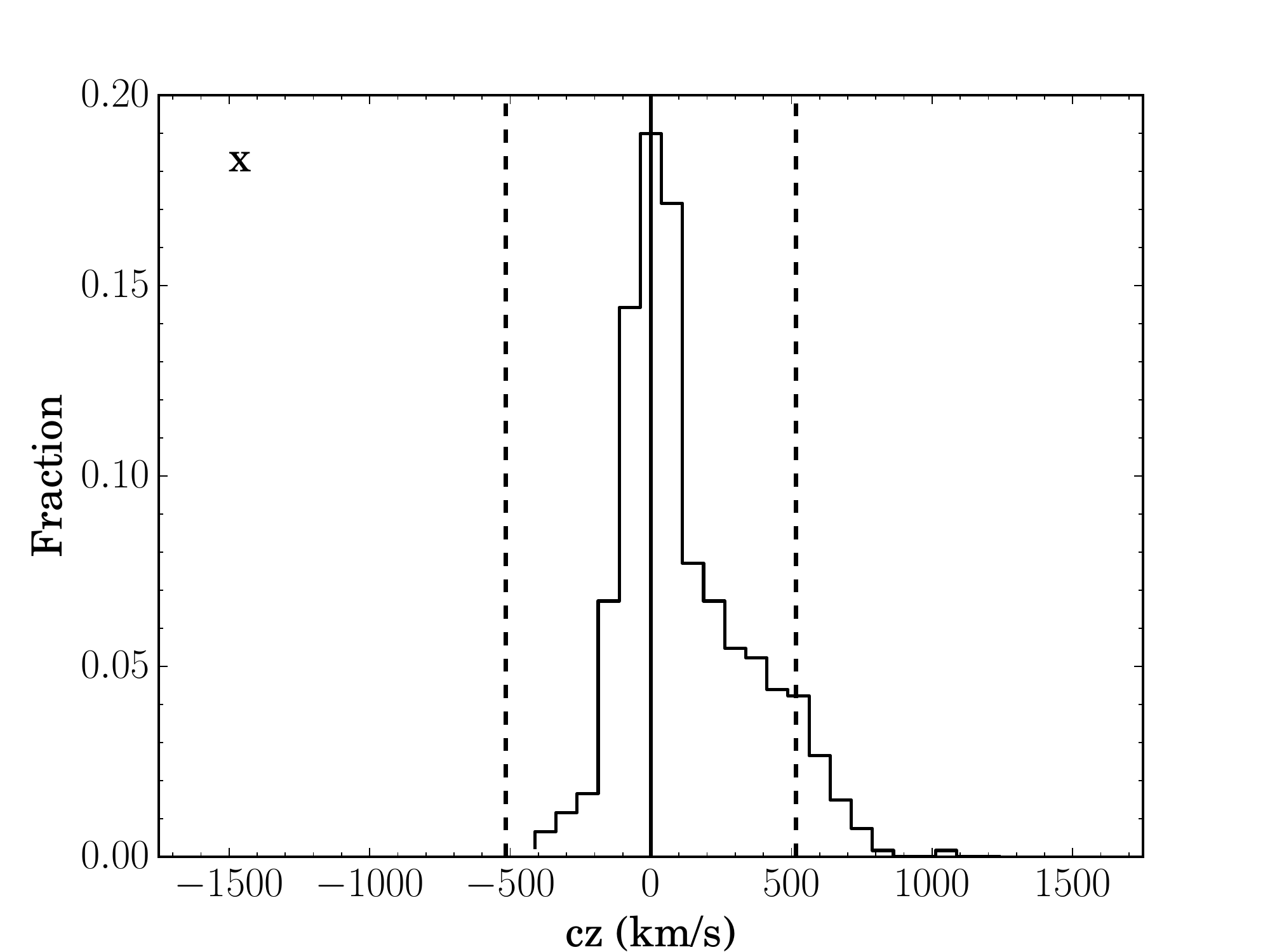}
\includegraphics[width=0.45\linewidth]{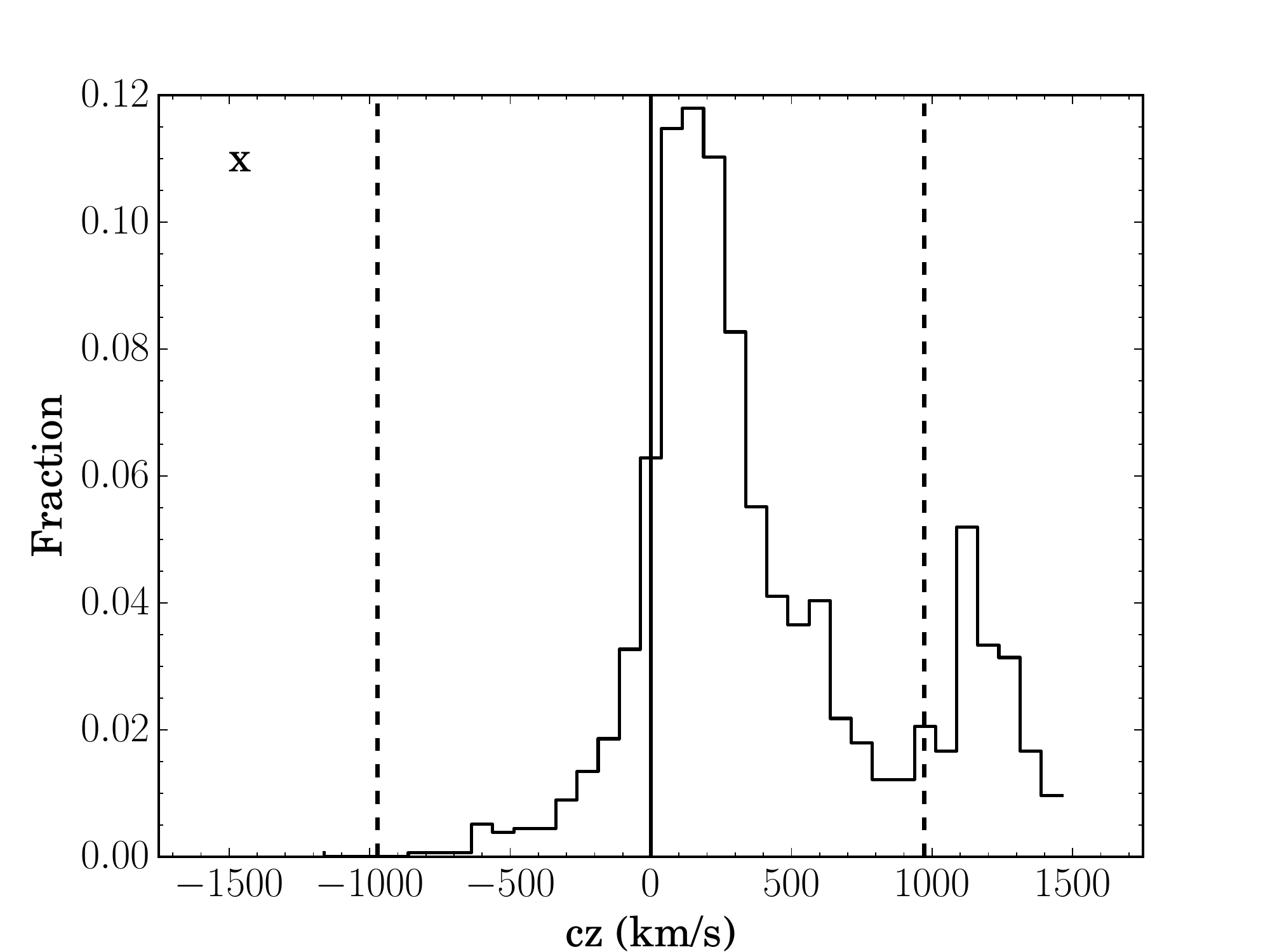}
\vspace{0.1cm}
\includegraphics[width=0.45\linewidth]{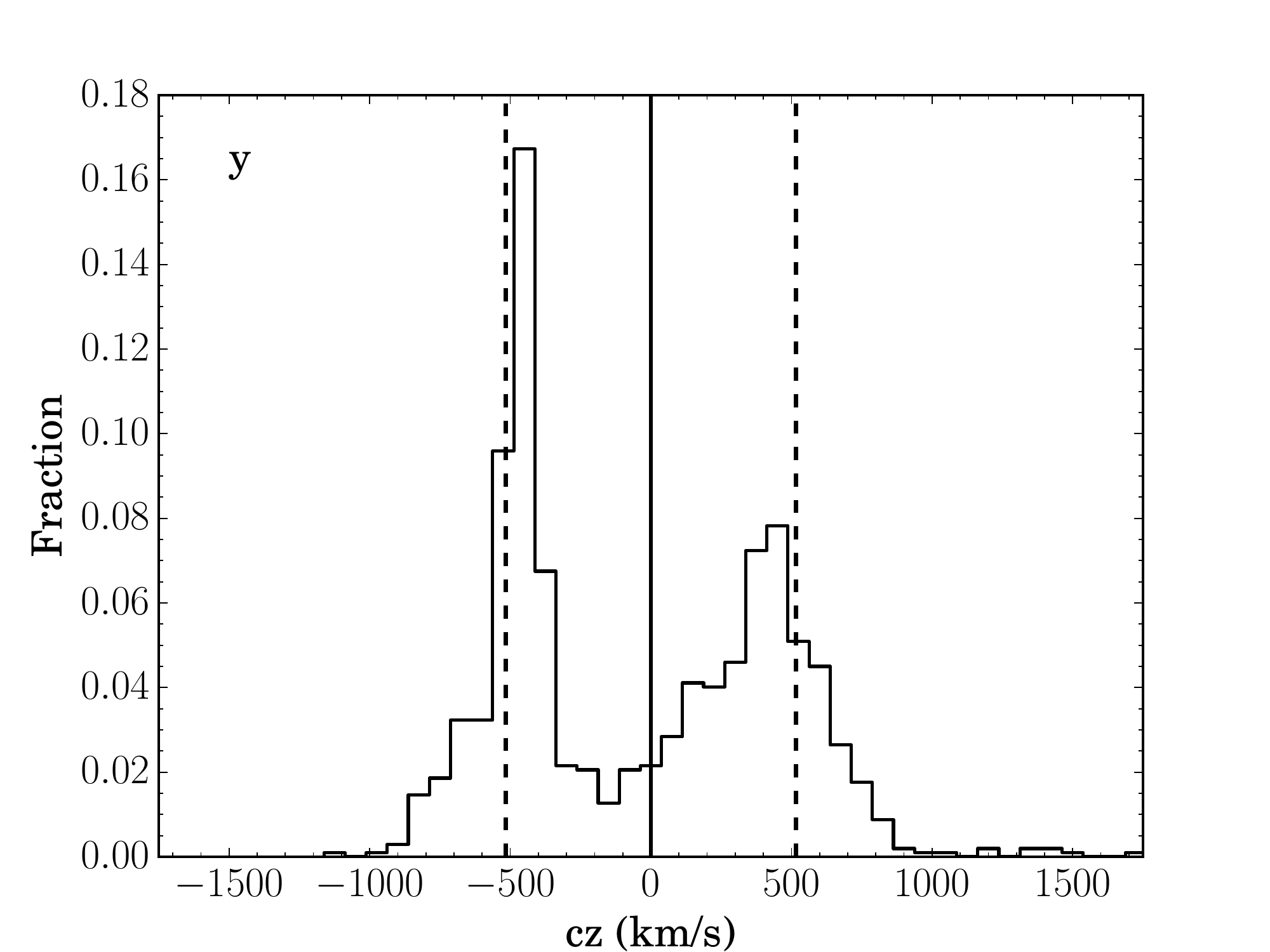}
\includegraphics[width=0.45\linewidth]{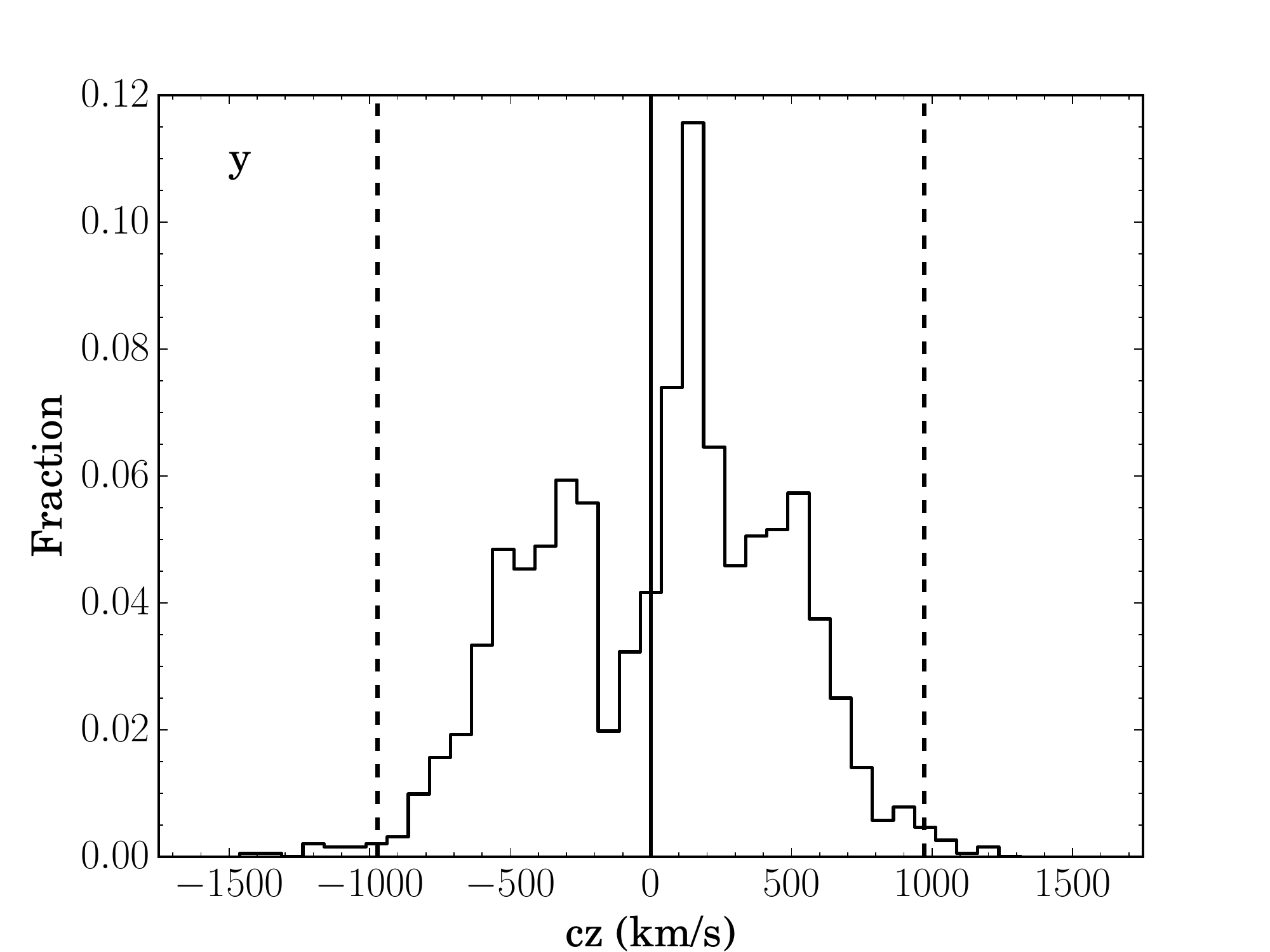}
\caption{Shown is the observed velocity distribution ($v_{\rm{los}} = v_{\rm{H}} + v_{\rm{pec}}$) for the identified absorbers in the Virgo-like (left) and Coma-like (right) simulated galaxy clusters along the z, x, and y viewing axes (from top to bottom). The z axis is the primary viewing direction used in this work. The velocities are shifted to the cluster redshift. The dark matter velocity dispersion for each cluster is given by the dashed vertical lines.}
\label{fig:vel disp}
\end{figure*}

Next, we examine the velocity distribution of the absorbers.  This is shown in Figure~\ref{fig:vel disp} for three different projections. The dashed lines in each panel give the dark matter velocity dispersion for each cluster as computed directly from the simulation dark matter particles. In most cases, the distribution of absorbers is not single-peaked, as would be expected for galaxies in galaxy clusters (and for dark matter particles). Instead, there is (usually) a clear bi-modality in the velocity distributions, with the trough between the peaks centered approximately on the cluster systemic velocity, indicating that the absorbers are avoiding the central cluster region in velocity space. Strictly speaking, however, this does not mean that the absorbers avoid the cluster center in physical space, since the distribution of velocities is subject to projection effects.

\begin{figure*}
\centering
\includegraphics[width=0.45\linewidth]{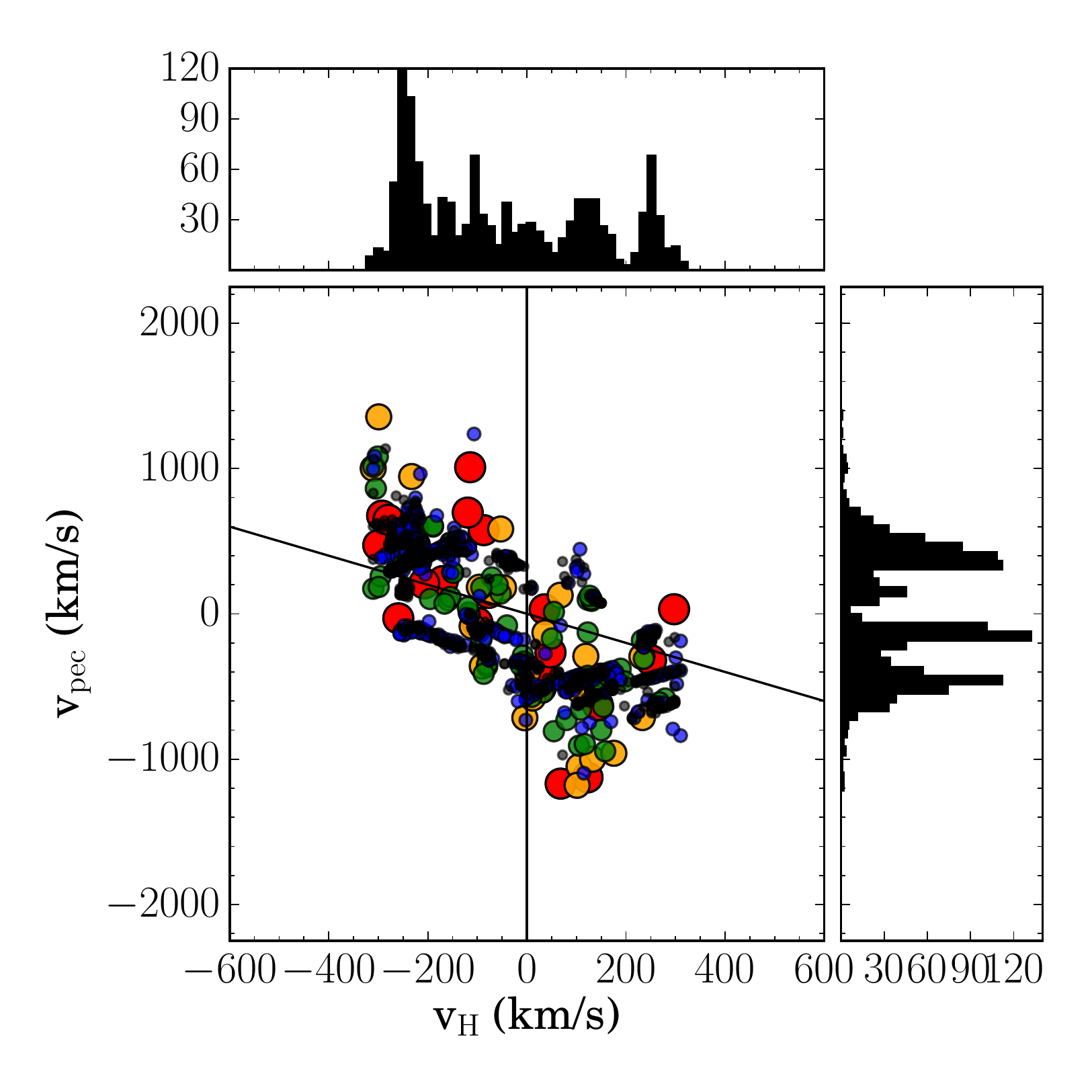}
\includegraphics[width=0.45\linewidth]{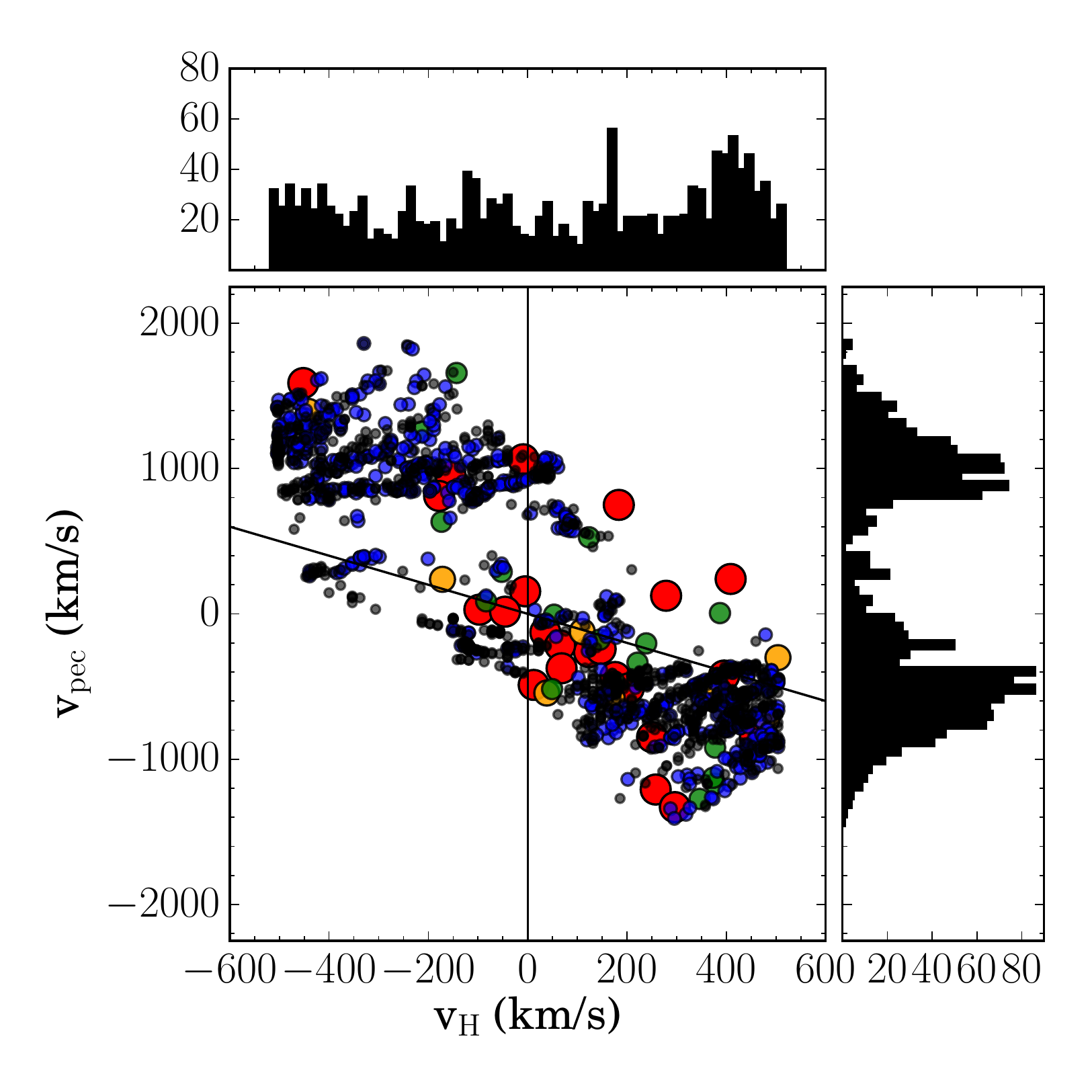}
\caption{Plotted is the peculiar velocity of each identified absorber against the Hubble flow velocity for the Virgo-like (left) and Coma-like (right) clusters. Note the difference in scales between the vertical and horizontal axes.  Absorbers are colored by their column density: \logN = 12-13 (black), 13-14 (blue), 14-15 (green), 15-16 (yellow), and 17+ (red). The horizontal axis is scaled such that cluster center has a Hubble flow velocity of zero (vertical line). The total observed $v_{\rm{los}}$ for each absorber is the sum of its peculiar velocity and Hubble flow velocity. Absorbers with an observed $v_{\rm{los}}$ of zero would lie on the diagonal line and would represent absorbers appearing at the cluster redshift in projection.}
\label{fig:absorber velocities}
\end{figure*}

To investigate the influence of projection effects, and the origin of the troughs in Fig.~\ref{fig:vel disp}, we plot the peculiar velocity projected along the line of sight ($v_{\rm{pec}}$) as a function of the Hubble flow velocity for each absorber ($v_{\rm{H}}$) in Figure~\ref{fig:absorber velocities}. \footnote{We note that the apparent hard cutoff of absorbers at large absolute Hubble velocities is simply because this corresponds to the edge of the sampled volume around each cluster. The difference in extent between the Virgo-like and Coma-like clusters along the horizontal axis is because the sampled volume was a function of the virial radius, with the Coma-like cluster being larger by a factor of $\sim 1.6$.} Again, the total observed line of sight velocity of any absorber is $v_{\rm{los}}$ = $v_{\rm{pec}} + v_{\rm{H}}$. We center $v_{\rm{H}}$ on the cluster frame, with the vertical line at $v_{\rm{H}} = 0$.  

The plots in Figure~\ref{fig:absorber velocities} are subdivided into four regions by the vertical black line at $v_{\rm{H}}$ = 0, and the diagonal black line at $v_{\rm{pec}}$ = -$v_{\rm{H}}$.  Absorbers with a negative $v_{\rm{H}}$ are physically located in front of the cluster, relative to the observer, and absorbers with a positive $v_{\rm{H}}$ are physically located behind the cluster, relative to the observer. Since the observed velocity of each absorber is the sum of its value on the vertical and horizontal axes ($v_{\rm{los}} = v_{\rm{H}} + v_{\rm{pec}}$), the diagonal line gives a fixed observed velocity of $v_{\rm{los}}$ = 0; if $v_{\rm{los}}$ = 0, the absorber would appear at cluster center in velocity space. This means that absorbers in the upper left quadrant are physically located in front of the cluster ($v_{\rm{H}} <$ 0), but have a positive projected kinematic velocity such that $v_{\rm{pec}} > v_{\rm{H}}$, causing these absorbers to appear to be behind the galaxy cluster in velocity space (or $v_{\rm{los}} > 0$). Likewise, absorbers in the lower right quadrant are physically behind the galaxy cluster, but appear to be located in front of the cluster in velocity space ($v_{\rm{los}} < 0$). The absorbers in the remaining two quadrants are physically located on the same side of the cluster as they appear in velocity space. 

As shown in Figure~\ref{fig:absorber velocities}, the majority of absorbers in both clusters are located in the upper left or lower right quadrants, indicating that most absorbers have large kinematic velocities pointed towards the cluster center --  they are infalling.

The exact fraction of absorbers in these two regions is different for these two clusters. This effect seems to be more significant in the Coma-like cluster (80\% of absorbers), compared to the Virgo-like cluster (68\%). The fraction shows no obvious trend in the Coma-like cluster with column density, although the fraction increases to $\sim$83\% for the Virgo-like cluster at \NHI\ $> 10^{15}$ cm$^{-2}$. 

The inset histograms in Figure~\ref{fig:absorber velocities} of the projected peculiar velocities and the Hubble flow velocities reveal a bimodal distribution in the projected peculiar velocities of absorbers. This is consistent with our earlier observation that the total velocity distribution of the identified absorbers is also bimodal (see Figure~\ref{fig:vel disp}). In fact, since the $v_{\rm{H}}$ distribution for both clusters is roughly uniform over the probed volume, this indicates that the observed bimodal distribution is mostly a result of the projected peculiar velocities: or a direct result of the fact that most of the absorbers are infalling onto the cluster. Although absorbers do tend to avoid the inner cluster ICM (see Sec.~\ref{sec:number density}), the troughs in Figure~\ref{fig:vel disp} are not indicative of this (since there is no trough in the v$_{\rm{H}}$ histogram of either cluster), but rather are a direct result of projected peculiar velocities which change depending on the orientation of the warm gas filaments relative to the observer. It is for this reason that the troughs can shift far from v$_{\rm{H}}$ = 0, and even disappear, in Figure~\ref{fig:vel disp}.

\begin{figure*}
\centering
\includegraphics[width=0.45\linewidth]{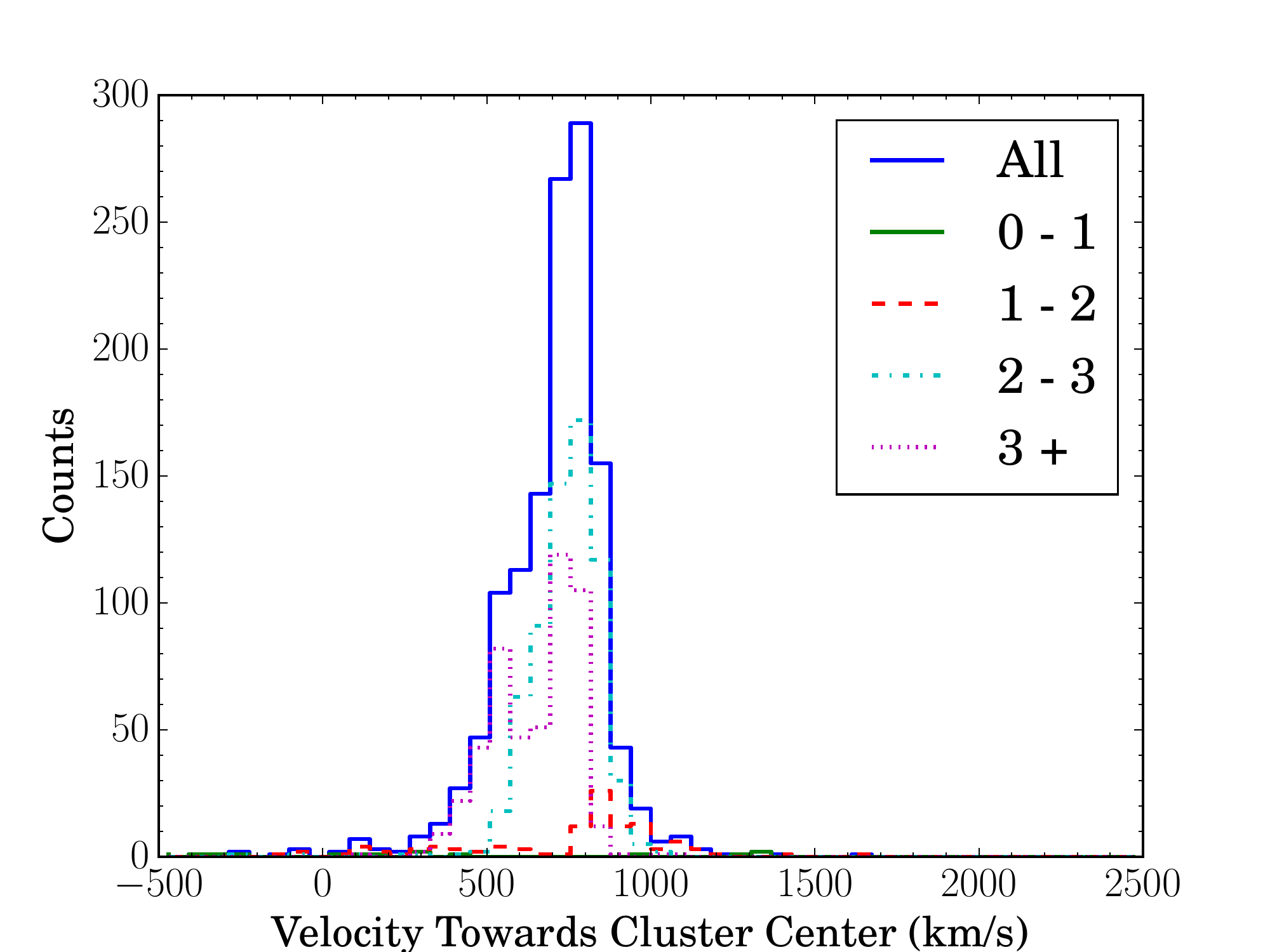}
\includegraphics[width=0.45\linewidth]{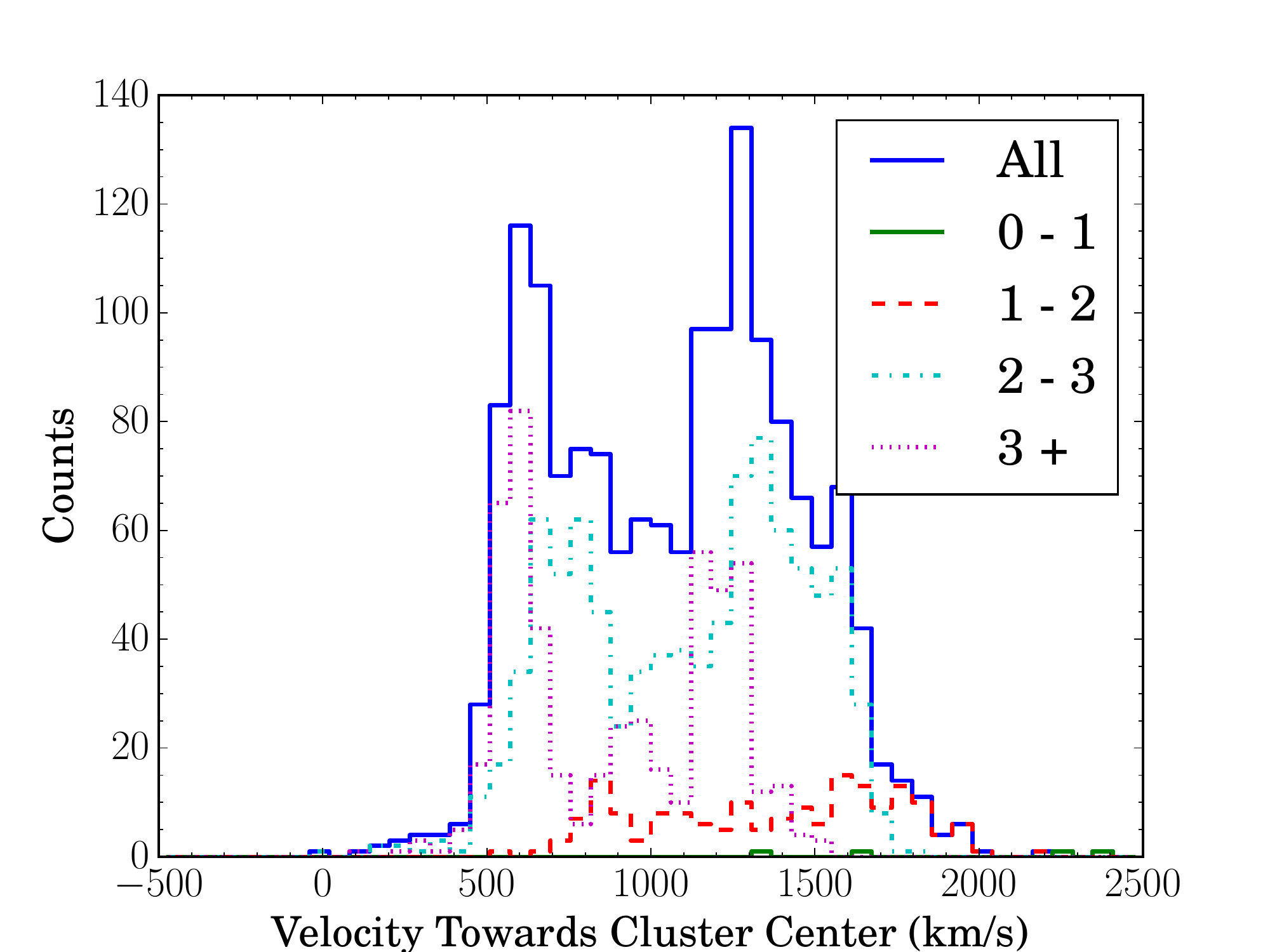}
\caption{The distribution of absorber velocities towards the cluster center (v$_{\rm{r}}$) in our Virgo-like (left) and Coma-like (right) simulations, with positive velocities indicating motion towards cluster center. The blue (solid) line shows all absorbers, while colored lines indicate the distribution split into radial distance bins, with 0 - 1 \rvir (green, solid), 1 - 2 \rvir (red,dashed), 2 - 3 \rvir (cyan, dot-dashed), and \rvir $>$ 3 (purple, dotted).}
\label{fig:radial velocity}
\end{figure*}

These results have all been based on velocities along a line-of-sight (i.e. relative to the observer, not the cluster); however, we can confirm the infall scenario by looking directly at the velocity of the absorbers relative to the cluster center (v$_{\rm{r}}$).  This is shown in Figure~\ref{fig:radial velocity}.  It is clear that a majority of the absorbers in each cluster represent infall at high velocities onto the cluster, as virtually all absorbers have large, positive v$_{\rm{r}}$.  As expected from its larger mass, the Coma-like cluster absorber v$_{\rm{r}}$ distribution generally has a higher mean velocity and larger dispersion than the Virgo-like cluster. 

Examining the distribution broken down by cluster distance in Figure~\ref{fig:radial velocity}, there is a noticeable trend: in general, moving towards the cluster center shifts the distributions towards the right (higher velocities). This is most obviously seen going from \rvir\ $>$ 3 to 2 - 3 \rvir\ (purple to cyan) in both clusters. The shift continues going to 1 - 2 \rvir, but less obviously. The bimodal peaks are still present in the 2 - 3 and \rvir\ $>$ 3 bins in the Coma-like cluster, but are absent for absorbers 1 - 2 \rvir\ from the cluster center, in favor of an almost uniform distribution spanning 500 km s$^{-1}$ to 2000 km s$^{-1}$. Interestingly, a second peak appears in the Virgo-like cluster distribution for absorbers \rvir\ $>$ 3 from the cluster center, which is somewhat responsible for the hump in the full distribution. The motion of the absorbers towards the cluster center is consistent with the idea that most of the observed warm gas represents first infall onto the cluster. If any absorbers were to survive passage through cluster center, there should be some subset of the distribution in Figure~\ref{fig:radial velocity} with negative v$_{\rm{r}}$. There are only a handful ($<$ 10) absorbers with negative velocities in the Virgo-like cluster, and none in the Coma-like cluster. Therefore, if any absorbers do survive passage through cluster center, their lifetimes must be short.

\subsection{Thermal Properties}
\label{sec:thermal properties}

\begin{figure*}
\centering
\includegraphics[width=0.45\linewidth]{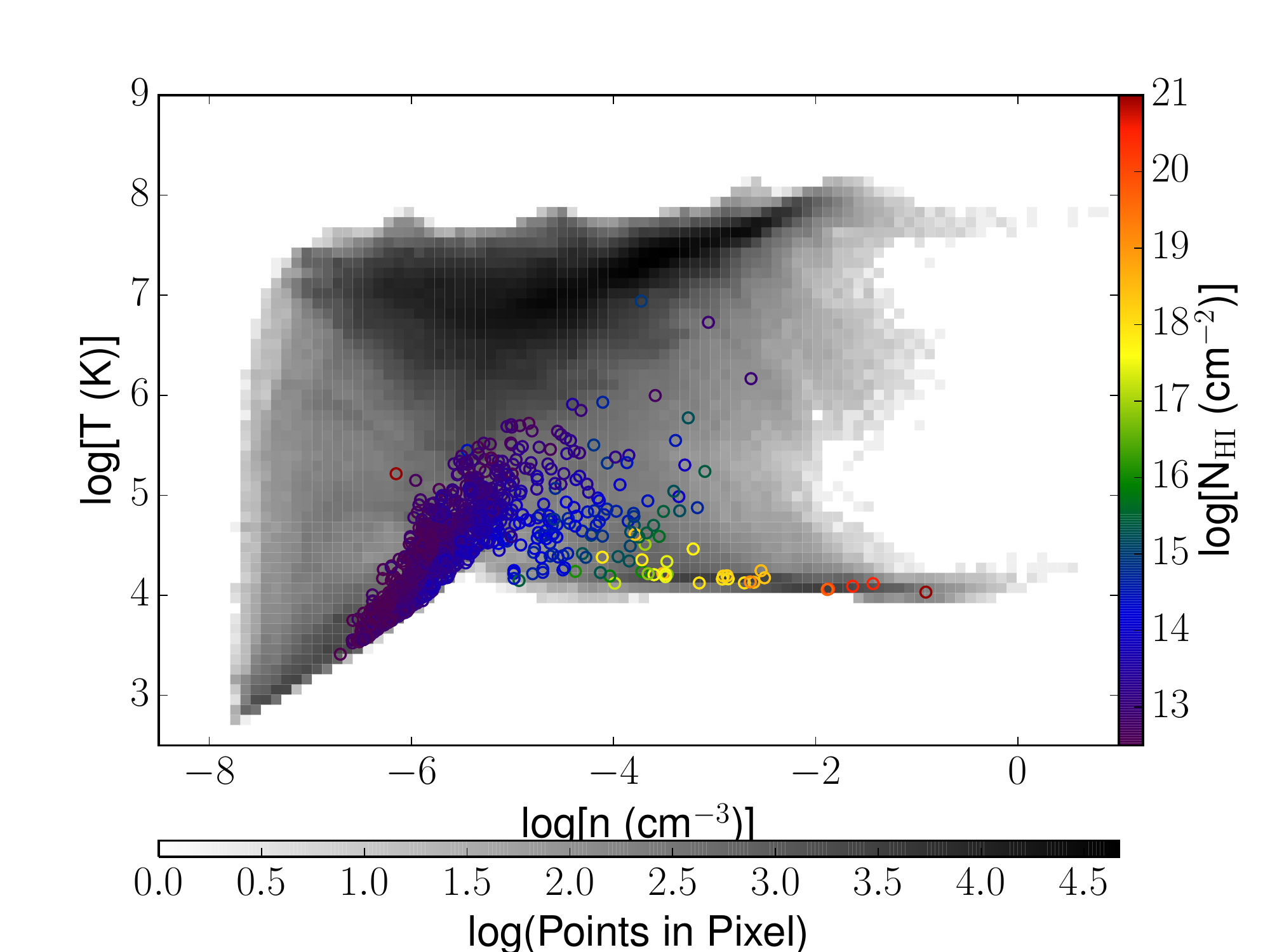}
\includegraphics[width=0.45\linewidth]{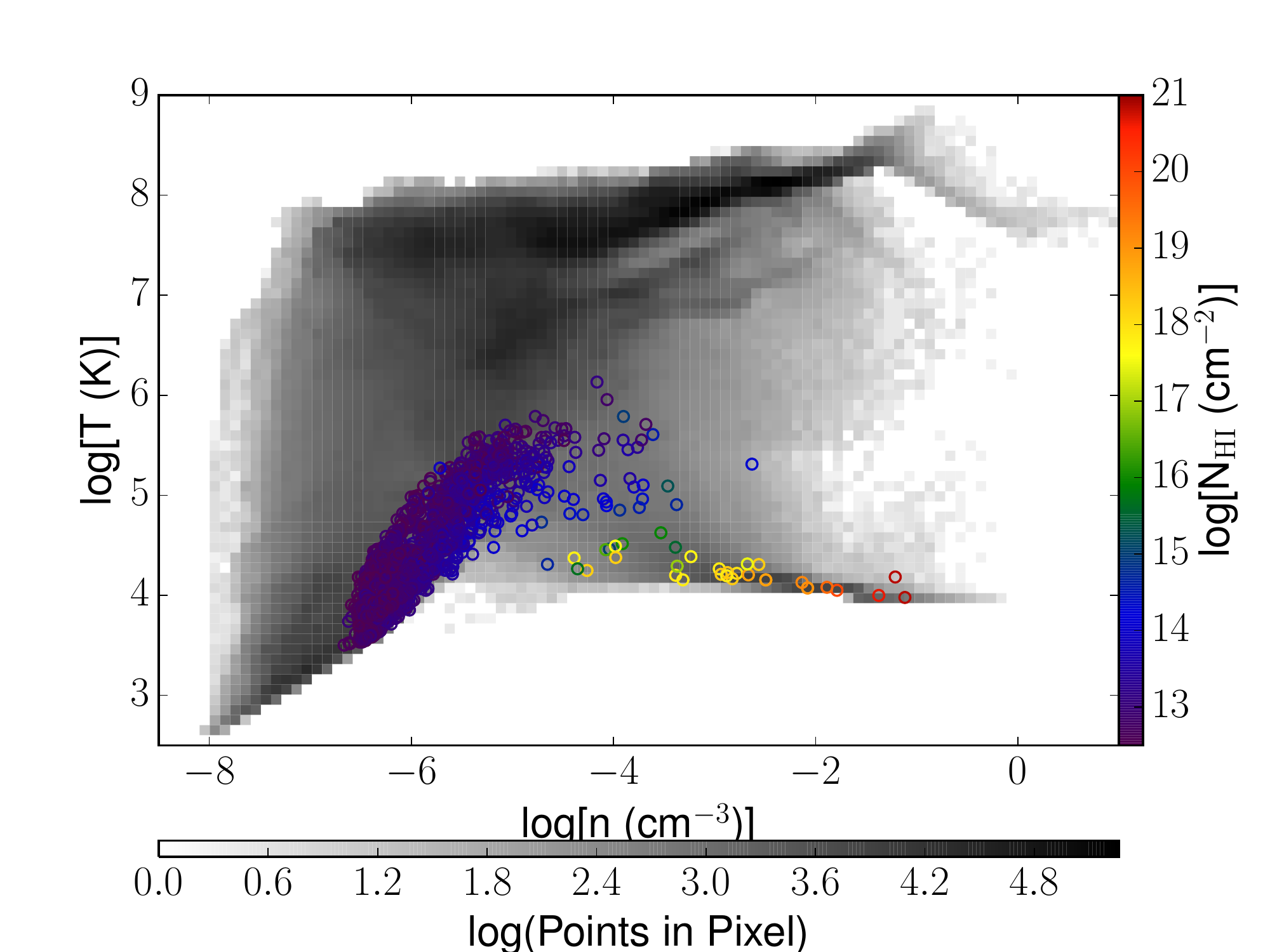}
\vspace{0.1cm}
\includegraphics[width=0.45\linewidth]{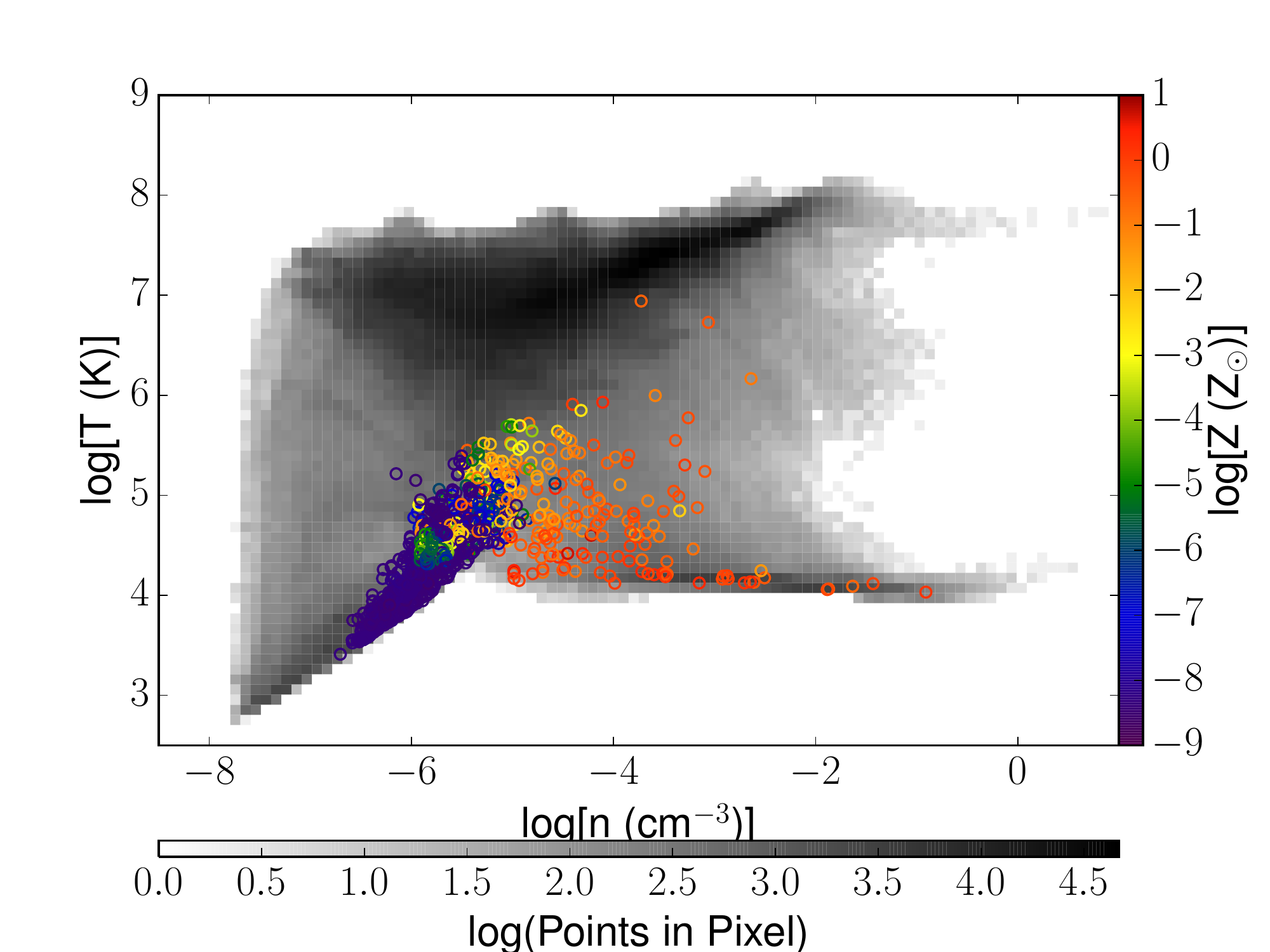}
\includegraphics[width=0.45\linewidth]{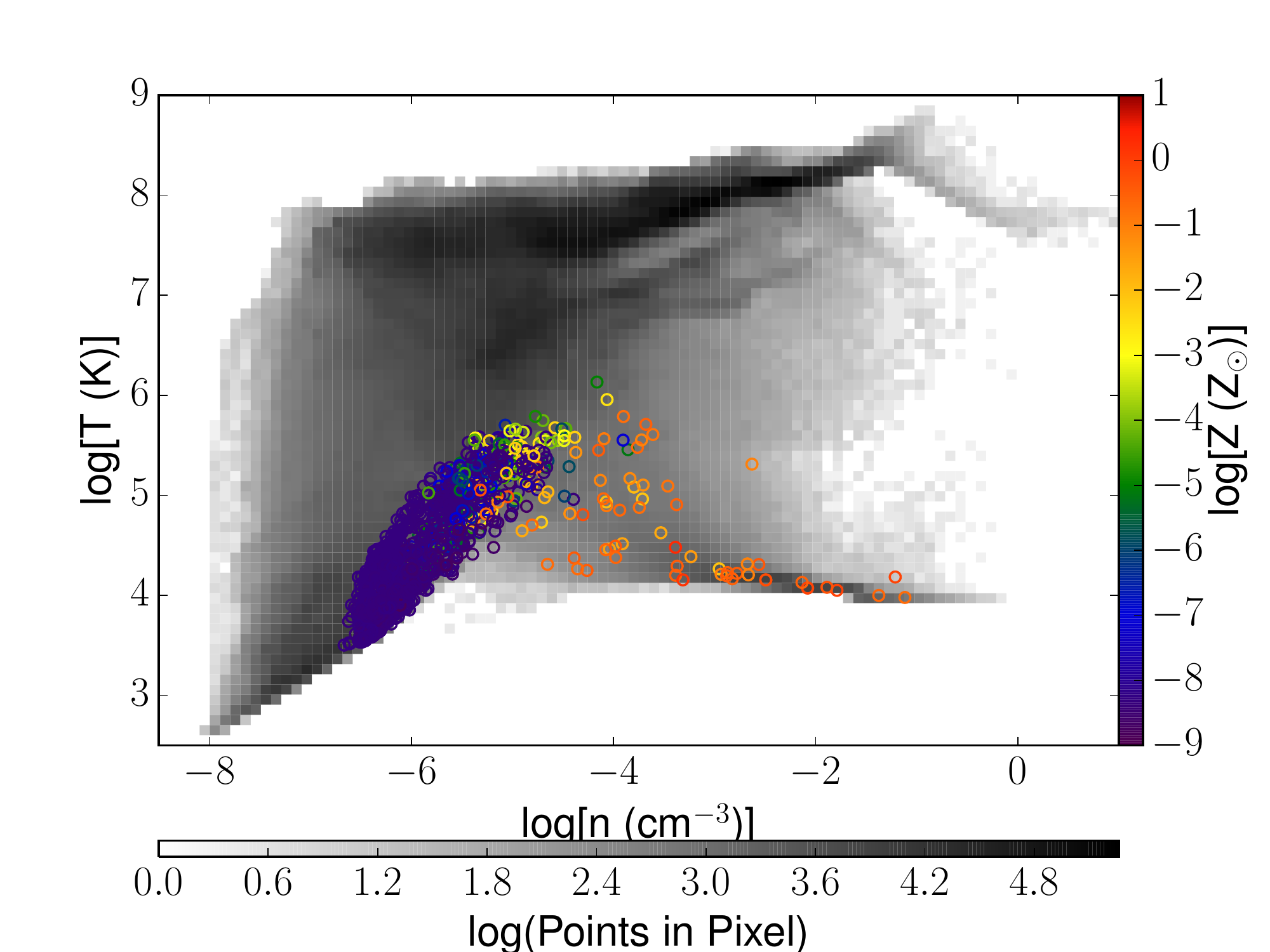}
\caption{Phase plots of all the gas (greyscale) and absorbers (colored circles) for the Virgo-like cluster (left) and Como-like simulations.  In the top set of panels, the absorbers are color coded by column density, while in the bottom, they are color-coded by metallicity.}
\label{fig:phase diagram}
\end{figure*}

In our final section on the properties of the absorbers, we turn to their thermal characteristics.  In Figure~\ref{fig:phase diagram}, we show the distribution of the absorbers in a temperature-density phase diagram (keeping in mind that this is the temperature and density of the cell with the largest contribution to the absorbing gas).  To provide context, we also show the distribution of all the gas as a gray scale, weighted by mass.

Looking first at the distribution of all the gas, we see that, as expected for such simulations \citep[e.g.,][]{Dave1999}, gas falls largely into three regions: (i) a low-density and low-temperature region, similar to the intergalactic medium (IGM), (ii) a high-temperature shocked ICM region, and (iii) cold, dense gas associated with galaxies.  As expected, the HI absorbers arise primarily from the first phase, with most absorbers below about \NHI\ $< 10^{14}$ cm$^{-2}$ arising from gas which shows a clear density-temperature relation \citep{HuiGnedin1997}.  For column densities above about $10^{17}$ cm$^{-2}$, the absorbers arise almost entirely from dense, cold gas which we expect to be in or near galaxies.  Interestingly, the population of intermediate absorbers come from gas with a wide range of properties, including some gas which is quite warm ($T \sim 10^5 $ K), indicating that it might be interesting to look at the width of the absorber profiles \citep[e.g.,][]{Richter2006}.

The metallicity distribution of the absorbers, shown in the bottom panels of Figure~\ref{fig:phase diagram}, indicates that, broadly speaking, the IGM-like absorbers have very low metallicities, while the absorbers from cold-dense gas are highly enriched. The population of absorbers in the transition region between warm, diffuse IGM and cold, condensed gas is more mixed, with a wide range of metallicities. However, the wide distribution is limited to above about  Z = 0.1 Z$_{\odot}$ for absorbers with N$_{\rm{HI}}$ $>$ 10$^{14}$ cm$^{-2}$. Absorber metallicity is discussed further in Sec.~\ref{sec:absorber metallicity}.

\section{Direct Comparison to Observation}
\label{sec:comparison to observation}

The previous section demonstrates that the majority of absorbers identified in our simulations come from gas which is of low density falling into the cluster for the first time.   We now turn to making direct comparisons and predictions for observations of absorbers in and around galaxy clusters.   It is important to keep in mind that we currently only have the Virgo Cluster observations to compare to directly and that the number of sightlines through a cluster will always be very limited compared to our simulation results.  We begin with the covering fraction, as this can be computed for small sample sizes.  We then go on to discuss the metallicities and the position-velocity distributions of the absorbers in the two simulated clusters.

\subsection{Covering Fraction and Column Density Distribution}
\label{sec:covering fraction}
\begin{figure}
\centerline{\includegraphics[width=\columnwidth]{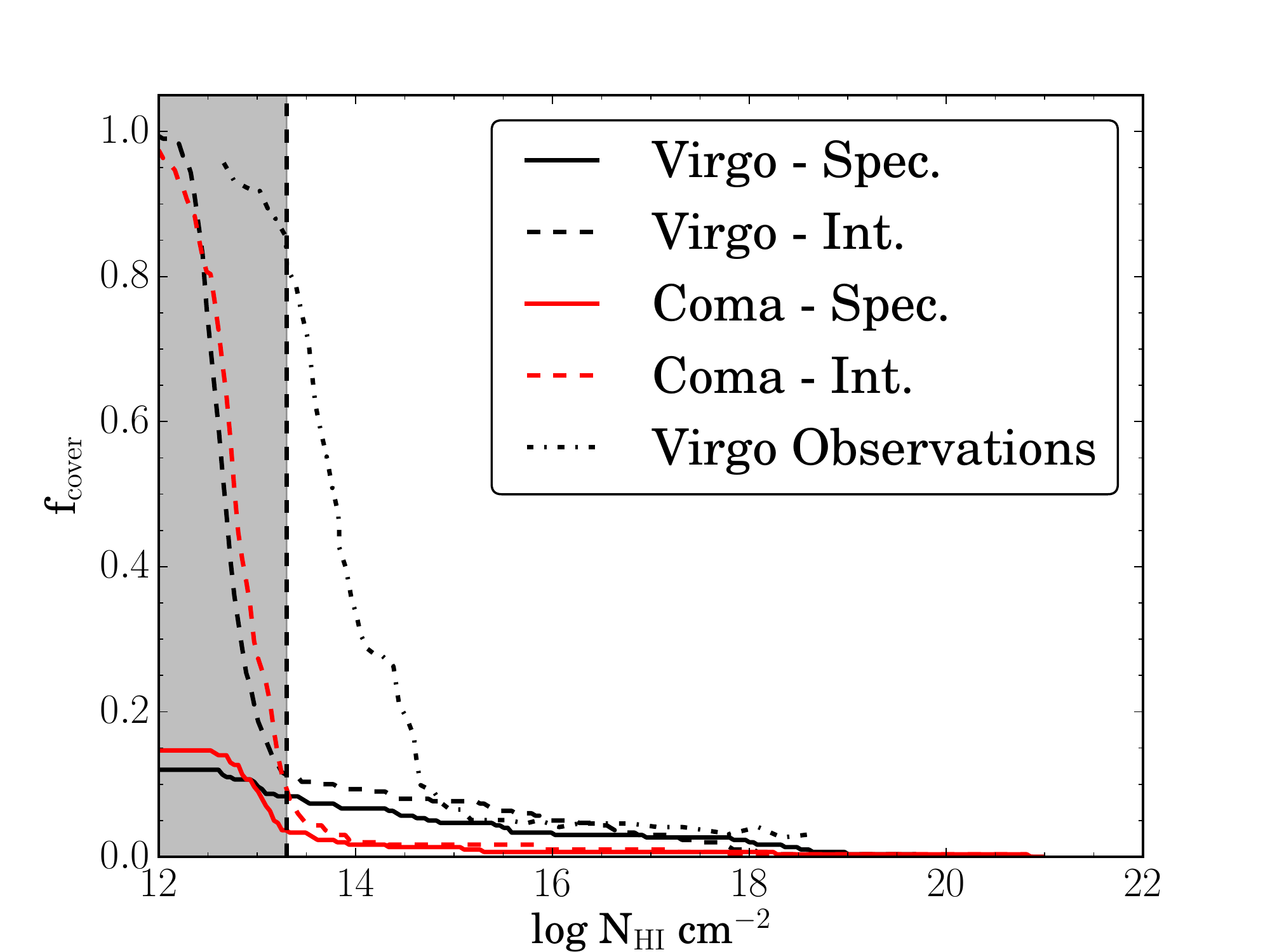}}
\caption{Comparison of covering fraction as a function of N$_{\rm{HI}}$. The covering fraction is defined as the fraction of lines of sight with identified absorption at or above the given N$_{\rm{HI}}$. The synthetic observations are shown with solid lines for the Virgo-like (black) and Coma-like (red) clusters, with the total (integrated through all cells along the line of sight) N$_{\rm{HI}}$ for each shown with dashed lines. These are compared to the Virgo observations (black, dash-dot). The shaded region marks the completeness limit at log( N$_{\rm{HI}}$ cm$^{-2}$ ) $<$ 13.3 given in Y12.}
\label{fig:cf}
\end{figure}

The covering fraction of absorbers, f$_{\rm{cover}}$, can be computed as the fraction of sight lines that contain a feature above a given column density.  We present the covering fraction for our Virgo-like and Coma-like clusters in Figure~\ref{fig:cf}, along with a direct comparison to the actual Virgo covering fraction from Y12. The gray region in this figure denotes the completeness limit of \NHI $> 10^{13.3}$ cm$^{-2}$ given in Y12. The covering fractions for our Virgo-like and Coma-like galaxy clusters are similar with each other. Yet, above the completeness limit the Coma-like cluster has a lower \fcover\ than the Virgo-like cluster, in agreement with the column density distribution in Figure~\ref{fig:cluster NfN}. Both simulated clusters, however, lie well below the actual covering fraction of the Virgo cluster at all column densities. The discrepancy varies by a factor of up to 7 at \NHI\ $< 10^{15}$ cm$^{-2}$, to about 3 at \NHI\ $> 10^{15}$ cm$^{-2}$. 

To ensure that our synthetic observation methods are not the source of this disagreement, we plot the covering fraction of the total HI content in the simulation integrated along each line of sight as the dashed lines in Figure~\ref{fig:cf}. The integrated HI covering fraction reproduces the relationship between the two clusters in that the Virgo-like cluster generally has more absorption than the Coma-like cluster (above the completeness limit). In addition, it reproduces the larger covering fraction of the Coma-like cluster at low column densities (below the completeness limit). Above the completeness limit of Y12 (which we tried to reproduce using noise comparable to the COS observations of Y12), the integrated \fcover\ corresponds well with the synthetic observations. Therefore, the missing HI content is not an artifact of the synthetic observational technique, but corresponds to a real lack of HI in the cluster simulations. 

To examine morphological differences between the two simulated clusters, and further compare to observations, we tabulate the covering fraction for each cluster at various column densities separated into \rvir\ bins in Table~\ref{table:cf}. The column densities correspond to those chosen in the Y12 Virgo results, corresponding to equivalent widths of W = 65, 100, 150, 200, and 300 m{\AA}, assuming a Doppler broadening of $b = 30$ km s$^{-1}$. The disagreement between the Virgo cluster and our synthetic clusters is clear here, with covering fractions of unity at \NHI $= 10^{13.123}$ cm$^{-2}$ in Virgo, and fractions ranging from several to 20\% in our simulated clusters. The general trend seen in the Virgo observations is that covering fraction decreases with increasing column density and increases away from the cluster center. This trend is seen also in our Coma-like cluster, and extends outside 2 \rvir, a region not probed by the Virgo observations. However, this behavior is not matched in our Virgo-like cluster at \NHI $> 10^{13.572}$ cm$^{-2}$, as the maximum covering fraction at these column densities occurs within the virial radius of the cluster (in projection). At all column densities, however, the fraction at 2 - 2.5 \rvir\ is greater than that at 1 - 2 \rvir. 

\begin{table*}
 \centering
 \begin{tabular}{ c | c c | c c c | c  c  c}
    \hline
    \hline
         & \multicolumn{2}{c}{Virgo} & \multicolumn{3}{c}{Virgo-like} & \multicolumn{3}{c}{Coma-like} \\
    \NHI & 0-1 \rvir & 1-2 \rvir & 0-1 \rvir & 1-2 \rvir & 2-2.5 \rvir & 0-1 \rvir & 1-2 \rvir & 2-2.5 \rvir \\
    \hline
    13.123 & 1.00$_{-0.15}$ & 1.00$_{-0.17}$ & 0.086 & 0.134 & 0.159 & 0.065 & 0.175 & 0.202 \\
    13.338 & 0.60$^{+0.13}_{-0.16}$ & 1.00$_{-0.14}$ & 0.083 & 0.093 & 0.106 & 0.033 & 0.090 & 0.102\\
    13.572 & 0.50$^{+0.15}_{-0.15}$ & 0.86$^{+0.08}_{-0.18}$ & 0.073 & 0.059 & 0.072 & 0.027 & 0.052 & 0.050\\
    13.758 & 0.27$^{+0.15}_{-0.10}$ & 0.75$^{0.11}_{-0.17}$ & 0.074 & 0.037 & 0.051 & 0.024 & 0.035 & 0.034\\
    14.109 & 0.08$^{+0.12}_{-0.05}$ & 0.56$^{+0.15}_{-0.16}$ & 0.067 & 0.030 & 0.033 & 0.017 & 0.018 & 0.022\\
    \hline     
 \end{tabular}
 \caption{Covering Fractions for the Virgo-like and Coma-like clusters in selected \NHI\ and \rvir\ ranges.}
 \label{table:cf}
\end{table*}

\begin{figure}
\centerline{\includegraphics[width=\columnwidth]{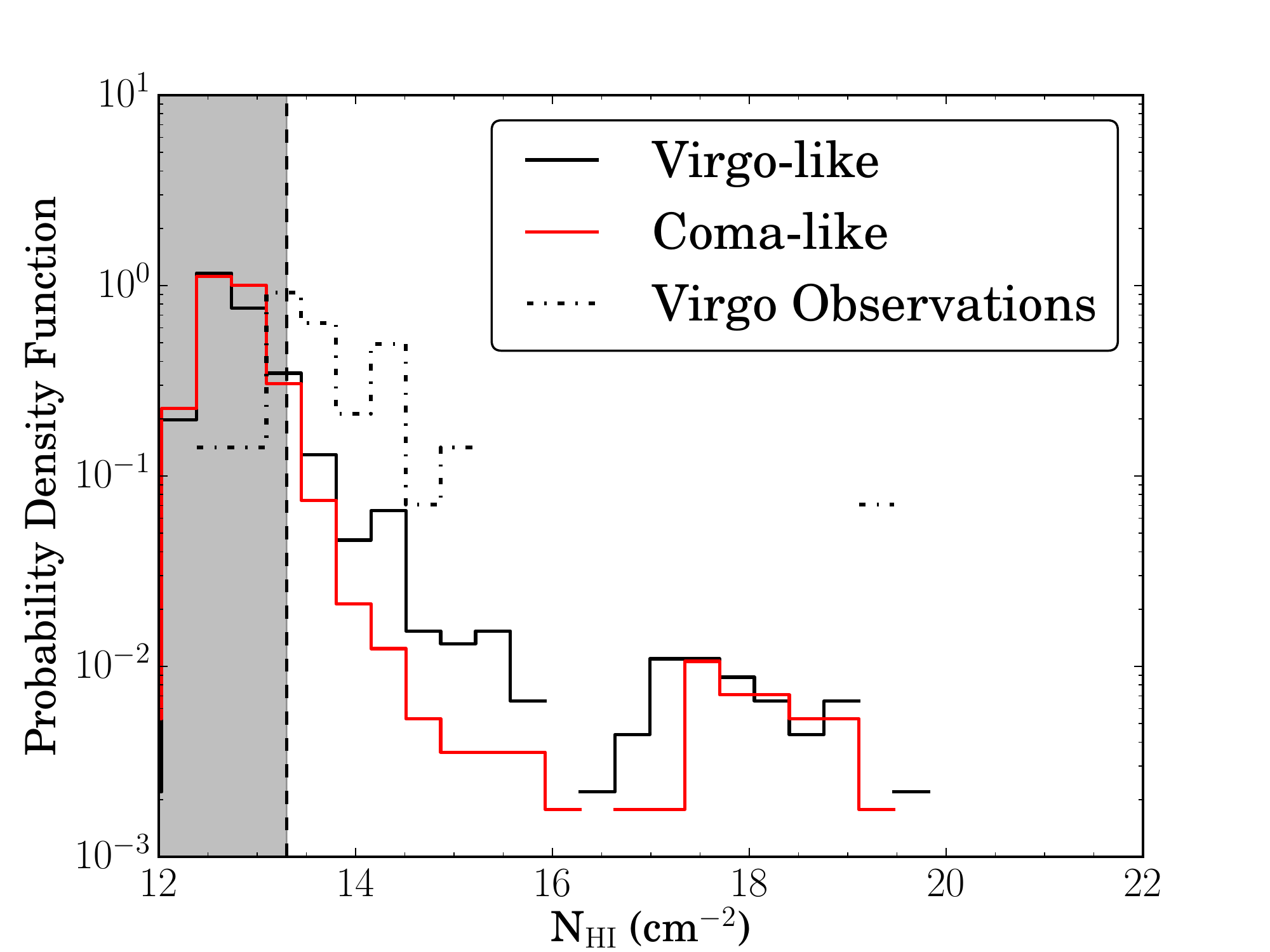}}
\caption{The normalized column density distribution for the Virgo-like (black) and Coma-like (red) clusters in comparison to the observed distribution in the Virgo cluster from Y12 (black, dash-dotted). The gray region marks the completeness limit of Y12 given as \NHI $= 10^{13.3}$ cm$^{-2}$.}
\label{fig:CDD observation}
\end{figure}

Consistent with what is shown by the covering fraction, the column density distribution of our simulated clusters lies below that given in Y12, as shown in Figure~\ref{fig:CDD observation}. We plot here the normalized probability density function, rather than the CDD given in Figure~\ref{fig:cluster NfN}. The Virgo-like cluster contains more absorbers compared to the Coma-like cluster except at very high and low column densities. The largest disagreement between the simulated clusters and Virgo is, again, near \NHI $= 10^{14}$ cm$^{-2}$, though the disagreement extends over all column densities. The limited sight lines through the Virgo cluster and small number statistics make rigorous comparison to our simulations at \NHI $> 10^{15}$ cm$^{-2}$ difficult. 

Finally, the typical number of absorbers per sight line in our cluster simulations also lies below that seen in Y12, even though in both cases the sightlines probe similar regions (in velocity space). In the Virgo cluster observations from Y12, 70\% of COS sightlines (6 of 9) had more than one absorber, and two sight lines contained five absorbers each.   The Y12 STIS and GHRS sightlines also show this percentage of multiple absorbers.  From the simulations, we find only 22.6\% of the sightlines with absorbers have more than one in the Virgo-like cluster and 24.7\% in the Coma-like cluster.  In addition, the maximum number of absorbers per sightline is only 3 (4.7\% in the Coma-like cluster, none in the Virgo-like). The differences between the Virgo observations and our simulated clusters show that the difference in absorber frequency is noticeable on an individual sightline basis. The differences do not manifest themselves solely in the global absorber properties (i.e. Figures~\ref{fig:cf} and~\ref{fig:CDD observation}, and Table~\ref{table:cf}).

In our comparison to observations from Y12, we note that it is possible Y12 included absorbers not directly associated with the Virgo cluster or its substructures. We compare against the full velocity range of Y12, or 700 $<$ cz $<$ 3000 km s$^{-1}$; for reference, M87 is located at about 1307 km s$^{-1}$, and the Virgo cluster has a velocity dispersion of $\sigma_{\rm{v}}$ = 544 km s$^{-1}$. As shown in Figure~\ref{fig:vel disp}, the absorbers in our simulated clusters mostly lie within about $\sigma_{\rm{v}}$ of cluster systemic velocity. Making a $2\sigma_{\rm{v}}$ and $1\sigma_{\rm{v}}$ velocity cut on the Y12 absorbers reduces the covering fraction from about 80\% at the completeness limit (\NHI $= 10^{13.3}$ cm$^{-2}$; see Figure~\ref{fig:cf}), to 72\% and 36\% respectively. At \NHI $= 10^{14}$ cm$^{-2}$, the covering fraction stays at 32\% for the $2\sigma_{\rm{v}}$ cut, but goes down to and 10\% at $1\sigma_{\rm{v}}$. These velocity cuts would bring the Y12 observations into much closer agreement to our Virgo-like cluster at \NHI $> 10^{14}$ cm$^{-2}$, but a disagreement by a factor of a few remains below \NHI $= 10^{14}$ cm$^{-2}$. One should keep in mind, however, that the Y12 observations are missing part of the velocity range of the Virgo cluster.

We also approach the discrepancy from the simulation side and investigate the possibility that the disagreement seen in Figures~\ref{fig:cf} and~\ref{fig:CDD observation} is due to the difference in observed and simulated cluster volume. Using H$_{\rm{o}}$ = 100 km s$^{-1}$, and ignoring peculiar velocities, the full Y12 velocity range corresponds to probing 23 Mpc of physical distance along the line of sight in Virgo. Our synthetic observations probe 5 \rvir\ along a line of sight, corresponding to 9.25 and 15.0 Mpc for the Virgo-like and Coma-like clusters respectively. We repeat our analysis using the same lines of sight in each cluster, but increasing their length to probe the entire high resolution region in each cluster, corresponding to 45 and 60 Mpc in the Virgo-like and Coma-like clusters respectively. As expected, the covering fraction and CDD increases with the larger observed volume, but the increase lies only for absorbers at \NHI $< 10^{13}$ cm$^{-2}$, by a factor of about 2.5 (f$_{\rm{cover}}$ = 0.33 for the Virgo-like cluster at \NHI $= 10^{12}$ cm$^{-2}$). Increasing the observed volume in the simulation (thereby including fore/background structures) does not appreciably bring the simulations closer into agreement with Y12. Ultimately, removing fore/background absorbers in observations presents a challenge in comparing between simulation and observation. Further observations will help clarify if the disagreement seen here is universal, or if it is due to peculiarities of the Virgo cluster.

\subsection{Absorber Metallicity}
\label{sec:absorber metallicity}
Since complete distance and kinematic information of the absorbing gas in unavailable in observations, absorber metallicity could be a valuable discriminator between the IGM and galactic origins of absorbers. Within the spectral range of the COS observations in Y12 and the ongoing Coma observations, the Si III [1206], Si II [1260], OI [1302], and CII [1334] lines can potentially be used to obtain an estimate of the absorber metallicity. The IGM at z = 0 has a low metallicity, Z $\la$ 0.1 Z$_{\odot}$ \citep{Danforth2014}, making it distinct from the typical metallicity of galactic outflows, Z $\ga$ 0.1 Z$_{\odot}$ up to Z $\sim$ Z$_{\odot}$ \citep{Creasey2015}. 

In our simulated galaxy clusters, we find an important distinction in absorber metallicity with both absorber column density and distance from the cluster center. In both clusters, almost all absorbers with column densities above about \NHI\ = 10$^{14}$ cm$^{-2}$ have Z $\ga$ 0.1 Z$_{\odot}$, and up to a few times solar, regardless of their location in the cluster environment. High column density absorbers, then, are gas processed through galaxies (galactic waste). Absorbers with column densities below 10$^{14}$ cm$^{-2}$ span the entire range of observed metallicities, from virtually pristine up to solar, yet most lie below 0.1 Z$_{\odot}$ with the greatest concentration at virtually pristine metallicities. Low column density gas is then easily associated with the IGM, yet some may still have been processed in galaxies.

Metallicity may also prove valuable in disentangling projection effects. In projection, absorbers with metallicities between galactic waste and pristine gas are only found outside a distance of about 1.5 \rvir.  Within this radius, absorbers appear to either be enriched or pristine (with very little middle ground).  Metallicity as a function of physical distance from the center of the simulated cluster provides a stark contrast.  We find that all absorbers within a physical distance of about 1.5 \rvir\ and 1.3 \rvir\ for the Virgo-like and Coma-like clusters respectively, have metallicities greater than Z $\sim$ 0.05 Z$_{\odot}$. It is only outside these radii that absorbers have low metallicity or are pristine. Therefore the infalling IGM absorbers do not penetrate inside the cluster virial radius without first being heated and/or ionized. Since the vast majority of the Virgo absorbers are low column density and no metal lines were detected, the majority are likely to lie beyond the virial radius as suggested by Y12.

\subsection{Velocity Distribution of Absorbers}
\label{sec:observed velocity}

The velocity distribution of the simulated absorbers in Figure~\ref{fig:vel disp} shows that most absorbers sit in front of or behind the cluster in velocity space.  Figures~\ref{fig:absorber velocities} and \ref{fig:radial velocity} show that the majority of the absorbers that appear in front of and behind the galaxy cluster in Figure~\ref{fig:vel disp} are in fact physically located on the opposite side of the cluster as they represent gas falling in along cosmic filaments. Unfortunately, we are unable to directly compare our velocity distributions to the Virgo cluster observations because the damping wings of the Milky Way render a large region of velocity space unobservable in the Virgo cluster (Y12). The observed distribution is truncated right around where a forward peak would lie (if it exists in the Virgo cluster) making it difficult to interpret the nature of the distribution. However, there is a comparative lack of absorbers at the Virgo cluster systemic velocity. The intuitive explanation for this trough is that, since a majority of absorbers avoid the hot ICM, there is a decrement in the central cluster regions in velocity space. However, since the redshift projection effects of these absorbers are dramatic, we examined the observed velocity distribution along two other, orthogonal viewing angles (`x' and `y') from our primary observations (along the `z' axis). These distributions are included in the bottom two rows of Figure~\ref{fig:vel disp}. For both clusters, all but one projection shows a bimodal distribution of absorber velocities. Given the bimodality is not universal, we conclude that the observed velocity distribution of absorbers in a galaxy cluster is biased by the unique nature of the infalling material, namely the location and size of infalling filamentary material relative to the observer. 

\subsection{Projected Distribution of Absorbers}
\begin{figure}
\centerline{\includegraphics[width=\columnwidth]{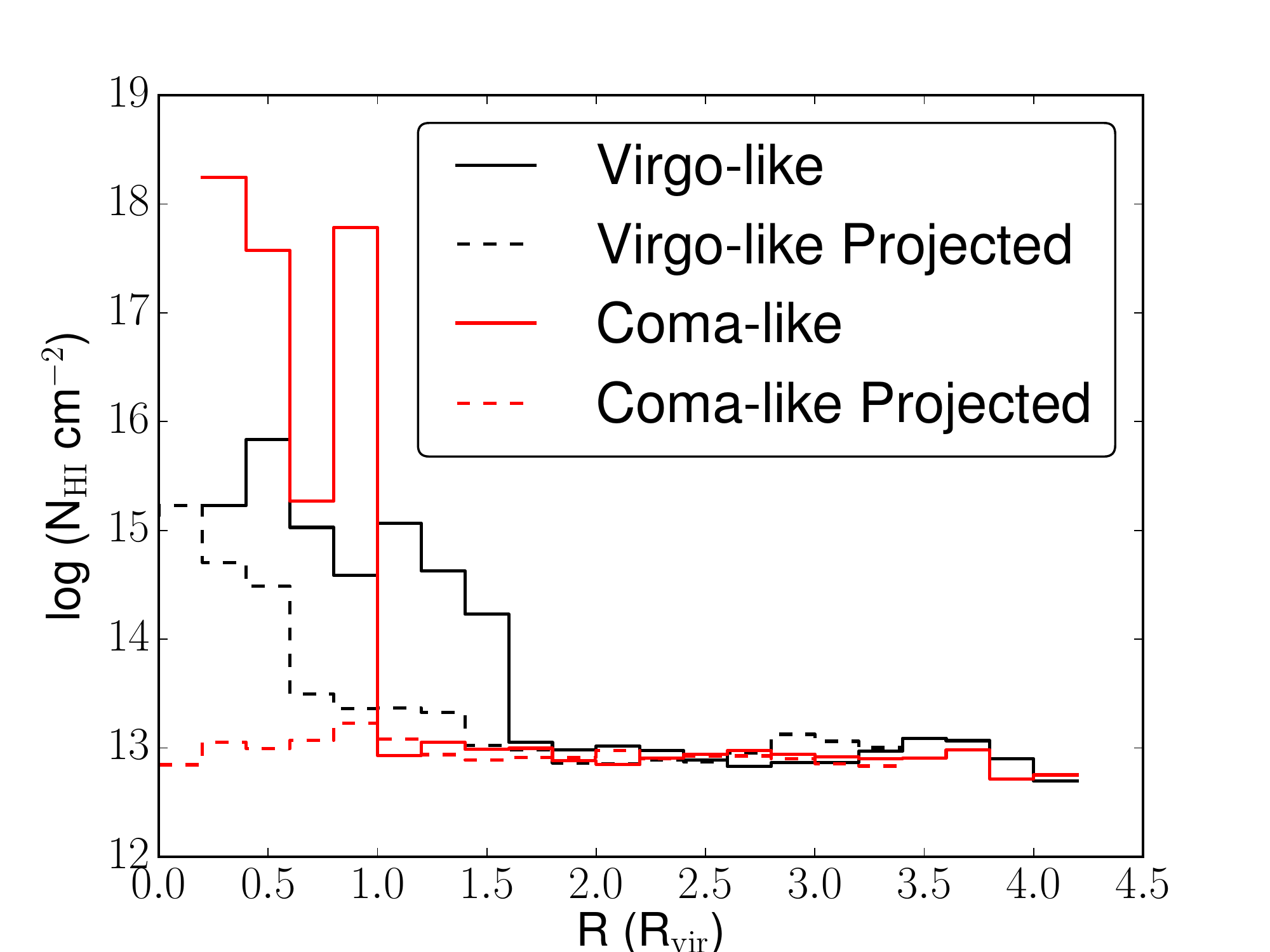}}
\caption{The median column density of absorbers as a function of radial (solid) and projected (dashed) distance from cluster center for the Virgo-like (black) and Coma-like (red) clusters. In each case, there is a negative trend with distance.}
\label{fig:N radial}
\end{figure}

Observations can examine the distribution of column density with projected distance from the cluster center. Y12 found tentative evidence for a positive correlation between absorber column density and projected distance from the cluster virial radius when absorbers associated with Virgo substructures were excluded. Our Virgo-like cluster does not show this behavior in projection (black, dashed) as shown in Figure~\ref{fig:N radial}, which gives the median absorber column density in radial bins; although we make no cuts to remove absorbers associated with substructures. The median column is greatest near the cluster center, at \NHI\ $> 10^{15}$ cm$^{-2}$, and decreases towards the cluster outskirts, remaining roughly constant at around $10^{13}$ cm$^{-2}$ at projected distances greater than 1.5 \rvir. The radial profile of the Virgo-like cluster (black, solid) shows the transition from cluster to outskirts more clearly, however, remaining near $10^{15}$ cm$^{-2}$ up to $R = 1.25$ \rvir, then dropping off quickly to \NHI\ $\sim 10^{13}$ cm$^{-2}$ at $R > 1.5$ \rvir. 

The Coma-like cluster radial profile (red, solid) is qualitatively similar to the Virgo-like cluster, yet has a significantly higher median column within \rvir\, dropping quickly to $10^{13}$ cm$^{-2}$ outside the virial radius. This is a dramatically sharper transition than is shown in the Virgo-like cluster. Interestingly, the projected profile of the Coma-like cluster is flat, at \NHI\ $\sim 10^{13}$ cm$^{-2}$, at all impact parameters. In projection, the few high column central absorbers in the Coma-like cluster are drowned out by the low column foreground/background absorbers.  This may be similar to what is observed for Virgo, as the distribution of column densities with impact parameter is relatively flat before the removal of sub-structure absorbers (Y12).

The Coma-like cluster stands in contrast to the observed relationship of \NHI\ with impact parameter for galaxies (e.g., \cite{Prochaska2011,Stocke2013,Tumlinson2013}). \NHI\ has been consistently found to increase with decreasing impact parameter for observed galaxies.  Although our Virgo-like cluster shows this relationship, the flat profile of the Coma-like cluster indicates there may be a change in this observed relationship for the most massive dark matter halos. This suggests there is some definite mass scale at which point the nature of gas accretion onto dark matter halos changes, and a majority of the accreted gas remains hot and unable to cool.  This will be interesting to test further with future observations and simulations of galaxy clusters.  

\section{Discussion}
\label{sec:discussion}
We discuss the implications of our study on the understanding of gas flows in and around galaxy clusters in Sec.~\ref{sec:gas flows}. The effects of resolution on studying the \lya\ absorbers in our clusters is discussed in Sec.~\ref{sec:resolution}. Finally, limitations of our simulations in capturing physical processes relevant to the formation and destruction of absorbers, including cluster specific processes, are summarized in Sec.~\ref{sec:uncertainties}.

\subsection{Gas Flows in Galaxy Clusters}
\label{sec:gas flows}
We find that warm, absorbing gas in the environment surrounding galaxy clusters is generally infalling in our simulations. The warm gas identified in absorption that is located within the virial radius of the cluster is most likely galactic in origin, generally having both a higher observed HI column density and a higher metallicity than an absorber located in the pristine, diffuse IGM. Outside of the virial radius, the total population is dominated by lower column density and metallicity absorbers. These are associated with the IGM filaments that are feeding the growth of our two simulated galaxy clusters. Like the absorbers within the virial radius of the cluster, these filamentary absorbers are fast moving, infalling material. However, the absorbers at large radii more clearly represent coherent gas flows from the IGM. 

Although we only examine two simulations in this study, we find that morphology and cluster dynamics are likely to affect the resulting presence and distribution of warm, neutral gas. Our more relaxed and massive Coma-like cluster exhibits noticeably less absorption within \rvir\ than the still merging Virgo-like cluster. As the Virgo-like cluster relaxes, it is likely that many of the galactic absorbers within the virial radius will reach thermal equilibrium with the hot ICM, relaxing to a state similar to that of the Coma-like cluster.  Due to its large mass, however, the Coma-like cluster has significantly more filamentary absorption than the Virgo-like cluster, cand these filaments have more physical volume and contained gas mass. Interestingly, the flat \NHI\ profile for the Coma-like cluster in Fig.~\ref{fig:N radial} suggests that there is a dark matter halo mass regime where the nature of warm, neutral infalling gas changes, when compared to the Virgo-like cluster, and similar observed profiles for \NHI\ around galaxies (e.g., \cite{Prochaska2011,Stocke2013,Tumlinson2013}).

\subsection{Resolution Study}
\label{sec:resolution}

The resolution in any simulation is important to consider in order to assess the validity of the results. In our case, too low a resolution could prevent proper accounting of high column density absorbers located in very overdense regions, and can spread diffuse absorbers too thin, reducing their observed column density (where again, N$_{\rm{HI}} \sim l n_{\rm{HI}}$ where $l$ is a given cell length, and $n_{\rm{HI}}$ is the neutral hydrogen number density). \cite{Luki2014} found a resolution of 20 $h^{-1}$kpc was necessary for 1\% convergence in the \lya\ forest at $z = 2$.  Over the region around each cluster probed by our observations, we have a minimum and maximum cell resolution of 7.3 and 233 kpc respectively. Averaging over all cells, the Virgo-like and Coma-like clusters have a resolution of 31 and 34 kpc, respectively.  However, a simple extrapolation like this is challenging as the gas over density giving rise to a given \NHI\ absorber is higher at low redshift, due to the evolution in the ionizing background \citep[e.g.,][]{Dave2010}.

It is computationally difficult to improve the resolution of our cluster simulations; however, we can test the resolution dependence of our results by examining the convergence of a smaller, uniform region.  In particular, we examine the column density distribution function f(N) in 3 unigrid simulations of a 20.0 $h^{-1}$ Mpc box using the same parameters and included physics as the cluster simulations described in Section~\ref{sec:computational methods}. Essentially, these simulations probe the $z = 0$ \lya\ forest in the IGM (sacrificing the higher column density absorbers which are due to dense gas which is not resolved in these unigrid calculations).  Although the grid in our cluster simulations is certainly more complex, with adaptive mesh refinement, these unigrid simulations, performed with 128$^{3}$, 256$^3$, and 512$^3$ grid cells and dark matter particles, provide a straightforward relationship between f(N) convergence and grid size in a computationally tractable manner.  To test the effect of AMR, we conduct a fourth AMR enabled simulation with an initial grid of 128$^{3}$. These simulations have a corresponding resolution at $z = 0$ of 232.9 kpc, 116.4 kpc, and 58.2 kpc, with the 128$^3$ AMR simulation reaching the same minimum and maximum resolution as our cluster runs (7.28 and 232.9 kpc), with an average resolution of 86 kpc. 

\begin{figure}
\centerline{\includegraphics[width=\columnwidth]{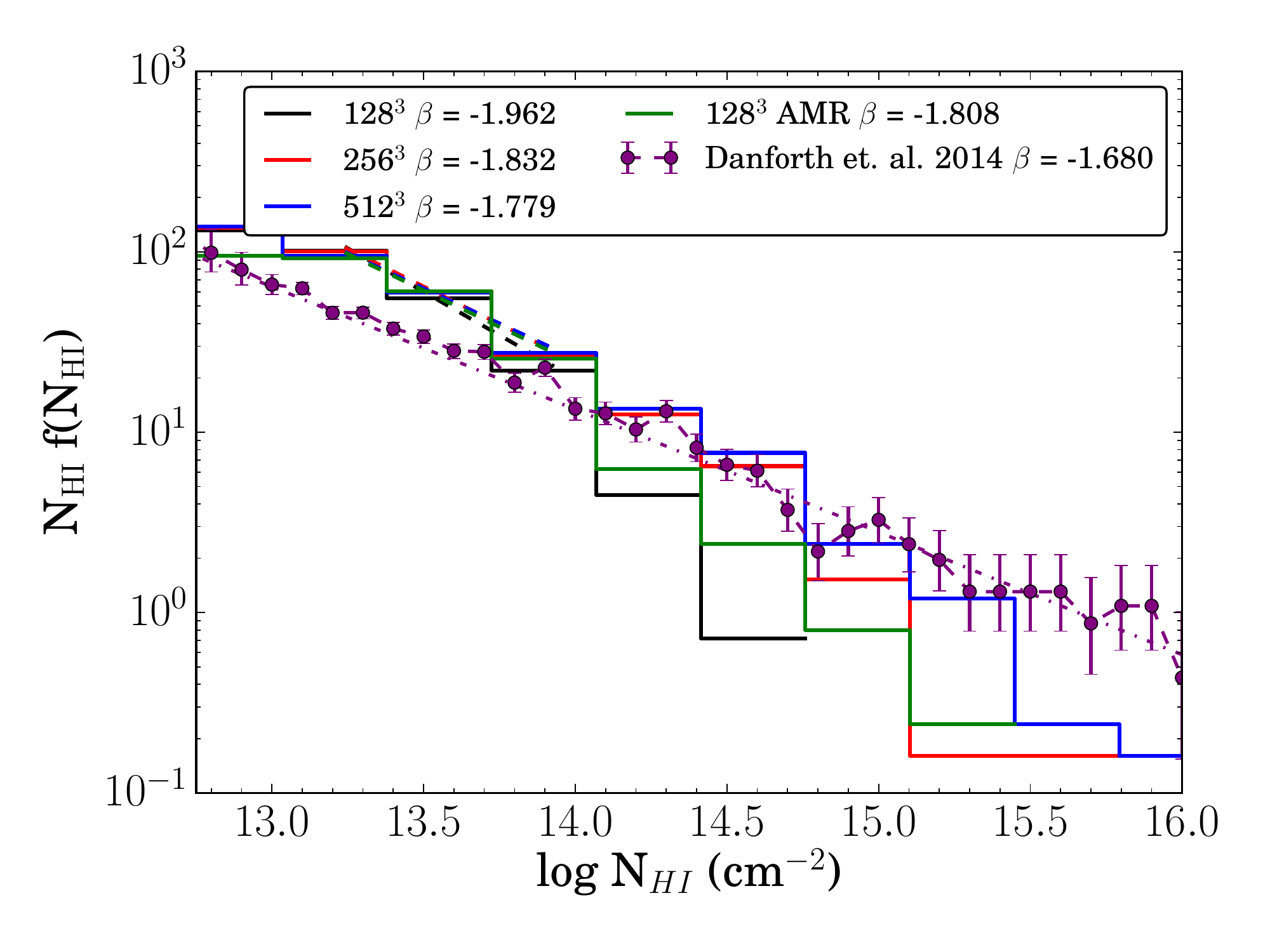}}
\caption{A resolution study of the column density distribution given as N$_{\rm{HI}}$ f(N$_{\rm{HI}}$) for our three unigrid simulations of a 20 $h^{-1}$Mpc box at 128$^3$ (black), 256$^3$ (red), and 512$^3$ (blue). Included also is the 128$^3$ run with  one level of AMR enabled (green). Observational results from \cite{Danforth2014} are shown in purple.}
\label{fig:resolution}
\end{figure}

In Figure~\ref{fig:resolution} we plot N$_{\rm{HI}}$f(N$_{\rm{HI}}$) for these simulations. As shown, the column density distribution is well converged for low column density absorbers (\NHI\ $< 10^{14}$ cm$^{-2}$) even at 128$^3$ (black) resolution (232.9 kpc grid size). However, the distribution is not well resolved at \NHI\ $\ga 10^{14}$ cm$^{-2}$ at this resolution, and is only converged up to \NHI\ $\sim 10^{14.5}$ at 512$^3$ (blue). At column densities above this point, the fixed grid resolution of 58.2 kpc is insufficient to completely capture these absorbers, and the CDD begins to deviate from a single power law. Fortunately, in our cluster simulations, we are not restricted to a fixed, uniform grid. 

For the Virgo-like and Coma-like clusters, 77.8\% and 67.2\% of the probed volume has resolution equal to or better than the 256$^3$ simulation. Likewise, 31.6\% and 21.7\% of each cluster has resolution at or better than the 512$^{3}$ simulation. Only overdense regions refined by the AMR in the cluster simulations reach resolution better than the 512$^{3}$ run, or 5.8\% and 3.9\% of the total cluster volume. Our cluster simulations are thus well converged for absorbers at \NHI\ $< 10^{14.0}$ cm$^{-2}$, and are fairly well converged at \NHI\ $< 10^{14.7}$ cm$^{-2}$. However, the results (like the 512$^3$ run) are probably not well converged at higher column densities of \NHI\ $> 10^{15}$ cm$^{-2}$. Most of the observed Virgo absorbers are below this column density and so overall comparisons between the simulations and observations will not change with resolution. However, the specific column densities of the high column, galactic absorbers may vary somewhat with increased resolution runs.

\subsection{Simulation Uncertainties}
\label{sec:uncertainties}
Although resolution is likely the largest source of uncertainty for high column density absorbers, it is not the dominant source of uncertainty at modest column densities as shown by the convergence in Sec.~\ref{sec:resolution}. We discuss additional uncertainties and missing physics in our simulations that could play a role in creating or destroying warm neutral gas that would be visible as \lya\ absorption. 

\subsubsection{Ionizing Background}
\label{sec:ionizing background}
The adopted ionizing background plays a significant role in establishing the neutral fraction at all redshifts. The ionization rate, $\Gamma$, is well constrained at high redshifts (z $>$ 2), with excellent agreement between the observed and simulated \lya\ forest \citep[e.g.,][]{Croft2002, Viel2005, Seljak2005}. However, the low redshift \lya\ forest is poorly constrained, and the z = 0 forest especially, is not well understood. Although pre-COS observations have placed constraints on the low redshift IGM \citep[][and references therein]{Dave2010}, COS has only recently allowed for the largest sample of low redshift \lya\ observations \citep{Danforth2014}. There are many simulations of the high redshift \lya\ forest, yet only a few works examine the forest in the local Universe \citep{Paschos2009,Dave2010,Egan2014,Juna}. Most of these employ the \cite{HM01} ionizing background and suggest fair agreement with observation, although generally underreporting the HI column density distribution by a factor of a few. To date, only one work has examined the low redshift \lya\ forest using the updated \cite{HM12} background, \cite{Juna}, which strongly suggests that the updated background results in \textit{too much} HI in the IGM at $z = 0$ at low to moderate column densities. This suggests a significant uncertainty in the sources of ionizing background radiation at low redshift, which could substantially change the neutral HI content in our simulations.

As shown in Figure~\ref{fig:resolution}, our highest resolution unigrid simulation, 512$^3$ (blue), agrees qualitatively with \cite{Juna}, with an overproduction of HI compared to the observations in \cite{Danforth2014} (purple), at least at low column densities. At \NHI\ $> 10^{14.5}$ cm$^{-2}$, our IGM simulation starts to drop below the observations, where resolution begins to have an effect. 

Although we only ran our cluster simulations with the HM12 background, we can analytically estimate how a change in the background would affect the  observed \NHI. Since the gas studied here has a large ionization fraction, we can assume ionization equilibrium. Considering only electrons and Hydrogen, the Hydrogen neutral fraction, and thus the HI column densities, will scale inversely with the choice of background ionization rate ($\Gamma$) (see \cite{Rauchlya} for a complete summary of this argument). For HM01, $\Gamma_{\rm{HI}}$ is roughly a factor of 5 higher at z = 0 than in HM12. Using the HM01 background instead of HM12, then, would reduce the neutral fraction by a factor of 5, effectively shifting the lines in Figure~\ref{fig:resolution} to the left.

\subsubsection{Galaxy Outflows}
\label{sec:outflows}
It is currently computationally prohibitive to run cosmological simulations of galaxy clusters to high enough resolution to completely capture the galaxy physics responsible for substantial outflows (indeed, this is challenging even for zoom calculations of individual galaxies). Relevant physics can include a multi-phase ISM, detailed effects of ram-pressure stripping of the ISM, star formation feedback, and AGN \citep[e.g.,][]{Springel2003, Bower2006, Governato2007}. However, with a maximum resolution of several kpc, the galactic outflow processes must be addressed with subgrid physics models.  As mentioned in Sec.~\ref{sec:enzo}, our simulations include star formation, which contains the ``distributed feedback" model for star formation and supernova feedback discussed in ~\cite{Smith2011}. Although this model still suffers from overcooling and hence overproduction of stars and metals, in general ~\cite{Smith2011} shows reasonable agreement with observations of the distribution of metals in the IGM. 

Although the influence of various feedback models on the \lya\ forest in \Enzo~simulations has not been tested, ~\cite{Dave2010} discusses the effects of various galactic wind models on their SPH simulations of the low redshift IGM. Their preferred wind model, a momentum driven wind, produces a distribution of low redshift \lya\ absorbers consistent with the then-available observations. In general, they found that turning on their wind model had no significant effect on low column density absorbers, those with \NHI\ $< 10^{14}$ cm$^{-3}$. The differences between wind models became significant for higher column density absorbers, especially for lines of sight with low galactic impact parameters \citep{Ford2013}. It is difficult to judge how much the cluster environment would modify these results obtained for the IGM. If much of the observed HI absorption in the Virgo cluster is galactic in origin, it is likely that feedback processes and the effects of the cluster environment on galaxies must be examined further as possible additional sources of warm HI. As demonstrated in \cite{Dave2010} and \cite{Ford2013}, however, this would likely only serve to increase the number of absorbers observed at \NHI\ $> 10^{14}$ cm$^{-2}$, and therefore would not be able to fully explain the differences between our simulations and the Virgo observations.

\subsubsection{AGN and Cool-core Clusters}
\label{sec:agn coolcore}
Finally, we note that the Virgo cluster is a cool-core cluster that exhibits a drop in ICM temperature towards cluster center, and strong signs of active AGN feedback. The Coma cluster does not have significant current AGN jet heating and is not a cool core cluster. The effect of cool-cores on the presence of warm gas within the ICM of the cluster is unknown, yet cluster properties may be dependent on the presence/absence of a cool-core \citep[e.g.,][]{Frank2013}. The physics required to generate a cool-core cluster in our simulations is not included, and is currently computationally prohibitive to employ in detail in cosmological simulations, although the effects of AGN in clusters has been studied before with \Enzo\ \citep{LiBryan2014}.  Inclusion of AGN on cool and warm diffuse gas is likely only to be significant (if at all) in the very central regions of the cluster, well within \rvir. Our ongoing observations of the Coma cluster will be one way to address observationally the comparative differences between warm gas in cool core and non-cool core clusters. 

\section{Summary}
The flow of gas into and throughout galaxy clusters is fundamental to their evolution. Warm gas traces gas flow across cluster scales, between galaxies and the ICM, and between the cluster environment and the IGM. Our work is the first systematic study of the warm gas content in simulated galaxy clusters in the context of \lya\ absorption.  These simulations provide a new understanding of the spatial and kinematic properties of the absorbing gas, as well as insight into their environment. Utilizing synthetic observations to carry out this study allows us to make direct comparisons to previous observations of the Virgo cluster, and to establish what we expect from future cluster observations. We summarize the main results of this work below.

\begin{enumerate}

\item In each of our clusters, the warm and cold gas mass fraction is small within \rvir\ (less than a few percent) but increases substantially (to about 30\% at R $>$ \rvir) transitioning away from the hot cluster environment into the IGM.  The population of absorbing material traces this gas, generally avoiding the hot cluster center (with the exception of some high column density absorbers, which correspond to galactic waste). In particular, we show that the numerous low column density absorbers trace the filaments of the cosmic web feeding gas into the cluster.

\item Low column density absorbers dominate the total absorber counts in each cluster, and dominate the number density of absorbers in each cluster outside 1 \rvir\ and 2 \rvir\ for the Coma-like and Virgo-like clusters respectively. The increased number density (in units of \rvir$^{-3}$) of low column absorbers in the Coma-like cluster indicates greater filamentary feeding in the more massive, more relaxed of our two clusters.

\item The absorbers in each of our simulated galaxy clusters are generally fast moving and infalling for the first time into the cluster, with peculiar velocities sufficiently larger than their Hubble velocity such that they are redshifted to the opposite side of the cluster central velocity.  This is consistent with the idea that absorbers trace infalling filaments which do not survive first infall.

\item In general, the observed velocity distribution of absorbers is bi-modal, with two peaks separated by a trough centered roughly on the cluster systemic velocity. The relative size of these peaks, and the width of the trough, is determined by projection effects. 

\item When compared to the Virgo cluster observations of Y12, both of our simulated clusters show less than the observed warm HI content. In terms of covering fraction, the disagreement is a factor of several, with f$_{\rm{cover}}$ $\sim$ 0.4 for Virgo and f$_{\rm{cover}}$ $\sim$ 0.05 for our simulations at \NHI\ = 10$^{14}$ cm$^{-2}$. It remains to be seen whether the \lya\ absorbing warm gas in the Virgo cluster is representative of  galaxy clusters in the nearby universe. Future observations will be important for further interpreting this disagreement.

\item Metallicity is an important discriminator of absorber origin. High column absorbers have metallicities characteristic of gas processed in galaxies, and low column absorbers are predominately pristine. No low metallicity absorbers are found within the cluster virial radius, suggesting that the infalling IGM absorbers heat up or ionize quickly, even before entering the cluster environment.

\item A resolution study indicates that our low-column density population \NHI\ $< 10^{14.5}$ cm$^{-2}$ is reasonably well-converged, but that we probably systematically under predict higher column-density absorbers. However, this is not enough to explain the above difference in f$_{\rm{cover}}$ between Virgo and our simulations.

\end{enumerate}

This work provides important insight into cluster warm gas content. Cluster \lya\  absorbers correspond to fast moving, filamentary infall, which is enhanced in more massive clusters. The presence/absence of absorbers within \rvir~ is dependent upon cluster morphology, and we expect dynamically young clusters to have enhanced absorption within \rvir. These simulations can be used to examine the correspondence between galactic absorbers and the cluster galaxies from which they came. In addition, higher resolution versions of these simulations can be used in detail to trace the origin and fate of the identified absorbers, examining the lifetimes of absorbers with different origins (filamentary vs. galactic). Future cluster observations, including the ongoing COS Coma observations, will be important to understanding how absorbers vary with cluster mass and morphology over time.

\section*{Acknowledgements}
We would like to thank Juna Kollmeier, Romeel Dav$\rm{\acute{e}}$, Joo Heon Yoon and David Weinberg for their insightful discussions, and Charles Danforth for sharing his data. 
A.E. is supported by a National Science Foundation Graduate Research Fellowship under Grant No. DGE-11-44155. 
G.L.B. acknowledges support from NASA grant NNX12AH41G and NSF-1008134, NSF-1210890, and NSF-1312888.  We gratefully recognize computational resources provided by NSF XSEDE and Columbia University. M.E.P acknowledges support from HST-GO-13382 and the Clare Boothe Luce Program. 
Computations described in this work were performed using the publicly-available \texttt{Enzo} code (http://enzo-project.org), and analyzed using \textit{yt} (http://yt-project.org). \texttt{Enzo} and \textit{yt} are the products of a collaborative effort of many independent scientists from numerous institutions around the world.  Their commitment to open science has helped make this work possible.  

\bibliographystyle{mn2e}
\bibliography{ms}

\label{lastpage}

\end{document}